\documentclass[aps,prd,twocolumn,showpacs,floatfix,nofootinbib,]{revtex4}

\usepackage{graphics}
\usepackage{graphicx}

\usepackage{latexsym}
\usepackage[cp866]{inputenc}
\usepackage[english]{babel} %%% ,russian

\begin{document}

\title{Light Scalar Mesons in Photon-Photon Collisions}
\author{N.N. Achasov\footnote{E-mail: achasov@math.nsc.ru}
and G.N. Shestakov\footnote{E-mail: shestako@math.nsc.ru}}
\affiliation{Laboratory of Theoretical Physics, S.L. Sobolev
Institute for Mathematics, 630090, Novosibirsk, Russia}

%\date{\today}

\begin{abstract} %{\it
The light scalar mesons, discovered over forty years ago, became a
challenge for the naive quark-antiquark model from the outset. At
present the nontrivial nature of these states is no longer denied
practically anybody. Two-photon physics has made a substantial
contribution to understanding the nature of the light scalar mesons.
Recently, it entered a new stage of high statistics measurements. We
review the results concerning two-photon production mechanisms of
the light scalars, based on the analysis of current experimental
data. %}
\end{abstract}

\pacs{12.39.-x, 13.40.-f, 13.60.Le, 13.75.Lb}

\maketitle

\noindent {\bf Outline} {\small
\begin{description}
\item{1.} Introduction.
\item{2.} Special place of the light scalar mesons in the hadron
world. Evidences of their four-quark structure.
\item{3.} Light scalar mesons in the light of photon-photon
collisions.\\
3.1. History of investigations. 3.2. Current experimental situation.
3.3. Dynamics of the reactions $\gamma\gamma\to\pi\pi$: Born
contributions and angular distributions. 3.4. Production mechanisms
of scalar resonances.
\item{4.}
Analysis of high statistics Belle data on the reactions $\gamma
\gamma\to\pi^+\pi^-$ and $\gamma\gamma\to\pi^0\pi^0$. Manifestations
of the $\sigma(600)$ and $f_0(980)$ resonances.
\item{5.}
Production of the $a_0(980)$ resonance in the reaction $\gamma\gamma
\to\pi^0\eta$.
\item{6.} Preliminary summary.
\item{7.} Future Trends.\\
7.1 The $f_0(980)$ and $a_0(980)$ resonances near $\gamma\gamma\to
K^+K^-$ and $\gamma\gamma\to K^0\bar K^0$ reaction thresholds. 7.2.
The $\sigma(600)$, $f_0(980)$, and $a_0(980)$ resonances in
$\gamma\gamma^*$ collisions. 7.3. Searches for the $J/\psi\to\omega
f_0(980)$ and $J/\psi\to\rho a_0(980)$ decays. 7.4 Inelasticity of
$\pi\pi$ scattering and $f_0(980)-a_0(980)$ mixing.
\item{8.} Appendix\\
8.1. $\gamma\gamma\to\pi\pi$. 8.2. $\gamma\gamma\to\pi^0\eta$. 8.3.
$\gamma\gamma\to K\bar K$.
\item{References.}
\end{description}}

%\section{Введение}
\vspace{0.3cm} \noindent{\large \bf 1. Introduction}\vspace{0.2cm}

\noindent The scalar channels in the region up to 1 GeV became a
stumbling block of QCD because both perturbation theory and sum
rules do not work in these channels. \,\footnote{The point is that,
in contrast to classic vector channels, in this region there are not
solitary resonances, i.e., scalar resonances, which are not
accompanied by a large inseparable from resonance background.
Particularly, in the case of the solitary  $a_0(980)$ and $f_0(980)$
resonances, the resonance peaks in the $\phi$\,$\to$\,$\gamma
a_0(980)$\,$\to$\,$\gamma\pi\eta$ and $\phi$\,$\to$\,$\gamma
f_0(980)$\,$\to$\,$\gamma\pi\pi$ decays would be not observed at all
because the differential probabilities of these decays vanish
proportionally cubic function of the photon energy in a soft photon
region for gauge invariance \cite{AI89, AG01, AG02YF,Ac03a,Ac04YF},
see Section 2. The principal role of the chiral background in the
fate of the $\sigma(600)$ resonance was demonstrated in the linear
$\sigma$ model \cite{AS94a,AS93,AS94a1,AS07}. The solitary resonance
approximation is nothing more than an academic exercise in the light
scalar meson case.} At the same time the question on the nature of
the light scalar mesons, $\sigma(600)$, $\kappa(800)$, $a_0(980)$,
and $f_0(980)$ \cite{PDG08,PDG10}, is major for understanding the
mechanism of the chiral symmetry realization, arising from the
confinement, and hence for understanding the confinement itself.

Hunting the light  $\sigma$ and $\kappa$ mesons had begun in the
sixties already and a preliminary information on the light scalar
mesons in Particle Data Group (PDG) reviews had appeared at that
time (see, for example, \cite{PDG65,PDG67,PDG69}). The theoretical
ground for a search for scalar mesons was the linear $\sigma$ model
(LSM) \cite{GL60,Ge64,Le67}, which takes into account spontaneous
breaking of chiral symmetry and contains pseudoscalar mesons as
Goldstone bosons. The surprising thing is that after ten years it
has been made clear that LSM could be the low energy realization of
QCD. At the end of the sixties and at the beginning of the seventies
\cite{PDG67,PDG71,PDG73} there were discovered the narrow light
scalar resonances, the isovector $a_0(980)$ and isoscalar
$f_0(980)$.\,\footnote{In 1977 Jaffe noted that in the MIT bag
model, which incorporates confinement phenomenologically, there
exists the nonet of the light scalar four-quark states \cite{Ja77}.
He suggested also that $a_0(980)$ and $f_0(980)$ might be these
states with symbolic structures: $a^+_0(980)=u\bar{d}s\bar{s}$,
$a^0_0(980)=(us\bar u\bar s-ds\bar d\bar s)/\sqrt{2}$, $a^-_0(980)=
d\bar{u}s\bar{s}$, and $f_0(980)=(us\bar u\bar s + ds\bar d\bar
s)/\sqrt{2}$. From that time $a_0(980)$ and $f_0(980)$ resonances
came into beloved children of the light quark spectroscopy.}

As for the $\sigma$ and $\kappa$ mesons, long-standing unsuccessful
attempts to prove their existence in a conclusive way entailed
general disappointment and an information on these states
disappeared from PDG reviews. One of principal reasons against the
$\sigma$ and $\kappa$ mesons was the fact that the $S$ wave phase
shifts, both $\pi\pi$ and $\pi K$ scattering, do not pass over
$90^0$ at putative resonance masses. Nevertheless, experimental and
theoretical investigation of processes, in which the $\sigma$ and
$\kappa$ states could reveal themselves, had been continued.

Situation changes when we showed \cite{AS94a} that in LSM there is a
negative background phase in the $\pi\pi$ scattering $S$ wave
amplitude with isospin $I$\,=\,0, which hides the $\sigma$ meson
with the result that the $\pi\pi$ $S$ wave phase shift does not pass
over $90^0$ at a putative resonance mass.
%----------------------------------------------------------------------------------------
It has been made clear that shielding wide lightest scalar mesons in
chiral dynamics is very natural. This idea was picked up and
triggered new wave of theoretical and experimental searches for the
$\sigma$ and $\kappa$ mesons, see, for example,
\cite{SS95,To95,Is96,HSS96,Is97,Bl98,Bl99,Is00}. As a result the
light $\sigma$ resonance, since 1996, and the light $\kappa$
resonance, since 2004, appeared in the PDG reviews
\cite{PDG96,PDG04}.

By now there is an impressive amount of data about the light scalar
mesons \cite{PDG08,PDG10,STA08,Am10}. The nontrivial nature of these
states is no longer denied practically anybody. In particular, there
exist numerous evidences in favour of their four-quark structure.
These evidences are widely covered in the literature
\cite{AI89,Ac03a,STA08,Am10,Mo83,ADS84a,Ac90,Ac98,Ac99,Ac02,Ac01,
Ac03b,Ac03c,Ac04a,Ac04b,Ac06,Ac08a,Ac08b,Ac09Bog,AS91,AS99,DS98,GN99,Tu01,
Tu03,CT02,AJ03,JW03,AT04,Ma04,Ja05,Ja07,Ka05,AKS06,AKS08,CCL06,Bu06,
FJS06,FJS08a,FJS08b,FJS09,Na06,Na08,To07,KZ07,MPR07,Pe07a,Pe07b,vBR07,
By08,Le08,tH08,IK08,EFG09,AS09China,AS-Q10}. They are presented also
in Sections 2--6.

One of them is the suppression of the $a_0(980)$ and $f_0(980)$
resonances in the $\gamma\gamma$\,$\to$\,$\pi^0\eta$ and
$\gamma\gamma$\,$\to$\,$\pi\pi$ reactions, respectively, predicted
in 1982 \cite{ADS82a,ADS82b} and confirmed by experiment
\cite{PDG08,PDG10}. The elucidations of the mechanisms of the
$\sigma(600)$, $f_0(980)$, and $a_0(980)$ resonance production in
the $\gamma\gamma$ collision and their quark structure are
intimately related. That is why the studies of the two-photon
processes are the important part of the light scalar meson physics.

\begin{figure}
\includegraphics[width=38mm]
{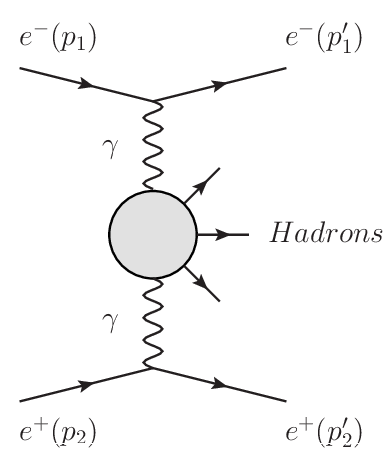}\caption{{\footnotesize The two-photon process of
hadron formation at $e^+e^-$ colliders; $p_1$, $p'_1$ and $p_2$,
$p'_2$ are the 4-momenta of electrons and positrons.
}}\label{ee-eeh}\end{figure}

It should be noted that the reactions of hadron production in
photon-photon collisions are measured at $e^+e^-$ colliders, i.e.,
the information on the transitions $\gamma\gamma$\,$\to$\,{\it
hadrons} is extracted from the data on the processes $e^+e^-$\,$\to
$\,$e^+e^-\gamma\gamma$\,$\to $\,$e^+e^-$\,{\it hadrons} (Fig.
\ref{ee-eeh}). The most statistics is obtained by the so-called
``non tag'' method when hadrons only are detected and the scattered
leptons are not. In this case the main contribution to the cross
section of $e^+e^-$\,$\to$\,$e^+e^-$\,{\it hadrons} is provided by
photons with very small virtualities. Therefore, this method allows
to extract data on hadron production in collisions of almost real
photons. The absolute majority of data on the inclusive channels
$\gamma\gamma$\,$\to$\,{\it hadrons} has been obtained with the use
of this method. If the scattered electrons are detected (which leads
to a loss of statistics), then one can investigate in addition the
$Q^2$ dependence of the hadron production cross sections in
$\gamma\gamma^*(Q^2)$ collisions, where $\gamma$ is a real photon
and $\gamma^*(Q^2)$ is a photon with virtuality $Q^2=(p_1-p'_1)^2
$.\,\footnote{Detailed formulae for experimental investigations of
the reactions $e^+e^-$\,$\to$\,$e^+e^- $\,{\it hadrons} may be found
in the reviews \cite{BGMS75,Ko84}.}

Recently a qualitative leap had place in the experimental
investigations of the $\gamma\gamma$\,$\to$\,$\pi\pi$ and $\gamma
\gamma$\,$\to$\,$\pi^0\eta$ processes \cite{Mo03,Mo07a,Mo07b,
Ue08,Ue09} that proved the theoretical expectations based on the
four-quark nature of the light scalar mesons \cite{ADS82a,ADS82b}.
The Belle Collaboration published the data on the cross sections for
the $\gamma\gamma$\,$\to$\,$\pi^+\pi^-$ \cite{Mo07a,Mo07b},
$\gamma\gamma$\,$\to$\,$\pi^0\pi^0$ \cite{Ue08}, and
$\gamma\gamma$\,$\to$\,$\pi^0\eta$ \cite{Ue09} reactions, statistics
of which are hundreds of times as large as statistics of all
previous data. The Belle Collaboration observed for the first time
the clear signals of the $f_0(980)$ resonance in the both charge
channels. The previous indications for the $f_0(980)$ production in
the $\gamma\gamma$ collisions \cite{Ma90,Bo90,Oe90,Be92,Bi92,
Bara00,Bra00} were rather indefinite.

In the given paper there are presented the results of the
investigation the mechanisms of the $\gamma\gamma$\,$\to$\,$\pi^+
\pi^-$, $\gamma\gamma$\,$\to$\,$\pi^0\pi^0$, and $\gamma\gamma$\,$
\to$\,$\pi^0 \eta$ reactions (see Sections 3--5) based on the
analysis \cite{AS05,AS07,AS08a,AS08b,AS09,AS10a,AS10b} of the Belle
data \cite{Mo03,Mo07a,Mo07b,Ue08,Ue09} and our previous
investigations of the scalar meson physics in the $\gamma\gamma$
collisions \cite{ADS82a,ADS82b,ADS85,AS88,AS92,AS94b,AS91}. We also
briefly (sometimes critical) survey analyses of other authors.

The joint analysis of the Belle high-statistics data on the
$\gamma\gamma$\,$\to$\,$\pi^+\pi^-$ and $\gamma\gamma$\,$\to$\,$
\pi^0\pi^0$ reactions is presented and the principal dynamical
mechanisms of these processes are elucidated in the energy region up
to 1.5 GeV. The analysis of the Belle high-statistics data on the
reaction $\gamma\gamma$\,$\to$\,$\pi^0\eta$ is presented too. It is
shown that the two-photon decays of the light scalar resonances are
the four-quark transitions caused by the rescatterings
$\sigma$\,$\to $\,$\pi^+\pi^-$\,$\to $\,$\gamma\gamma$,
$f_0(980)$\,$\to $\,$(K^+K^-+\pi^+\pi^-)$\,$\to $\,$\gamma\gamma$,
and $a_0(980)$\,$\to $\,$(K\bar K+\pi^0\eta+\pi^0\eta')$\,$\to
$\,$\gamma\gamma$ in contrast to the two-photon decays of the
classic $P$ wave tensor $q\bar q$ mesons $a_2(1320)$, $f_2(1270)$
and $f'_2(1525)$, which are caused by the direct two-quark
transitions $q\bar q$\,$\to$\,$\gamma\gamma$ in the main. As for the
direct coupling constants of the $\sigma(600)$, $f_0(980)$, and
$a_0(980)$ resonances with the $\gamma\gamma$ system, they are
small. It is obtained the two-photon widths averaged over resonance
mass distributions $\langle\Gamma_{f_0\to
\gamma\gamma}\rangle_{\pi\pi}$\,$ \approx $\,0.19 keV, $\langle
\Gamma_{a_0\to\gamma\gamma}\rangle_{\pi\eta}$\,$\approx $\,0.4 keV
and $\langle\Gamma_{\sigma \to\gamma\gamma}\rangle_{\pi\pi}$\,$
\approx$\,0.45 keV.

In Section 7, we attend to the additional possibilities of the
investigation of the $a_0(980)$ and $f_0(980)$ resonances in the
reactions $\gamma\gamma$\,$\to$\,$K^+K^-$ and $\gamma\gamma$\,$\to
$\,$K^0\bar K^0$, which are as yet little studied experimentally,
and also to the promising possibility of investigating the nature of
the light scalars $\sigma(600)$, $f_0(980)$, and $a_0(980)$ in
$\gamma\gamma^*(Q^2)$ collisions.

\vspace{0.4cm} \noindent{\large \bf 2. Special place of the light
scalar mesons in the hadron world. Evidences of their four-quark
structure}\vspace{0.2cm}

%----------------------------------------------------------------------------------------

\noindent Even a cursory examination of PDG reviews gives an idea of
the four-quark structure of the light scalar meson nonet\,\footnote{
To be on the safe side, notice that the linear $\sigma$ model does
not contradict to non-$q\bar q$ nature of the low lying scalars
because Quantum Fields can contain different virtual particles in
different regions of virtuality.}, $\sigma(600)$, $\kappa(800)$,
$a_0(980)$, and $f_0(980)$,
\begin{equation}
\begin{array}{rrcll}
a_0^-& &a_0^0/f_0& &a_0^+ \\
& & & &  \\
& \{\kappa\}& &\{\kappa\} &  \\
& & & &  \\
& & \sigma & &
\end{array}
\label{4q-nonet}\end{equation} inverted \cite{AS99} in comparison
with the classical $P$ wave $q\bar q$ tensor meson nonet
$f_2(1270)$, $a_2(1320)$, $K_2^\ast(1420)$, and $f_2^\prime (1525)$
\begin{equation}
\begin{array}{rrcll}
& &f'_2& & \\
& & & & \\
&\{K^*_2\}& &\{K^*_2\}& \\
& & & & \\
\ a^-_2& &a^0_2/f_2& & a^+_2\ , \\
\end{array}
\label{2q-nonet}\end{equation} \noindent or also in comparison with
the classical $S$ wave vector meson nonet $\rho(770)$,
$\omega(782)$, $K^*(892)$, and $\phi(1020)$.\,\footnote{In Eqs.
(\ref{4q-nonet}) and (\ref{2q-nonet}) the mass and isotopic spin
third component of states increase bottom-up and from left to right,
respectively.} In the naive quark model such a nonet cannot be
understood  as the $P$ wave $q\bar q$ nonet, but it can be easy
understood as the $S$ wave $q^2\bar q^2$ nonet, where $\sigma(600)$
has no strange quarks, $\kappa(800)$ has the $s$ quark, $a_0(980)$
and $f_0(980)$ have the $s\bar s$ pair.

The scalar mesons $a_0(980)$ and $f_0(980)$, discovered about forty
years ago, became the hard problem for the naive $q\bar q$ model
from the outset.\,\footnote{Note here a series of important
experiments of seventies in which the $f_0(980)$ and $a_0(980)$
resonances were investigated \cite{Fl72,Pr73,Hy73,Gr74,Hy75,Ga76},
as well as a few theoretical analyses of scalar meson properties
relevant to this period \cite{Mo74,Fl76,MOS77,Pe77,Ja77,Es79,
ADS79}. In the last-named paper there was theoretically discovered
the fine threshold phenomenon of the $a_0(980)-f_0(980)$ mixing
which breaks the isotopic invariance (see also \cite{ADS81b}). Now a
rebirth of interest in the $a_0(980)-f_0(980)$ mixing takes place
and there appear new suggestions on search for this phenomenon (see,
for example, \cite{AS04a,AS04b,WZZ07,WZ08} and references in these
papers) as well as the first indications for its manifestation in
the $f_1(1285)$\,$\to$\,$\pi^+\pi^-\pi^0$ decay, which is measured
with the help of the VES detecor at IHEP in Protvino
\cite{Do07,Dor08,Ni09}, and in the decays $J/\psi\to\phi
f_0(980)\to\phi a_0(980)\to\phi\eta\pi$ and $\chi_{c1}\to\pi^0
a_0(980)\to\pi^0 f_0(980)\to\pi^+\pi^-\pi^0$, which are being
investigated with the BESIII detector at BEPCII in Chine
\cite{Har10}.} Really, on the one hand the almost exact degeneration
of the masses of the isovector $a_0(980)$ and isoscalar $f_0(980)$
states revealed seemingly the structure $a^+_0(980)$\,=\,$u\bar{d}$,
$a^0_0(980)$\,=\,$(u\bar{u}$\,-\,$ d\bar{d})/\sqrt{2}$,
$a^-_0(980)$\,=\,$d\bar{u}$ and $f_0(980)$\,=\,$(u\bar{u}$\,+\,$d
\bar{d})/\sqrt{2}$ similar to the structure of the vector $\rho$ and
$\omega$  or tensor $a_2(1320)$ and $f_2(1270)$ mesons, but on the
other hand, the strong coupling of the $f_0(980)$ with the $K\bar K$
channel as if suggested a considerable part of the strange pair
$s\bar s$ in the wave function of the $f_0(980)$.

At the beginning of eighty it was demonstrated  in a series of
papers \cite{ADS80a,ADS80b,ADS81a,ADS81b, ADS81c,ADS84a,ADS84b} that
data on the $f_0(980)$ and $a_0(980)$ resonances, available at that
time, can be interpreted in favour of the $q^2\bar q^2$ model, i.e.,
can be explained by using coupling constants of the $f_0(980)$ and
$a_0(980)$ states with pseudoscalar mesons superallowed by the
Okubo-Zweig-Iizuka (OZI) rule as it is predicted by the $q^2\bar
q^2$ model. In particular, in these papers there were obtained and
specified formulae for scalar resonance propagators with taking into
account corrections for finite width in case of strong coupling with
two-particle decay channels. Late on, these formulae were used in
fitting data of a series of experiments on the $f_0(980)$ and
$a_0(980)$ resonance production (see, for example,
\cite{Ach98a,Ach98b,Ach00a,Ach00b,Akh99a,Akh99b,Al02a,Al02b,Du03,
Am06,Am07a,Am07b,Bo07,Cav07,Mo07a,Mo07b,Ue08,Bo08}). Recently, it
was shown that the above scalar resonance propagators satisfy the
K\"allen-Lehmann representation in the domain of coupling constants
usually used \cite{AK04}.

At the end of eighties it was shown that the study of the radiative
decays $\phi\to\gamma a_0\to\gamma\pi\eta$ and $\phi\to\gamma f_0\to
\gamma\pi\pi$ can shed light on the problem of $a_0(980)$ and
$f_0(980)$ mesons \cite{AI89}. Over the next ten years before
experiments (1998) the question was considered from different points
of view \cite{BGP92,CIK93,LN94,Ac95,AGS97,AGS97YF,AGShev97,
AGShev97MP,AGShev97YF,AG97,AG98YFa,AcGu98,AG98YFb}.

Now these decays have been studied not only theoretically but also
experimentally with the help of the SND \cite{Ach98a,Ach98b,Ach00a,
Ach00b} and CMD-2 \cite{Akh99a,Akh99b} detectors at Budker Institute
of Nuclear Physics in Novosibirsk and the KLOE detector at the
DA$\Phi$NE $\phi$-factory in Frascati \cite{Al02a,Al02b,Am06,Am07a,
Am07b,Amb09a,Ambr09b,Bin08,Bo08}.

These experimental data called into being a series of theoretical
investigations \cite{AG01,AG02YF,Ac03a,Ac04YF,Ac03b,AK03,AK04YF,
AK06,AK07a} in which evidences for the four-quark nature of the
$f_0(980)$ and $a_0(980)$ states were obtained. Note the clear
qualitative one. The isovector $a_0(980)$ resonance is produced in
the radiative $\phi$ meson decay as intensively as the isoscalar
$\eta'(958)$ meson containing $\approx 66\% $ of $s\bar s$,
responsible for the $\phi\approx s\bar s\to\gamma s\bar
s\to\gamma\eta'(958)$ decay. In the two-quark model,
$a^0_0(980)$\,=\,$(u\bar{u}-d\bar{d})/\sqrt{2}$, the
$\phi$\,$\approx$\,$s\bar s$\,$\to$\,$\gamma a_0(980)$ decay should
be suppressed by the OZI rule. So, experiment, probably, indicates
for the presence of the $s\bar s$ pair in the isovector $a_0(980)$
state, i.e., for its four-quark nature.

\begin{figure}\begin{tabular}{ccc}
\includegraphics[width=5pc]{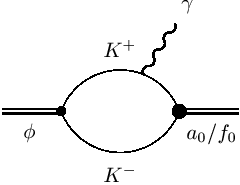}\
\raisebox{-3.6mm}{$\includegraphics[width=5pc]{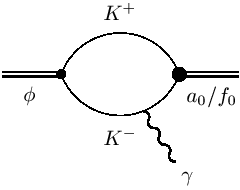}$}\
\includegraphics[width=5pc]{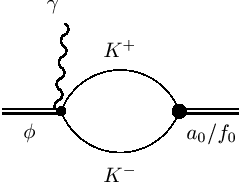}\\
\end{tabular}
\caption{{\footnotesize The $K^+K^-$ loop mechanism of the radiative
decays $\phi(1020)$\,$\to$\,$\gamma (a_0(980)/f_0(980))$.}}
\label{KK-loop-mod}\end{figure}

\begin{figure}\begin{tabular}{cc}
\includegraphics[height=6.1pc]{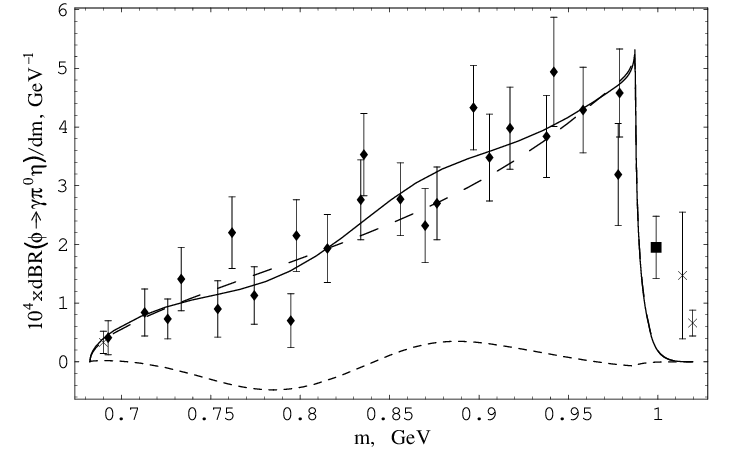}\
\includegraphics[height=6.1pc]{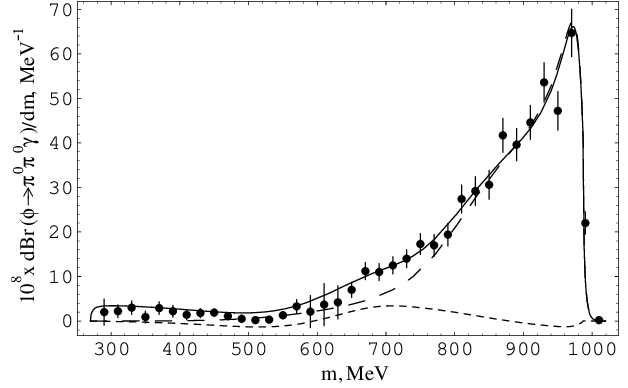}
\end{tabular}\\
\caption{{\footnotesize The left and right plots illustrate the fit
to the KLOE data for the $\pi^0\eta$ and $\pi^0\pi^0$ mass spectra
in the $\phi$\,$\to$\,$\gamma\pi^0\eta$ \cite{Al02a} and
$\phi$\,$\to$\,$\gamma\pi^0\pi^0$ \cite{ Al02b} decays,
respectively. See for details \cite{AK03,AK04YF,AK06,AK07a}}}.
\label{KLOE-spectra}\end{figure}

\begin{figure}
\includegraphics[height=7pc]{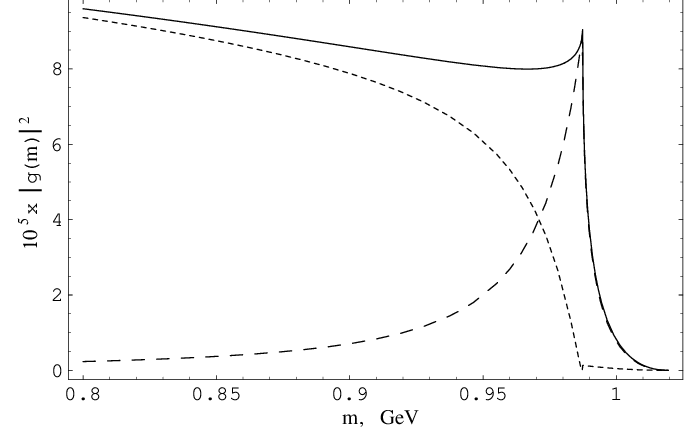}
\caption{{\footnotesize A new threshold phenomenon in $\phi\to
K^+K^-$\,$\to$\,$\gamma R$ decays. The universal in the $K^+K^-$
loop model function $|g(m)|^2$\,=\,$ |g_R(m)/g_{RK^+K^-}|^2$ is
drawn with the solid line. The contributions of the imaginary and
real parts of $g(m)$ are drawn with the dashed and dotted lines,
respectively.}}\label{M-phi-gR}\end{figure}

When basing the experimental investigations \cite{AI89}, it was
suggested  the kaon loop model $\phi$\,$\to$\,$K^+K^-$\,$\to$\,$
\gamma a_0(980)$\,$\to$\,$\gamma\pi^0\eta$ and $\phi$\,$\to$\,$
K^+K^-$\,$\to$ $\gamma f_0(980)$\,$\to$\,$\gamma\pi\pi$, see Fig.
\ref{KK-loop-mod}. This model is used in the data treatment and is
ratified by experiment \cite{Ach98a,Ach98b,Ach00a,Ach00b,Akh99a,
Akh99b,Al02a, Al02b,Am06, Am07a,Am07b,Amb09a,Ambr09b,Bo08,Bin08,
DiMi08,SVP09,A-C10}, see Fig. \ref{KLOE-spectra}.
%----------------------------------------------------------------------------------------
The key virtue of the kaon loop model has the built-in nontrivial
threshold phenomenon, see Fig. \ref{M-phi-gR}. To describe the
experimental mass distributions $dBR(\phi\to\gamma R\gamma ab;\,m
)/dm\sim |g(m)|^2\omega(m)$,\,\footnote{Here $m$ is the invariant
mass of the $ab$-state, $R= a_0(980)$ or $ f_0(980)$, $ab=\pi^0\eta$
or $\pi^0\pi^0$, the function $g(m)$ describes the
$\phi\to\gamma[a_0(m)/f_0(m)]$ transition vertex.} the function
$|g(m)|^2$ should be smooth at $m\leq 0.99$ GeV. But gauge
invariance  requires that $g(m)$ is proportional to the photon
energy $\omega(m)$. Stopping the impetuous increase of the function
$(\omega (m))^3$ at $\omega (990\,\mbox{MeV})$=29 MeV is the crucial
point in the data description. The $K^+K^-$ loop model solves this
problem in the elegant way \cite{AG01,AG02YF,Ac01,Ac03a,Ac04YF,
Ac03b}, see Fig. \ref{M-phi-gR}. In truth this means that $a_0(980)$
and $f_0(980)$ resonances are seen in the radiative decays of $\phi$
meson owing to the  $K^+K^-$ intermediate state. So the mechanism of
the $a_0(980)$ and $f_0(980)$ mesons production in the $\phi$
radiative decays is established at a physical level of proof at
least.

Both real and imaginary parts of the $\phi\to\gamma R$ amplitude are
caused by the $K^+K^-$ intermediate state. The imaginary part is
caused by the real  $K^+K^-$ intermediate state while the real part
is caused by the virtual compact  $K^+K^-$ intermediate state, i.e.,
we are dealing here with  the four-quark transition
\cite{Ac01,Ac03a, Ac04YF,Ac03b}. Needless to say, radiative
four-quark transitions can happen between two  $q\bar q$ states as
well as between  $q\bar q$ and  $q^2\bar q^2$ states but their
intensities depend strongly on a type of the transition. A radiative
four-quark transition between two $q\bar q$ states requires creation
and annihilation of an additional  $q\bar q$ pair, i.e., such a
transition is forbidden according to the  OZI rule, while a
radiative four-quark transition between  $q\bar q$ and $q^2\bar q^2$
states requires only creation of an additional  $q\bar q$ pair,
i.e., such a transition is allowed according to the  OZI rule. The
consideration of this question from the large $N_C$ expansion
standpoint \cite{Ac03a,Ac04YF} supports a suppression of a radiative
four-quark transition between two  $q\bar q$ states in comparison
with a radiative four-quark transition between  $q\bar q$ and
$q^2\bar q^2$ states. So, both intensity and mechanism of the
$a_0(980)$ and $f_0(980)$ production in the radiative decays of the
$\phi(1020)$ meson indicate for their four-quark nature.

Note also that the absence of the decays $J/\psi\to\gamma f_0(980)$,
$J/\psi\to a_0(980)\rho$, $J/\psi\to f_0(980)\omega$ against a
background of the rather intensive decays into the corresponding
classical $P$ wave tensor $q\bar q$ resonances $J/\psi\to\gamma
f_2(1270)$ (or even $J/\psi\to\gamma f'_2(1525)$), $J/\psi\to
a_2(1320)\rho$, $J/\psi\to f_2(1270)\omega$ intrigues against the
$P$ wave $q\bar q$ structure of the $a_0(980)$ and $f_0(980)$ states
\cite{Ac98,Ac99,Ac02, Ac03c}.

\vspace{0.4cm} \noindent{\large \bf 3. Light scalars in the light of
two-photon collisions}\vspace{0.2cm}

\noindent{\bf 3.1. History of investigations}

\noindent Experimental investigations of light scalar mesons in the
$\gamma\gamma$\,$\to$\,$\pi^+\pi^-$, $\gamma\gamma$\,$\to$\,$\pi^0
\pi^0$ and $\gamma\gamma$\,$\to$\,$\pi^0\eta$ reactions with the
$e^+e^-$-colliders began in eighties and have continued up to now.
In first decade many groups, DM1, DM1/2, PLUTO, TASSO, CELLO, JADE,
Crystal Ball, MARK II, DELCO, and TPC/2$\gamma$, took part in that.
Only Crystal Ball and JADE could studied the $\pi^0\pi^0$ and
$\pi^0\eta$ channels, the others (and JADE) the $\pi^+\pi^-$
channel. For those, who wish to read more widely in the contribution
of this impressive period in light scalar meson physics, one can
recommend the following reviews and papers:
\cite{Fi81,Hi81,We81,ADS82a,ADS82b,ADS85,Ed82,Ol83,Me83,Ko84,Kol85,
KZ87,Ko88,Co85,Er85,Bar85,KaSer,An86,Jo86,Po86,BW87,MP87,MP88,AS88,
AS91,Ch88,PDG92,MPW94}.

First results on the $f_0(980)$ resonance production are collected
in Tables \ref{TabI} and \ref{TabII}.

It is reasonable that first conclusions had a qualitative character
and data on the  $f_0(980)$\,$\to$\,$\gamma\gamma$ decay width had
large errors or were upper bounds. Note as a guide that the TASSO
and Crystal Ball results, see Table \ref{TabII}, based on the
integral luminosity equals to 9.24\,pb$^{-1}$ and 21\,pb$^{-1}$,
respectively.

\begin{table}
\caption{ %Table I.
First conclusions on the $f_0(980)$ production in $\gamma\gamma$\,$
\to$\,$\pi\pi$ (see reviews \cite{Fi81,Hi81,We81}).}
\begin{tabular}{|l|l|} \hline
  \ Experiments\, & \,$\ \ \ \ \qquad$ Conclusions\\ \hline
  \ Crystal Ball & \ No significant $f_0(980)$\\
  \ CELLO & \ Hint of $f_0(980)$\\
  \ JADE & \ No evidence for $f_0(980)$\\
  \ TASSO & \ Good fit to data book values\\
  & \ for $f_2(1270)$ includes $f_0(980)$\\ & \ (3\,$\sigma$ effect)\\
  \ MARK II & \ No significant $f_0(980)$ signal\\
  \hline
\end{tabular}\label{TabI}\end{table}%\\[0.1cm]
\begin{table}
\caption{ %Table II.
First results on the $\gamma\gamma$ width of the $f_0(980)$ (see
reviews \cite{Fi81,Hi81,Ed82,Ol83,Ko84,Ko88}).}
\begin{tabular}{|l|l|} \hline
  \ Experiments\, & \,$\ \ \ \ \qquad \Gamma_{f_0\to\gamma\gamma}$ [keV]\\ \hline
  \ TASSO & \
  (1.3\,$\pm$\,0.4\,$\pm$\,0.6)/$B(f_0$\,$\to$\,$\pi^+\pi^-)$\\
  \ Crystal Ball & \ $<0.8/B(f_0\to\pi\pi)$ (95\% C.L.)\\
  \ JADE & \ $<0.8$ (95\% C.L.)\\
  \ Kolanoski (1988) \cite{Ko88} & \ 0.27\,$\pm$\,0.12 (average value) \\
  \hline
\end{tabular}\label{TabII}\end{table}%\\[0.1cm]

\begin{figure}
\includegraphics[height=9pc]{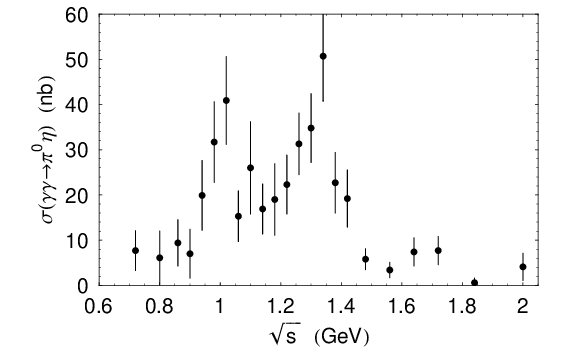}
\caption{{\footnotesize Cross section for $\gamma\gamma$\,$\to$\,$
\pi^0\eta$ as a function of $\sqrt{s}$ for $|\cos\theta| \leq0.9$,
where $\sqrt{s}$ is the invariant mass of $\pi^0\eta$ and $\theta$
is the polar angle of the produced $\pi^0$ (or $\eta$) meson in the
$\gamma\gamma$ center-of-mass system. The data are from the Crystal
Ball Collaboration \cite{An86}.}}\label{CB-gg-pieta}
\end{figure}

As for the $a_0(980)$ resonance, it was observed in the
$\gamma\gamma$\,$\to$\,$\pi^0\eta$ reaction only in three
experiments. The Crystal Ball group \cite{An86} collected during two
years the integral luminosity of 110\,pb$^{-1}$, selected at that
336 events relevant to the $\gamma\gamma$\,$\to$\,$\pi^0\eta$
reaction in the $a_0(980)$ and $a_2(1320)$ region, see Fig.
\ref{CB-gg-pieta}, and published in 1986 the following result:
$\Gamma_{a_0\to\gamma\gamma}B(a_0$\,$ \to$\,$\pi^0\eta)$\,=\,$(0.19
\pm0.07^{+0.10}_{-0.07})$ keV, where $\Gamma_{a_0\to\gamma\gamma}$
is the width of the $a_0(980)$\,$\to$\,$\gamma\gamma$ decay and
$B(a_0$\,$\to$\,$\pi^0\eta)$ is the branching ratio of the
$a_0(980)$\,$\to$\,$\pi^0\eta$ decay. The measured value of
$\Gamma_{a_0\to\gamma\gamma}B(a_0$\,$\to$\,$ \pi^0\eta)$
characterizes the intensity of $a_0(980)$ production in the channel
$\gamma\gamma$\,$\to$\,$a_0(980)$\,$\to$\,$\pi^0\eta$. The
prehistory of this result see in Refs. \cite{Ed82,Co85,Er85}. After
four years, the JADE group \cite{Oe90} (see also \cite{Ko88})
obtained $\Gamma_{a_0\to\gamma\gamma}B(a_0$\,$ \to$\,$\pi^0\eta)
$\,=\,$(0.28\pm0.04\pm0.10)$ keV based on the integral luminosity
149\,pb$^{-1}$ and 291 $\gamma\gamma$\,$\to$\,$\pi^0\eta$ events.
The Crystal Ball \cite{An86} and JADE \cite{Oe90} data on the
$a_0(980)$\,$\to$ \,$\gamma\gamma$ decay have aroused keen interest,
see, for example, \cite{Bar85,Barn92,Kol85,Ko88,Ko91,KZ87,BW87,AS88,
Ch88,PDG92}. Late on, need for high-statistic data arose. But until
very recently, there are no new experiments on the
$\gamma\gamma$\,$\to$\,$\pi^0 \eta$ reaction. According to the PDG
reviews from 1992 to 2008, the average value for $\Gamma_{a_0\to
\gamma\gamma}B(a_0$\,$\to$\,$\pi^0 \eta)$\,=\,$ (0.24^{+0.8}_{-0.7}
)$ keV \cite{PDG92,PDG08}. Only in 2009, the Belle Collaboration
obtained new high-statistics data on the reaction $\gamma\gamma$\,$
\to$\,$\pi^0\eta$ at the KEKB $e^+e^-$ collider \cite{Ue09}. The
statistics collected in the Belle experiment is 3 orders of
magnitude higher than in the earlier Crystal Ball and JADE
experiments. The detailed analysis of the new Belle data we present
in Section 5. Here we only point out the value for
$\Gamma_{a_0\to\gamma\gamma}B(a_0$\,$\to$\,$\pi^0 \eta)$\,=\,$
(0.128^{+0.003+0.502}_{-0.002-0.043})$ keV obtained by the authors
of the experiment \cite{Ue09} and the average value for
$\Gamma_{a_0\to\gamma\gamma}B(a_0$\,$\to$\,$\pi^0 \eta)$\,=\,$
(0.21^{+0.8}_{-0.4})$ keV from the last PDG review \cite{PDG10}.

The JADE group \cite{Oe90} measured also the
$\gamma\gamma$\,$\to$\,$\pi^0\pi^0$ cross section and having
(60\,$\pm$\,8)-events in the $f_0(980)$ region (and, for comparison,
(2177\,$\pm$\,47) events in the $f_2(1270)$ region) obtained for the
$f_0$\,$ \to$\,$\gamma\gamma$ decay width
$\Gamma_{f_0\to\gamma\gamma}$\,=\,$(0.42 \pm0.06^{+0.08}_{-0.18})$
keV (that corresponds to $\Gamma_{f_0\to\gamma\gamma}<0.6$ keV at
95\% C.L.).

\begin{figure}\begin{tabular}{c}
\includegraphics[height=14pc]{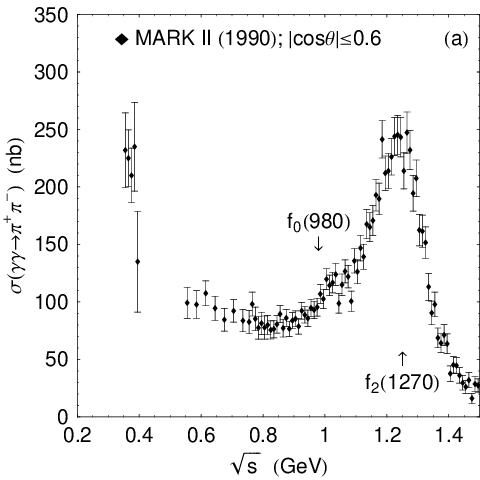}\\
\includegraphics[height=14pc]{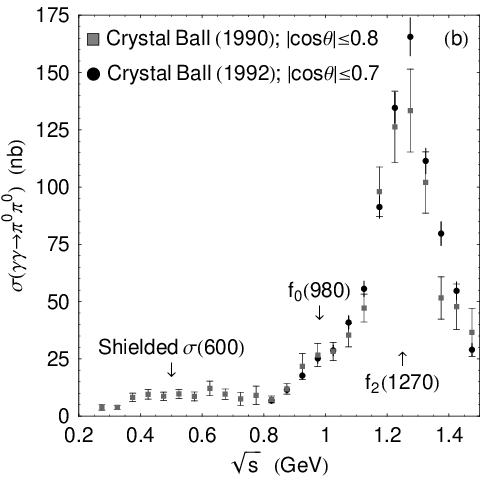}
\end{tabular}\\
\caption{{\footnotesize Cross sections for $\gamma\gamma$\,$\to$\,$
\pi^+\pi^-$ (a) and $\gamma\gamma$\,$\to$\,$ \pi^0\pi^0$ (b) as
functions of the invariant mass $\sqrt{s}$ of $\pi\pi$. The data
correspond to limited angular ranges of the registration of the
final pions; $\theta$ is the polar angle of the produced $\pi$ meson
in the $\gamma\gamma$ center-of-mass system.}}
\label{MII-CB-gg-pipi}
\end{figure}

In addition, in 1990 the MARK II group in experiment on the
$\gamma\gamma$\,$\to$\,$\pi^+\pi^-$ reaction with the integral
luminosity 209\,pb$^{-1}$ \cite{Bo90} and the Crystal Ball group in
1990--1992 in experiments on the $\gamma\gamma$\,$\to$\,$\pi^0\pi^0$
reaction with the integral luminosities 97\,pb$^{-1}$ \cite{Ma90}
and 255\,pb$^{-1}$ \cite{Ka91,Bi92} obtained also similar results
for $\Gamma_{f_0\to\gamma\gamma}$. All data are listed together in
Table \ref{TabIII}, and Figs. \ref{MII-CB-gg-pipi}(a) and
\ref{MII-CB-gg-pipi}(b) illustrate the manifestations of the
$f_0(980)$ and $f_2(1270)$ resonances observed by MARK II and
Crystal Ball in the cross sections for $\gamma\gamma$\,$\to$\,$
\pi\pi$.

Although the statistical significance of the $f_0(980)$ signal in
the cross sections and the invariant $\pi\pi$ mass resolution left
much to be desired, the existence of a shoulder in the $f_0(980)$
resonance region in the $\gamma\gamma$collision might be thought as
established, see Fig. \ref{MII-CB-gg-pipi}

\begin{table}%Table III.
\caption{1990--1992 data on the $\gamma\gamma$ width of the
$f_0(980)$ (see the text).}
\begin{tabular}{|l|l|} \hline
  \ Experiments\, & \,$\ \ \ \Gamma_{f_0\to\gamma\gamma}$ [keV]\\ \hline
  \ Crystal Ball (1990)& \ 0.31\,$\pm$\,0.14\,$\pm$\,0.09\\
  \ MARK II (1990) & \ 0.29\,$\pm$\,0.07\,$\pm$\,0.12\\
  \ JADE (1990)& \ 0.42\,$\pm0.06^{+0.08}_{-0.18}$\\
  \ Karch (1991) & \ 0.25\,$\pm$\,0.10\\
  \ Bienlein (1992) & \ 0.20\,$\pm$\,0.07\,$\pm$\,0.04\\
  \  & \ $\leq$\,0.31 (90\% CL)\\
  \hline
\end{tabular}\label{TabIII}\end{table}%\\[0.1cm]

The experiments of eighties and beginning of nineties showed that
the two-photon widths of the scalar $f_0(980)$ and $a_0(980)$
resonances are small in comparison with the two-photon widths of the
tensor $f_2(1270)$ and $a_2(1320)$ resonances, for which there were
obtained the following values $\Gamma_{f_2\to\gamma\gamma}$\,$
\approx$\,$2.6-3$\,keV \cite{Ma90,Bo90,Oe90,Be92} (see also
\cite{PDG08,PDG10}) and $\Gamma_{a_2\to\gamma\gamma}$\,$\approx
$\,1\,keV \cite{An86,Oe90} (see also \cite{PDG08,PDG10}). This fact
pointed to the four-quark nature of the $f_0(980)$ and $a_0(980)$
states \cite{An86,Kol85,Ko88,Ko91,Bar85,BW87,KZ87,Ch88,Ma90,Oe90,
PDG92,Cah89,FH90}.

As mentioned above, in the beginning of eighties it was predicted
\cite{ADS82a,ADS82b} that, if the $a_0(980)$ and $f_0(980)$ mesons
are taken as four-quark states, their production rates should be
suppressed in photon-photon collisions by a factor ten in relation
to the $a_0(980)$ and $f_0(980)$ mesons taken as two-quark $P$ wave
states. The estimates obtained for the four-quark model were
\cite{ADS82a,ADS82b}
\begin{equation}\label{Estimate4q}
\Gamma_{a_0\to\gamma\gamma}\sim\Gamma_{f_0\to\gamma\gamma}\sim0.
27\,\mbox{keV},\end{equation} which were supported by experiments.
As for the $q\bar q$ model, it predicted that
%----------------------------------------------------------------------------------------
\begin{equation}\label{Estimate2q}
\frac{\Gamma_{0^{++}\to\gamma\gamma}}{\Gamma_{2^{++}\to\gamma\gamma}}=
\frac{15}{4}\times corrections\approx1.3-5.5\end{equation} for the
$P$ wave states with $J^{PC}$\,=\,0$^{++}$ and 2$^{++}$ from the
same family, see, for example, \cite{BF73,BGK76,Ja76,BK79,Bar85,
Barn92,BW87, AS88,Ch88,Ma90,MP90,Li91, Ko91,Mu96,Ac98,Penn07c,
BHS83}. The factor $\frac{15}{4}$ is obtained in the
non-relativistic quark model according to which
$\Gamma_{0^{++}\to\gamma\gamma}=(256/3)\alpha^2|R'(0)|^2/M^4$ and
$\Gamma_{2^{++}\to\gamma\gamma}=(1024/45)\alpha^2|R'(0)|^2/M^4$,
where $R'(0)$ is the derivative of the $P$ state radial wave
function with a mass $M$ at the origin.
$\Gamma_{2^{++}\to\gamma\gamma}$ differs from
$\Gamma_{0^{++}\to\gamma\gamma}$ by the product of the
Clebsch-Gordan spin-orbit coefficient squared ($\frac{1}{2}$) and
value of $\sin^4\vartheta$ averaged over the solid angle
($\frac{8}{15}$); see Ref. \cite{Ja76} for details. This suggested
that $\Gamma_{f_0\to\gamma\gamma}\geq3.4$\,keV and $\Gamma_{a_0
\to\gamma\gamma}\geq1.3$\,keV.

One dwells else on predictions of the molecule model in which the
$a_0(980)$ and $f_0(980)$ resonances are non-relativistic bound
states of the $K\bar K$ system \cite{WI82,WIs90}. As the $q^2\bar
q^2$ model, the molecule one explains the state mass degeneracy and
their strong coupling with the $K\bar K$ channel. As in the
four-quark model, in the molecular one no questions arise with the
small rates $B[J/\psi$\,$\to$\,$a_0(980)\rho]/B[J/\psi$\,$
\to$\,$a_2(1320)\rho$] and $B[J/\psi$\,$\to$\,$f_0(980)\omega]
/B[J/\psi$\,$\to$\,$f_2(1270)\omega]$ (see specialities in Refs.
\cite{Ac98,Ac02}). However, the predictions of this model for the
two-photon widths \cite{Bar85,Barn92},
\begin{equation}\label{EstimateKK} \Gamma_{a_0(K\bar
K)\to\gamma\gamma}=\Gamma_{f_0(K\bar
K)\to\gamma\gamma}\approx0.6\,\mbox{keV},\end{equation} are rather
big, within two standard deviations contradict the experiment data
from Table \ref{TabIII}. More than that, the widths of $K\bar K$
molecules must be smaller (strictly speaking, much smaller) than the
binging energy $\epsilon\approx10$\, MeV. Recent data \cite{PDG10},
however, contradict this, $\Gamma_{a_0}$\,$\sim$\,$(50-100)$\,MeV
and $\Gamma_{f_0}$\,$\sim$\,$(40-100)$\,MeV. The $K\bar K$ molecule
model predicted also \cite{AG97,AGS97} that $B[\phi$\,$\to$\,$\gamma
a_0(980)]$\,$\approx$\,$B[\phi$\,$\to$\,$\gamma f_0(980)]$\,$
\sim$\,10$^{-5}$ that contradicts experiment \cite{PDG10}. In
addition, recently \cite{AK07b,AK08} it was shown that the kaon loop
model, ratified by experiment, describes production of a compact
state and not an extended molecule. Finally, experiments in which
the $a_0(980)$ and $f_0(980)$ mesons were produced in the
$\pi^-p$\,$\to$\,$\pi^0\eta n$ \cite{Dz95,Al99} and
$\pi^-p$\,$\to$\,$\pi^0\pi^0n$ \cite{Al95,Al98,Gu01} reactions
within a broad range of four-momentum transfer squared,
0\,$<$\,$-t$\,$<$\,1\,GeV$^2$, have shown that these states are
compact, e.g. as two-quark $\rho$, $\omega$, $a_2(1320)$,
$f_2(1270)$ and other mesons and not as extended molecule ones with
form factors determined by the wave functions. These experiments
have left no chances for the $K\bar K$ molecule model.\,\footnote{A
$K\bar K$ formation of unknown origin with the average relativistic
Euclidean momentum squared $<k^2>\approx 2$ GeV$^2$ was considered
recently and named ``a $K\bar K$ molecule'' \cite{BGL08}. Such a
free use of the molecule term can mislead readers considering a
molecule as an extent non-relativistic bound system.} As to
four-quark states, they are as compact as two-quark
states.\,\footnote{An additional argument against the molecular
model for the $a_0(980)$ resonance is presented in Section 5.}

The Particle Data Group gives information on an average value of
$\Gamma_{f_0\to\gamma\gamma}$ beginning from 1992. Note that no new
experimental data on $\Gamma_{f_0\to\gamma\gamma}$ emerged from 1992
up to 2006, nevertheless, its average value, adduced by PDG, evolved
noticeably in this period. Based on the data in Table \ref{TabIII},
the $\Gamma_{f_0\to\gamma\gamma}$ value would be (0.26\,$\pm
$\,0.08)\,keV. In 1992 PDG \cite{PDG92} obtained the average value
$\Gamma_{f_0\to\gamma\gamma} $\,=\,(0.56\,$\pm$\,0.11)\,keV
combining the JADE result (1990) \cite{Oe90}, see Table
\ref{TabIII}, with the value $\Gamma_{f_0\to\gamma\gamma}
$\,=\,(0.63$\pm$0.14)\,keV, which was found by Morgan and Pennington
(1990) \cite{MP90} as a result of a theoretical analysis of the MARK
II (1990) \cite{Bo90} and Crystal Ball (1990) \cite{Ma90} data. In
1999 Boglione and Pennington carried out a new theoretical analysis
\cite{BP99} of the situation and halved value, $\Gamma_{f_0\to\gamma
\gamma} $\,=\,(0.28\,$^{+0.09}_{-0.13} $)\,keV (see also
\cite{Pe99}). The Particle Data Group noted that the Boglione and
Pennington (1999) result replaces the Morgan and Pennington (1990)
one but used both results coupled with the JADE (1990) one for
calculation of the average $f_0$\,$\to$\,$\gamma \gamma$ decay
width. In this way the value $\Gamma_{f_0\to\gamma
\gamma}$\,=\,(0.39\,$^{+0.10}_{-0.13}$)\,keV emerged in the PDG
review (2000) \cite{PDG00}.

In 2003 preliminary super-statistics Belle data on
$\gamma\gamma$\,$\to$\,$\pi^+\pi^-$ were reported. They contain a
clear signal from the $f_0(980)$ resonance \cite{Mo03}. In  2005
there emerged our first response \cite{AS05} to these data. It has
become clear that $\Gamma_{f_0\to\gamma\gamma}$ is bound to be
small. In 2006 PDG excluded the Morgan and Pennington (1990) result,
$\Gamma_{f_0\to\gamma\gamma}$\,=\,$(0.63 \pm0.14)$\,keV, from its
sample and using only the JADE (1990) data and the Boglione and
Pennington (1999) result obtained a new guide
$\Gamma_{f_0\to\gamma\gamma}$\,=\,(0.31\,$^{+0.08}_{-0.11}$)\,keV
\cite{PDG06}. To the effect that happened later to the average value
of $\Gamma_{f_0\to\gamma\gamma}$ and can else happens to the one, we
are going to tell in the following subsections 3.2 and
3.4.\vspace{0.2cm}

\noindent{\bf 3.2. Current experimental situation}

\begin{figure}
\includegraphics[width=21pc]{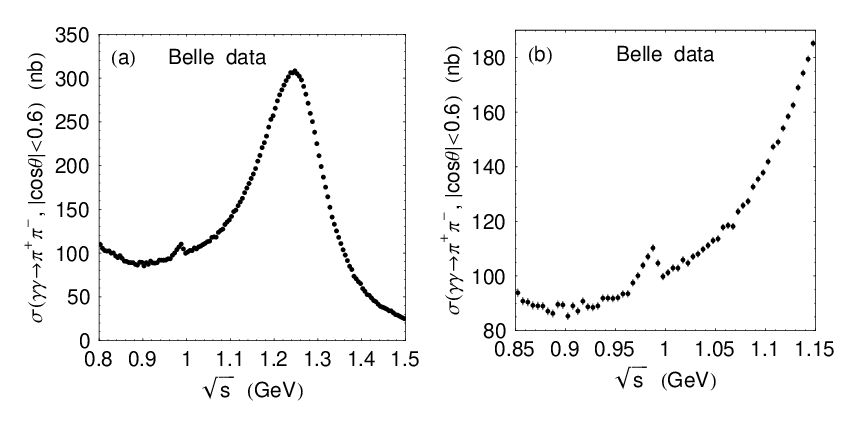}
\caption{{\footnotesize (a) The high statistics Belle data on the
$\gamma\gamma$\,$\to$\,$\pi^+\pi^-$ reaction cross section for
$|\cos\theta|\leq0.6$ \cite{Mo07b}. Plot (b) emphasizes the region
of the $f_0(980)$ peak. Errors shown include statistics only. They
are approximately equal to 0.5\%--1.5\%. The $\sqrt{s}$ bin size in
the Belle experiment has been chosen to be 5 MeV, with the mass
resolution of about 2 MeV.}}\label{Belle-gg-pipi}
\end{figure}

\noindent In 2007 the Belle collaboration published the data on
cross section of the $\gamma\gamma$\,$\to$\,$\pi^+\pi^-$ reaction in
the region of the $\pi^+\pi^-$ invariant mass, $\sqrt{s}$, from 0.8
up 1.5 GeV based on the integral luminosity 85.9\,fb$^{-1}$
\cite{Mo07a,Mo07b}. These data are shown on  Fig.
\ref{Belle-gg-pipi}. Thanks to the huge statistics and high energy
resolution in the Belle experiment, the clear signal of the
$f_0(980)$ resonance was detected for the first time. Its value
proved to be small that agrees qualitatively with the four-quark
model prediction \cite{ADS82a,ADS82b}. The visible height of the
$f_0(980)$ peak amounts of about 15 nb over the smooth background
near 100 nb. Its visible (effective) width proved to be about 30--35
MeV, see Fig. \ref{Belle-gg-pipi}.

%----------------------------------------------------------------------------------------

\begin{figure}
\includegraphics[height=10.2pc]{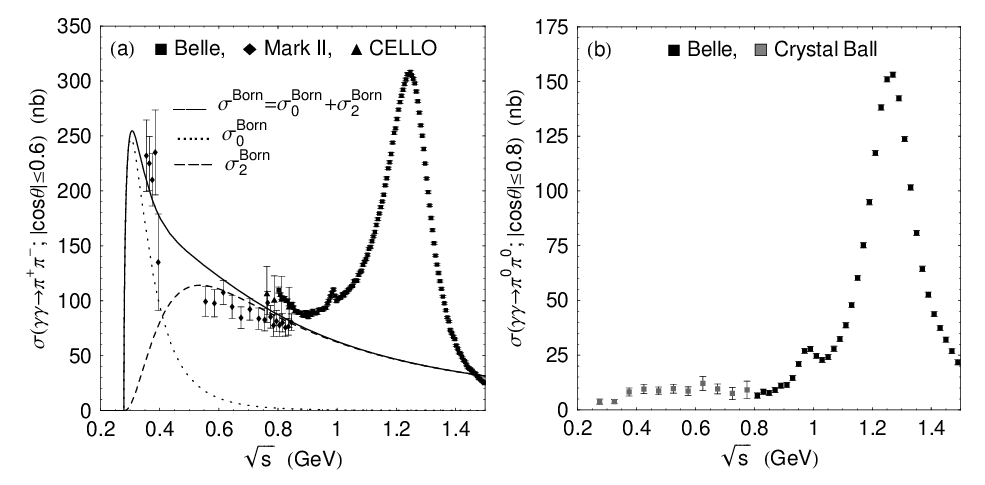}
\caption{{\footnotesize (a) The data on the $\gamma\gamma\to\pi^+
\pi^-$ reaction cross section from Mark II \cite{Ma90} and CELLO
\cite{Be92}, for $\sqrt{s}\leq 0.85$\,GeV, and from Belle
\cite{Mo07b}, for 0.8\,$ \leq\sqrt{s}\leq1.5$\,GeV. (b) The data on
the $\gamma\gamma\to \pi^0\pi^0$ reaction cross section from Crystal
Ball \cite{Bo90}, for $\sqrt{s}<0.8$\,GeV, and from Belle
\cite{Ue08}, for 0.8\,$\leq\sqrt{s}\leq1.5$\,GeV. Plots (a), for
$\sqrt{s}>0.85$\,GeV, and (b), for $\sqrt{s}>0.8$\,GeV, show
exclusively the Belle data to emphasize the discovered miniature
signals from the $f_0(980)$ resonance. The theoretical curves, shown
on plot (a), correspond to the cross sections for the process
$\gamma\gamma\to\pi^+\pi^-$ for $|\cos\theta|\leq0.6$ caused by the
electromagnetic Born contribution from the elementary one pion
exchange: the total integrated cross section $\sigma^{\mbox
{\scriptsize{Born}} }$\,=\,$\sigma^{\mbox {\scriptsize{Born}}
}_0$\,+\,$\sigma^{\mbox {\scriptsize{Born}}}_2$ and the integrated
cross sections $\sigma^{\mbox{\scriptsize{Born}}}_\lambda$ with
helicity $\lambda$\,=\,0 and 2.}}
\label{Belle-gg-pipi-pi0pi0}\end{figure}

Then the Belle collaboration published the data on cross section for
the $\gamma\gamma$\,$\to$\,$\pi^0\pi^0$ reaction in the region of
the $\pi^+\pi^-$ invariant mass, $\sqrt{s}$, from 0.6 to 1.6 GeV
based on the integral luminosity 95\,fb$^{-1}$ \cite{Ue08}; see also
\cite{Abe07,Nak08,Ada08}. Here also the clear signal of  the
$f_0(980)$ resonance was detected for the first time. Note that the
background conditions for the manifestation of the $f_0(980)$ in the
$\gamma\gamma$\,$ \to$\,$\pi^0\pi^0$ channel are more favourable
than in the $\gamma \gamma$\,$\to$\,$\pi^+\pi^-$ one.

Figures \ref{Belle-gg-pipi-pi0pi0}(a) and
\ref{Belle-gg-pipi-pi0pi0}(b) illustrate  a general picture of data
on the cross sections of the $\pi^+\pi^-$ and $\pi^0\pi^0$
production in photon-photon collisions from the $\pi\pi$ threshold
up to 1.5 GeV after the Belle experiments. It is instructive to
compare these results with a previous picture illustrated by Figs.
\ref{MII-CB-gg-pipi}(a) and \ref{MII-CB-gg-pipi}(b).

\begin{table}%Table IV.
\caption{The current data on the $f_0(980)\to\gamma\gamma$ decay
  width.}
  \begin{tabular}{|l|l|} \hline
  \,Experiments\, & \,$\ \ \ \Gamma_{f_0\to\gamma\gamma}$ [keV]\\ \hline
  \,$\gamma\gamma$\,$\to$\,$\pi^+\pi^-$ Belle (2007) \cite{Mo07a} &\,$0.205^{+0.095 +0.147}_{-0.083-0.117}$\\
  \,$\gamma\gamma$\,$\to$\,$\pi^0\pi^0$\ \ Belle (2008) \cite{Ue08} &\,$0.286\pm0.017^{+0.211}_{-0.070}$\\ \hline
  \,PDG average value \cite{PDG08,PDG10} &\,$0.29^{+0.07}_{-0.06}$ \\ \hline
\end{tabular}\label{TabIV}\end{table}
The current information about $\Gamma_{f_0\to\gamma \gamma}$ are
adduced in Table \ref{TabIV}. The Belle collaboration determined
$\Gamma_{f_0\to\gamma\gamma}$ (see Table \ref{TabIV}) as a result of
fitting the mass distributions (see Figs. \ref{Belle-gg-pipi}(b) and
\ref{Belle-gg-pipi-pi0pi0}(b)) taking into account the $f_0(980)$
and $f_2(1270)$ resonance contributions and smooth background
contributions, which are a source of large systematic errors in
$\Gamma_{f_0\to \gamma\gamma}$ (see for details in Refs.
\cite{Mo07a,Mo07b,Ue08}).\vspace{0.2cm}

\noindent{\bf \boldmath 3.3. Dynamics of the reactions
$\gamma\gamma\to\pi\pi$: Born contributions and angular
distributions}

\begin{figure}\includegraphics[width=18pc]{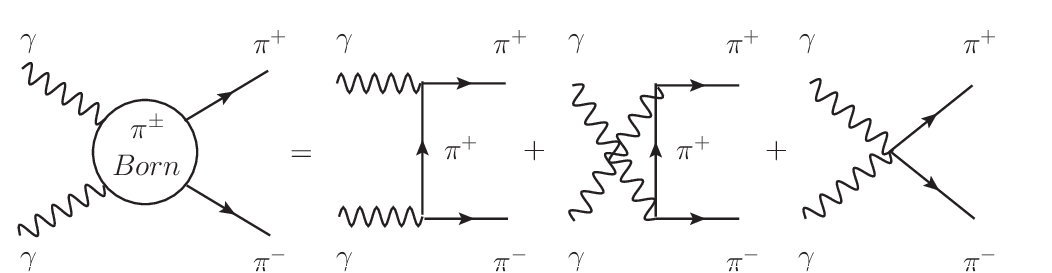} % height=7pc
\caption{{\footnotesize The Born diagrams for $\gamma
\gamma$\,$\to$\,$\pi^+\pi^-$.}}\label{Born-gg-pipi}\end{figure}

\noindent To feel the values of the cross sections measured by
experiment, in Fig. \ref{Belle-gg-pipi-pi0pi0}(a) the total Born
cross section of the $\gamma\gamma$\,$\to$\,$\pi^+\pi^-$ process,
$\sigma^{\mbox{\scriptsize{Born}}}$\,=\,$\sigma^{\mbox
{\scriptsize{Born}}}_0$\,+\,$\sigma^{\mbox{\scriptsize{Born}}}_2$,
and the partial helicity ones, $\sigma^{\mbox{\scriptsize{Born}}}_
\lambda$, are adduced as a guide, where $\lambda$\,=\,0 or 2 is the
absolute value of the photon helicity difference. These cross
sections are caused by the elementary one pion exchange mechanism,
see Fig. \ref{Born-gg-pipi}. By the Low theorem\,\footnote{
According to this theorem \cite{Low54,GMG54,AbG68}, the Born
contributions give the exact physical amplitude of the crossing
reaction $\gamma\pi^\pm\to \gamma\pi^\pm$ close to its threshold.}
and chiral symmetry\,\footnote{Chiral symmetry guarantees weakness
of the $\pi\pi$ interaction at low energy.}, the Born contributions
should dominate near the threshold region of the
$\gamma\gamma$\,$\to$\,$\pi^+\pi^-$ reaction. As shown in Fig.
\ref{Belle-gg-pipi-pi0pi0}(a), this anticipation does not contradict
the current data near threshold, but, certainly, errors leave much
to be desired. In additional, one can consider the Born
contributions as an reasonable approximation of background
(non-resonance) contributions in the $\gamma\gamma$\,$\to$\,$\pi^+
\pi^-$ amplitudes in  all the resonance region, including the
$f_2(1270)$ one. The Born contributions are also the base for a
construction of amplitudes, including strong interactions in final
state, see, for example, \cite{Me83,Ly84,Lyt85,Jo86,MP87,MP88,MP91,
BC88,DHL88, DH93,OO97,AS05,AS07,Pe06}.

\begin{figure}
\includegraphics[height=10.2pc]{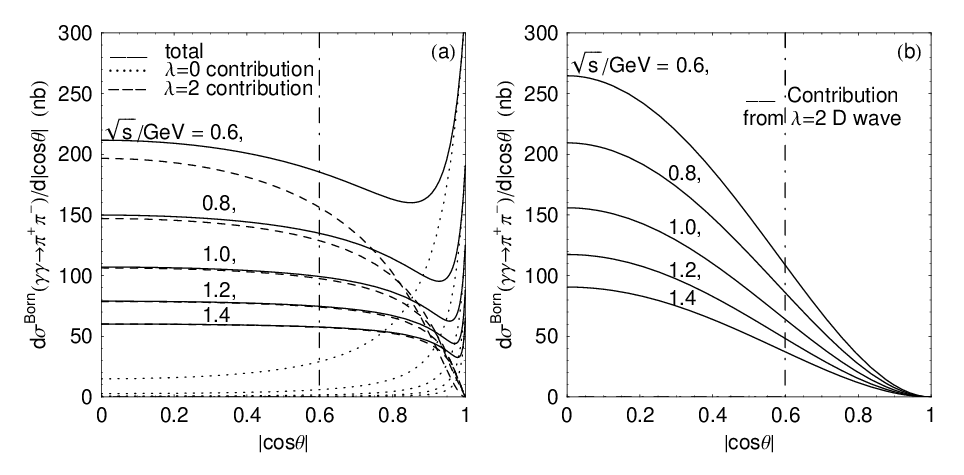}
\caption{{\footnotesize Plots (a) and (b) show the
$\gamma\gamma$\,$\to$\,$\pi^+\pi^-$ differential cross section in
the Born approximation (i.e., for the elementary one pion exchange
mechanism) and its components for different values of
$\sqrt{s}/\mbox{GeV}$. The vertical straight lines $|\cos\theta|=0.6
$ show the upper boundary of the region available for measurements.
}}\label{Born-diff-CS}
\end{figure}

The Born contributions have the following particular qualities.
First, $\sigma^{\mbox{\scriptsize{Born}}}$ has a maximum at
$\sqrt{s}\approx0.3$ GeV, where $\sigma^{\mbox{\scriptsize{Born}}}
\approx\sigma^{\mbox{\scriptsize{Born}}}_0$, then
$\sigma^{\mbox{\scriptsize {Born} }}_0$ falls with increasing
$\sqrt{s}$, so that the $\sigma^{\mbox{\scriptsize{Born}}}_2$
contribution dominates in $\sigma^{\mbox{\scriptsize{Born}}}$ at
$\sqrt{s}>0.5$ GeV, see Fig. \ref{Belle-gg-pipi-pi0pi0}(a). Second,
although  the $\sigma^{\mbox{\scriptsize{Born}}}_2$ value is
approximately 80\% caused by the $D$ wave amplitude, its
interference with the contribution of higher waves are considerable
in the differential cross section $d\sigma^{\mbox{\scriptsize{Born}}
}(\gamma\gamma$\,$\to$\,$\pi^+\pi^-)/d|\cos\theta|$, compare Figs.
\ref{Born-diff-CS}(a) and \ref{Born-diff-CS}(b). The interference,
destructive in the first half of the $|\cos\theta|\leq0.6$ interval
and constructive in the second one, flattens out the $\theta$ angle
distribution in this interval, so that this effect increases with
increasing $\sqrt{s}$, see Fig. \ref{Born-diff-CS}(a).

Since the first resonance with  $I^G(J^{PC})=0^+(4^{++})$ has the
mass near 2 GeV \cite{PDG08,PDG10}, then seemingly the $S$ and $D$
wave contributions only should dominate at $\sqrt{s}\leq$\,1.5\,GeV
and the differential cross section of the
$\gamma\gamma$\,$\to$\,$\pi^+ \pi^-$ process could be represented as
\cite{Mo07b}
\begin{equation}d\sigma(\gamma\gamma\to\pi^+\pi^-)/d\Omega=
|S+D_0Y^0_2|^2+|D_2Y^2_2|^2\,,\label{SD0D2-gg-pipi}\end{equation}
where $S$, $D_0$, and $D_2$ are the $S$ and $D_\lambda$ wave
amplitudes with the helicity $\lambda$\,=\,0 and 2, $Y^m_J$ are the
spherical harmonics.\,\footnote{Eq. (\ref{SD0D2-gg-pipi})
corresponds the situation ``untagget'' when the dependence on the
pion azimuth $\varphi$ is not measured, that took place in all above
experiments.} But, the above discussion shows that the smooth
background contribution in the $\gamma\gamma $\,$\to$\,$\pi^+\pi^-$
cross section contains the high partial wave due to the one pion
exchange, so that the smooth background can imitate the large $S$
wave at $|\cos\theta|$\,$\leq$\,0.6.

The one-pion exchange is absent in the $\gamma\gamma$\,$\to$\,$\pi^0
\pi^0$ channel and the representation of the cross section of this
reaction similar to Eq. (6) is a good approximation at
$\sqrt{s}\leq$\,1.5\, GeV
\begin{equation} d\sigma(\gamma\gamma\to\pi^0\pi^0)/d\Omega=
|\widetilde{S}+\widetilde{D}_0Y^0_2|^2+|\widetilde{D}_2Y^2_2|^2\,,
\label{SD0D2-gg-pi0pi0}\end{equation} where $\widetilde{S}$,
$\widetilde{D}_0$, and $\widetilde{D}_2$ are the $S$ and $D_\lambda$
wave amplitudes with the helicity $\lambda$\,=\,0 and 2.
Nevertheless, the partial wave analysis of the
$\gamma\gamma$\,$\to$\,$\pi^0\pi^0$ events, based on Eq.
(\ref{SD0D2-gg-pi0pi0}), is not prevented from difficulties for the
relation $\sqrt{6}|Y^2_2|$\,=\,$\sqrt{5}Y^0_0$\,--\,$Y^0_2$, which
gives no way of separating the partial waves when using only the
data on the differential cross section \cite{Oe90,Mo07b,Ue08}. So,
the separation of the contributions with the different helicities
requests some guesswork, for example, the domination of the helicity
2 in the $f_2(1270)$ resonance production \cite{KrKr78,Kra78,KV81,
Li91} that agrees rather well with the experimental angle
distribution.

The $d\sigma(\gamma\gamma$\,$\to$\,$\pi^0 \pi^0)/d\Omega$
differential cross section in Eq. (\ref{SD0D2-gg-pi0pi0}) is a
polynomial of the second power of $z$\,=\,$\cos^2\theta$, which can
be expressed in terms of its roots  $z_1$ and $z^*_1
$,\,\footnote{Such a procedure is the base of the determination all
solutions when carrying out partial wave analyses, see, for example,
\cite{Ger69,Barr72,Pe77,Al98,Al99,Sad99,Gu01}.}
\begin{equation} d\sigma(\gamma\gamma\to\pi^0\pi^0)/d\Omega=C
(z-z_1)(z-z^*_1)\,,\end{equation} where $C$ is a real quantity. So,
from fitting experimental data on the differential cross section one
can determine only three independent parameters, for example, $C$,
Re$z_1$, and Im$z_1$ up to the sign and not four ones,
$|\widetilde{S}|$, $|\widetilde{D}_0|$, $|\widetilde{D}_2|$, and
$\cos\delta$ ($\delta$ is a relative phase between the
$\widetilde{S}$ and $\widetilde{D}_0$ amplitudes), as one would
like.

\begin{figure}
\includegraphics[width=21pc]{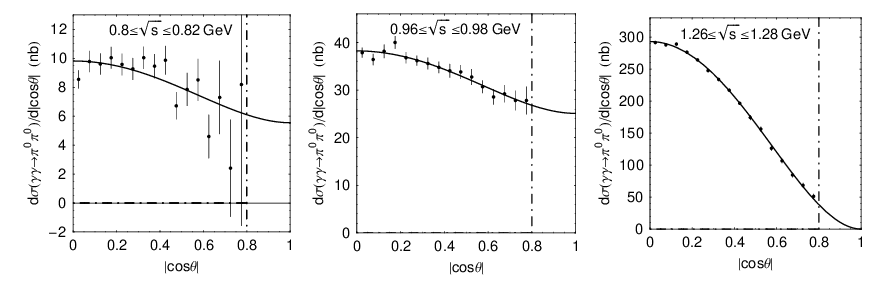}
\caption{{\footnotesize The Belle data on the angular distributions
for $\gamma\gamma$\,$\to$\,$\pi^0\pi^0$ \cite{Ue08}. The solid lines
are the approximations. The vertical straight lines $|\cos\theta|
$\,=\,0.8 show the upper boundary of the region available for
measurements.}}\label{Belle-diff-CS-gg-pi0pi0}
\end{figure}

In Fig. \ref{Belle-diff-CS-gg-pi0pi0} the Belle data on the angular
distributions in $\gamma\gamma$\,$\to$\,$\pi^0\pi^0$ are adduced at
three values of $\sqrt{s}$. All of them are described very well by
the simple two-parameter expression $|a|^2$\,+\,$|b\,Y^2_2|^2$
\cite{AS08b}. This suggests that the $\gamma\gamma$\,$\to$\,$\pi^0
\pi^0$ cross section  is saturated only by the $\widetilde{S}$ and
$\widetilde{D}_2$ partial wave contributions at $\sqrt{s}<1.5$\,GeV.
\vspace{0.2cm}

\noindent{\bf 3.4. Production mechanisms of scalar resonances}

\noindent Expectation of the Belle data and their advent have called
into being a whole series of theoretical papers which study dynamics
of the $f_0(980)$ and $\sigma(600)$ production in the $\gamma\gamma
$\,$\to$\,$\pi\pi$ processes by various means and discuss the nature
of these states \cite{AS05,AS07,AS08a,AS08b,Ac08a,Ac08b,AKS08,AS09,
AS10a, Pe06,Pe07a,Pe07b,Penn07c,Penni08,PMUW08,MNO07,MeNO08a,
MeNaO08b,Men08, Na08,ORS08,OR08,BKR08,KaV08,MWZZZ09,MNW10,GMM10}.

The main lesson from the analysis of the production mechanisms of
the light scalars in $\gamma\gamma $ collisions is the following
\cite{Ac08a,Ac08b}.

The classical $P$ wave tensor $q\bar q$ mesons $f_2(1270)$,
$a_2(1320)$, and $f'_2(1525)$ are produced in $\gamma\gamma$
collisions due to the direct $\gamma\gamma$\,$\to $\,$q\bar q$
transitions in the main, whereas the light scalar mesons
$\sigma(600)$, $f_0(980)$, and $a_0(980)$ are produced by the
rescatterings $\gamma\gamma\to\pi^+\pi^-\to\sigma$, $\gamma\gamma\to
K^+K^-\to f_0$, $\gamma\gamma\to(K^+K^-,\pi^0\eta)\to a_0$, and so
on, i.e., due to the four quark transitions. As to the direct
transitions $\gamma\gamma\to\sigma$, $\gamma\gamma\to f_0$, and
$\gamma\gamma\to a_0$, they are strongly suppressed, as it is
expected in four-quark model.

This conclusion introduces a new seminal view of the
$\gamma\gamma$\,$\to$\,$\pi\pi$ reaction dynamics at low energy. Let
us dwell on this point.

Recall  elementary ideas of interactions of  $C$ even mesons with
photons based on the quark model \cite{Ko84,Ko88,KZ87,PDG10}.
Coupling of the $\gamma\gamma $ system with the classical  $q\bar q$
states, to which the light pseudoscalar ($J^{PC}=0^{-+}$) and tensor
($2^{++}$) mesons belong, are proportional to four power of charges
of constituent quarks.

Only the width of the $\pi^0$\,$\to $\,$\gamma\gamma$ decay is
evaluated from the first principles \cite{Ad69,BeJ69, BFGM72,Leu98}.
$\Gamma_{\pi^0\to\gamma\gamma}$ is determined completely by the
Adler-Bell-Jackiw axial anomaly and in this case the theory (QCD) is
in excellent agreement with the experiment \cite{IO07,Ber07}. The
relations between the widths of the $\pi^0$\,$\to$\,$\gamma\gamma$,
$\eta$\,$\to$\,$\gamma \gamma$, and $\eta'$\,$\to$\,$\gamma\gamma$
decays are obtained in the $q\bar q$ model with taking into account
the effects of the $\eta-\eta'$ mixing and the $SU(3)$ symmetry
breaking \cite{KZ87,Leu98,Fe00}.

As for the tensor mesons, in the ideal mixing case, i.e., if
$f_2=(u\bar u+d\bar d)/\sqrt{2}$ and $f'_2=s\bar s$, the quark model
predicts the following relations for the coupling constant squared:
\begin{equation} g^2_{f_2\gamma\gamma}:g^2_{a_2\gamma\gamma}:
g^2_{f'_2\gamma\gamma}=25:9:2\,.\label{g2-Tensor-gg}\end{equation}
Though absolute values of the two-photon widths of the tensor meson
decays cannot be obtained from the first principles
\cite{Ko84,KZ87,BK79,BW87,BF73,BR76,Ros81,Be80} (see also references
herein), the $q\bar q$ model prediction (\ref{g2-Tensor-gg}),
underlying the relations between the widths of the $f_2(1270)$\,$\to
$\,$\gamma\gamma$, $a_2 (1320)$\,$\to$\,$\gamma\gamma$, and
$f'_2(1525)$\,$ \to$\,$\gamma\gamma$ decays, are used with taking
into account the effects of a deviation from the ideal mixing and
the $SU(3)$ symmetry breaking \cite{Ko84,Po86,KZ87,PDG10,Alb90,
LYS01}. Roughly speaking, the $q\bar q$ model prediction
(\ref{g2-Tensor-gg}) is borne out by experiment.
% As for the
% practically model-independent prediction of the $q\bar q$ model
% $g^2_{f_2\gamma\gamma}:g^2_{a_2\gamma\gamma}=25:9$,\,\footnote{The
% deviation from the ideal $f_2(1270)-f'_2(1525)$ mixing increases
% $g_{f_2\gamma\gamma}$ by less than 3\%  and the small difference of
% the $f_2(1270)$ and $a_2(1320)$ masses allows to expect for its
% small effect also.} this agrees with experiment rather well (Amsler,
% 2008).
Among other things, this implies that the final state interaction
effects are small, in particular, the contributions of the
$f_2(1270)$\,$\to$\,$\pi^+\pi^-$\,$\to$\,$\gamma \gamma$
rescattering type are small in comparison with the contributions of
the direct $q\bar q(2^{++})$\,$\to$\,$\gamma\gamma$ transitions.
% \,\footnote{ Evaluations, see below, show  such is indeed the case.}

The observed smallness of the  $a_0(980)$ and $f_0(980)$ meson
two-photon widths in comparison with the two-photon tensor meson
ones and thus the failure of the $q\bar q$ model prediction of the
relation (\ref{Estimate2q}) between the widths of the direct
$0^{++}$ and $2^{++}$\,$\to$\,$\gamma\gamma$ transitions point to
that $a_0(980)$ and $f_0(980)$ are not the quark and antiquark bound
states.
% \,\footnote{As to the ideal $q\bar q$ model prediction
% $g^2_{f_0\gamma\gamma}:g^2_{a_0\gamma\gamma}=25:9$, this excluded by
% experiment also.}
If the $q\bar q$ component is practically absent in the wave
functions of the light scalars and in their $q^2\bar q^2$ component
the white neutral vector meson pairs are practically absent too, as
in the MIT bag model \cite{ADS82a,ADS82b}, then the
$\sigma(600)$\,$\to$\,$ \gamma\gamma$,
$f_0(980)$\,$\to$\,$\gamma\gamma$, and
$a_0(980)$\,$\to$\,$\gamma\gamma$ decays could be the four-quark
transitions caused by the rescatterings
$\sigma(600)$\,$\to$\,$\pi^+\pi^-$\,$\to$\,$\gamma\gamma$,
$f_0(980)$\,$\to$\,$K^+K^-$\,$\to$\,$\gamma\gamma$, and
$a_0(980)$\,$\to$\,$(K^+K^-,\pi^0\eta)$\,$\to$\,$ \gamma\gamma$.
Already in 1998 we considered such a scenario extensively
\cite{AS88} analyzing the Crystal Ball data \cite{An86} on the
$a_0(980)$ resonance production in the $\gamma\gamma$\,$\to
$\,$\pi^0\eta$ reaction; see also the discussion of the
$\gamma\gamma $\,$\to$\,$K\bar K$ reaction mechanisms in Refs.
\cite{AS92,AS94b}. Fifteen years later, when the preliminary high
statics Belle data \cite{Mo03} on the $f_0(980)$ resonance
production in the $\gamma\gamma $\,$\to $\,$\pi^+\pi^-$ reaction
were reported, we studied what role the rescattering mechanisms, in
particular, the $\gamma\gamma$\,$\to$\,$K^+K^-
$\,$\to$\,$f_0(980)$\,$\to$\,$ \pi^+\pi^-$ mechanism, could play in
this process \cite{AS05}. As a result we showed that just this
mechanism gives a reasonable scale of the $f_0(980)$ manifestation
in the $\gamma\gamma$\,$\to $\,$\pi^+\pi^-$ and $\gamma\gamma$\,$\to
$\,$\pi^0\pi^0$ cross sections.

Then in the $SU(2)_L\times SU(2)_R$ linear $\sigma$ model frame we
showed that the $\sigma$ field are described by its four-quark
component at least in the energy (virtuality) region of the $\sigma$
resonance and the $\sigma(600)$ meson decay into $\gamma\gamma$ is
the four-quark transition $\sigma(600)$\,$\to$\,$\pi^+
\pi^-$\,$\to$\,$\gamma\gamma$ \cite{AS07}. We also emphasized that
the $\sigma$ meson contribution in the $\gamma\gamma$\,$\to$\,$\pi
\pi$ amplitudes is shielded due to its strong destructive
interference with the background contributions as in the
$\pi\pi$\,$\to $\,$\pi\pi$ amplitudes,\,\footnote{As already noted
in Introduction, the presence of the large background, which shields
the $\sigma$ resonance in $\pi\pi$\,$\to$\,$\pi\pi$, is a
consequence of chiral symmetry.} i.e., the $\sigma$ meson is
produced in the $\gamma\gamma$ collisions accompanied by the great
chiral background due to the rescattering mechanism
$\gamma\gamma$\,$\to $\,$\pi^+\pi^-$\,$\to $\,$(\sigma+\mbox{\it
background})$\,$\to $\,$\pi\pi$, that results in the modest
$\gamma\gamma$\,$\to $\,$\pi^0\pi^0$ cross section near (5--10)\,nb
in the $\sigma$ meson region, see Fig.
\ref{Belle-gg-pipi-pi0pi0}(b). The details of this shielding are
given in the next Section.

The above considerations about dynamics of the $\sigma(600)$,
$f_0(980)$, and $f_2(1270)$ resonance production were developed in
analyzing the final high-statistics Belle data \cite{AS08a,AS08b} on
the $\gamma\gamma$\,$\to$\,$\pi^+\pi^-$ and $\gamma\gamma$\,$
\to$\,$\pi^0 \pi^0$ reactions, to a discussion of which we proceed.

\vspace{0.4cm} \noindent{\large \boldmath \bf 4. Analysis of high
statistics Belle data on the reactions $\gamma \gamma\to\pi^+\pi^-$
and $\gamma\gamma\to\pi^0\pi^0$. Manifestations of the $\sigma(600)$
and $f_0(980)$ resonances}\vspace{0.2cm}

\noindent As noted above, the $S$ and $D_{\lambda =2}$ partial wave
contributions dominate in the Born cross sections $\sigma^{\mbox
{\scriptsize{Born}}}_0$ and $\sigma^{\mbox{\scriptsize{Born}}}_2$,
respectively, in region of interest, $\sqrt{s}<1.5$\,GeV, and the
$\pi\pi$ interaction is strong also in the $S$ and $D$ waves only in
this region, that is why the final-state strong interaction modifies
these Born contribution in $\gamma\gamma$\,$\to$\,$\pi^+\pi^-$
essentially.\,\footnote{It is reliably established by experiment
that the $S$ and $D$ wave contribution dominate in the $\pi\pi$
scattering cross sections in the isospin $I$\,=\,0 and 2 channels at
$\sqrt{s}<1.5 $\,GeV (see, for example, data
\cite{Pr73,Hy73,Hy75,Duru73,Hoo77, Pe77,Al95,Al98,Gu01}). The
$\pi\pi$ partial wave amplitudes $T^I_J(s)$\,=\,$
\{\eta^I_J(s)\exp[2i\delta^I_J(s)]-1\}/[2i \rho_{\pi^+}(s)]$ with
$J$\,=\,0,\,2 and $I$\,=\,0 (where $\delta^I_J(s)$ and $\eta^I_J(s)$
are the phase and inelasticity for the $J$ wave in the $\pi\pi$
scattering channel with the isospin $I$; $\rho_{\pi^+}(s)$\,=\,$
(1-4m^2_{\pi^+} /s)^{1/2}$) reach their unitarity limits at some
values of $\sqrt{s}$ in the region of interest and demonstrate both
the smooth energy dependence and the sharp resonance oscillations.
The $T^0_2(s)$ amplitude is dominated by the $f_2(1270)$ resonance
contribution. The $T^0_0(s)$ amplitude contains the $\sigma_0(600)$
and $f_0(980)$ resonance contributions. The $\sigma_0(600)$
resonance contribution is compensated strongly by the chiral
background near the $\pi\pi$ threshold to provide for the observed
smallness of the $\pi\pi$ scattering length $a^0_0$ and the Adler
zero in $T^0_0(s)$ at $s$\,$\approx$\,$m^2_\pi/2$ \cite{AS94a,AS07,
AK06,AK07a}. $|T^0_0(s)|$ reaches the unitary limit in the
0.85--0.9\,GeV region and has the narrow deep (practically up to
zero) right under the $K\bar K$ threshold caused by the destructive
interference of the $f_0(980)$ resonance contribution with the large
smooth background. It is established also that the $\pi\pi$
scattering in the $I$\,=\,0 channel is elastic up to the $K\bar K$
channel threshold in the very good approximation, but directly above
this the inelasticity $\eta^0_0(s)$ shows the sharp jump due to the
production of the $f_0(980)$ resonance coupled strongly with the
$K\bar K$channel.} In addition, the inelastic
$\gamma\gamma$\,$\to$\,$K^+K^-$\,$\to$\,$\pi\pi$ rescattering plays
the important role in the $f_0(980)$ resonance region (for the first
time this process was noted in Refs. \cite{ADS82a,ADS82b}).

\begin{figure}
\includegraphics[width=18pc]{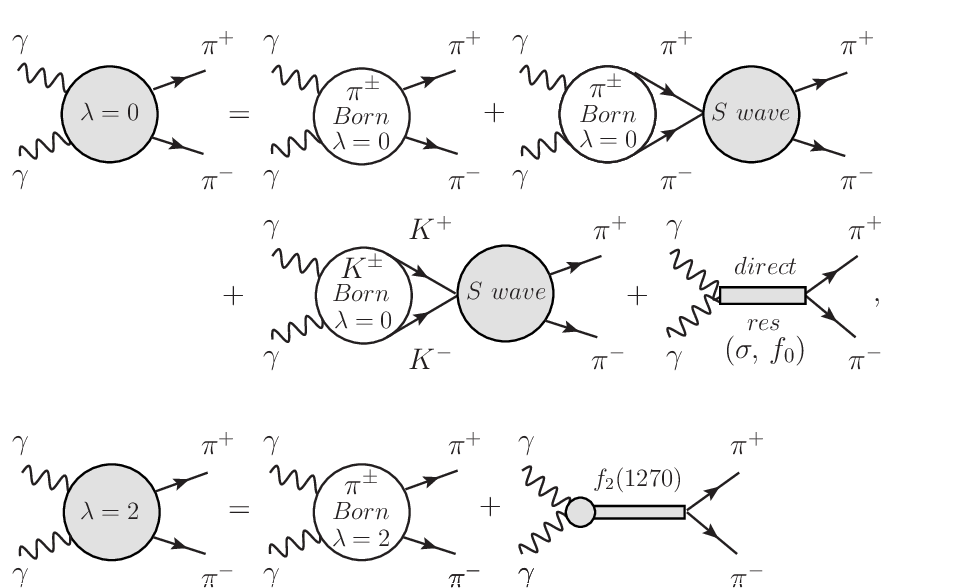}
\caption{{\footnotesize The diagrams corresponding to the helicity
amplitudes (\ref{eq:10}) and (\ref{eq:11}) for the $\gamma\gamma
$\,$\to$\,$\pi^+\pi^-$ reaction.}}\label{Diagr-gg-pipi}
\end{figure}

\begin{figure}
\includegraphics[width=18pc]{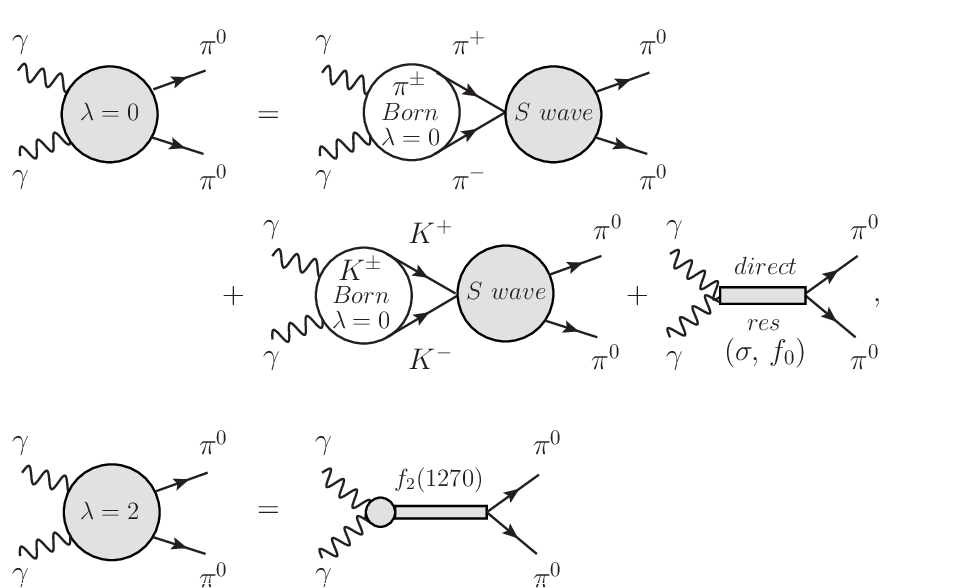}
\caption{{\footnotesize The diagrams corresponding to the helicity
amplitudes (\ref{eq:12}) and (\ref{eq:13}) for the $\gamma\gamma
$\,$\to$\,$\pi^0\pi^0$ reaction.}}\label{Diagr-gg-pi0pi0}
\end{figure}

\begin{figure}\includegraphics[width=18pc]{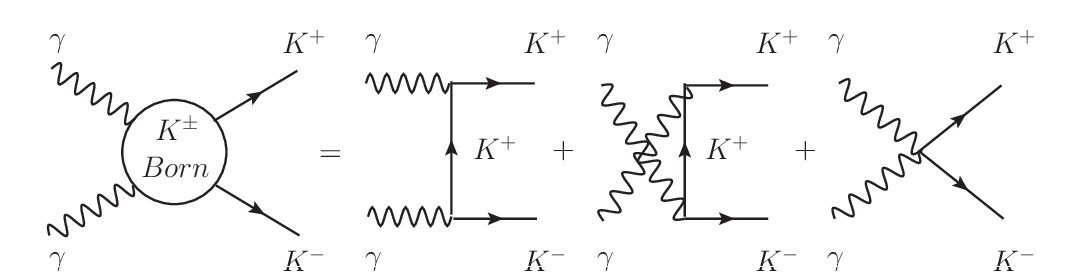} % height=7pc {MKBornFig.eps}
\caption{{\footnotesize The Born diagrams for $\gamma
\gamma$\,$\to$\,$K^+K^-$.}}\label{Born-gg-KK}\end{figure}

So, we use the model for the helicity, $M_\lambda$, and partial,
$M_{\lambda J}$, amplitudes of $\gamma\gamma$\,$\to$\,$\pi\pi$ in
which the Born charged $\pi$ and $K$ exchanges modified by the
strong final-state interactions in the $S$ and $D_2$ waves and the
direct transitions of the resonances in two photons are taken into
account (see, in addition, \cite{AS88,AS92,AS94b,AS05,AS07,AS08a,
AS08b,Me83,AcGu98,PMUW08,OR08,MeNO08a}),
\begin{eqnarray}
&
M_0(\gamma\gamma\to\pi^+\pi^-;s,\theta)=M^{\mbox{\scriptsize{Born}}
\,\pi^+}_0(s,\theta)+ & \nonumber\\ &
+\widetilde{I}^{\pi^+}_{\pi^+\pi^-}
(s)\,T_{\pi^+\pi^-\to\pi^+\pi^-}(s)+ & \nonumber\\ &
+\widetilde{I}^{K^+}_{K^+K^-}(s)\,T_{K^+K^-\to\pi^+\pi^-}(s)+M^{\mbox
{\scriptsize{direct}}}_{\mbox{\scriptsize{res}}}(s), & \label{eq:10}
\end{eqnarray}
\begin{eqnarray}
& M_2(\gamma\gamma\to\pi^+\pi^-;s,\theta)=M^{\mbox{\scriptsize{Born}}
\,\pi^+}_2(s,\theta)+ & \nonumber\\
&+80\pi d^2_{20}(\theta) M_{\gamma\gamma\to
f_2(1270)\to\pi^+\pi^-}(s),\ \ & \label{eq:11}
\end{eqnarray}
\begin{eqnarray}
& M_0(\gamma\gamma\to\pi^0\pi^0;s,\theta)=M_{00}(\gamma\gamma\to
\pi^0 \pi^0;s)= & \nonumber\\ & =
\widetilde{I}^{\pi^+}_{\pi^+\pi^-}(s)\,T_{\pi^+\pi^-\to\pi^0\pi^0}(s)+ & \nonumber\\
& +\widetilde{I}^{K^+}_{K^+K^-}(s)\,T_{K^+K^-
\to\pi^0\pi^0}(s)+M^{\mbox{\scriptsize{direct}}}_{\mbox{\scriptsize{res}}}
(s), & \label{eq:12}
\end{eqnarray}
\begin{eqnarray}&
M_2(\gamma\gamma\to\pi^0\pi^0;s,\theta)= & \nonumber\\
& =5d^2_{20}(\theta)M_{22}(\gamma \gamma\to\pi^0\pi^0;s)= &
\nonumber\\ & =80\pi d^2_{20}(\theta)M_{\gamma\gamma\to
f_2(1270)\to\pi^0\pi^0}(s)\,, & \label{eq:13}
\end{eqnarray} where $d^2_{20}(\theta)$\,=\,$(\sqrt{6}/4)\sin^2 \theta$.
The diagrams of the above amplitudes are adduced in Figs.
\ref{Born-gg-pipi}, \ref{Diagr-gg-pipi}, \ref{Diagr-gg-pi0pi0}, and
\ref{Born-gg-KK}.

The first terms in the right sides of Eqs. (\ref{eq:10}) and
(\ref{eq:11}) are the Born helicity amplitudes
$\gamma\gamma$\,$\to$\,$ \pi^+\pi^-$ corresponding to the elementary
one pion exchange mechanism (see Fig. \ref{Born-gg-pipi}). Their
explicit forms are adduced in Appendix 8.1. The terms in Eqs.
(\ref{eq:10}) and (\ref{eq:12}), containing the
$T_{\pi^+\pi^-\to\pi^+\pi^-}(s)$\,=\,$ [2T^0_0(s)+T^2_0(s)]/3$,
$T_{\pi^+\pi^-\to\pi^0\pi^0}(s)$\,=\,$ 2[T^0_0(s)-T^2_0(s)]/3$, and
$T_{K^+K^-\to\pi^+\pi^-}(s)$\,=\,$ T_{K^+K^-\to\pi^0\pi^0}(s)$
amplitudes, take into account the strong final-state interactions in
the $S$ wave. Eqs. (\ref{eq:10}) and (\ref{eq:12}) imply that
$T_{\pi^+\pi^-\to\pi\pi}(s)$ and $T_{K^+K^-\to\pi\pi}(s)$ in the
loops of the $\gamma\gamma$\,$\to$\,$\pi^+\pi^-$\,$\to$\,$\pi\pi$
and $\gamma\gamma$\,$\to$\,$K^+K^-$\,$\to$\,$\pi\pi$ rescatterings
(see Figs. \ref{Diagr-gg-pipi} and \ref{Diagr-gg-pi0pi0}) are on the
mass shell. In so doing the $\widetilde{I}^{\pi^+}_{\pi^+\pi^-}(s)$
and $\widetilde{I}^{K^+}_{K^+K^-}(s)$ functions are the amplitudes
of the triangle loop diagrams describing the transitions
$\gamma\gamma$\,$\to$\,$\pi^+\pi^-$\,$\to$\,($scalar\ state\ with\
a\ mass = \sqrt{s}$) and $\gamma\gamma$\,$\to$\,$K^+K^-$\,$\to$\,($
scalar\ state\ with\ a\ mass = \sqrt{s}$), in which the meson pairs
$\pi^+\pi^-$ and $K^+K^-$ are produced by the electromagnetic Born
sources, see Figs. \ref{Born-gg-pipi} and \ref{Born-gg-KK}. Their
explicit forms are adduced in Appendixes 8.1 and 8.3. The amplitude
$M^{\mbox{\scriptsize{direct}}}_{\mbox {\scriptsize{res}}}(s)$ in
Eqs. (\ref{eq:10}) and (\ref{eq:12}) caused by the direct coupling
constants of the $\sigma_0(600)$ and $f_0(980)$ with photons, and
the $f_2(1270)$ production amplitude $M_{\gamma\gamma\to
f_2(1270)\to\pi^+\pi^-}(s)=M_{\gamma\gamma\to
f_2(1270)\to\pi^0\pi^0}(s)$ in Eqs. (\ref{eq:11}) and (\ref{eq:13})
are specified below.



Let us show by the example of the $S$ wave amplitudes
$M_{00}(\gamma\gamma $\,$\to$\,$\pi^+ \pi^-;s)$ and
$M_{00}(\gamma\gamma$\,$\to$\,$ \pi^0\pi^0;s)$ that the unitary
condition requirement or the Watson theorem \cite{Wat52} about
interaction in final-state holds in the model under consideration.
First of all note that the 4$\pi$ and 6$\pi$ channel contributions
are small for $\sqrt{s}$\,$<$\,1\,GeV \cite{Pr73,Hy73,Gr74} and
consequently $T_{\pi^+\pi^-\to K^+K^-
}(s)$\,=\,$e^{i\delta^0_0(s)}|T_{\pi^+ \pi^-\to K^+K^- }(s)|$ and
$M^{\mbox {\scriptsize{direct}}}_{\mbox {\scriptsize{res}}}(s)$
=\,$\pm e^{i\delta^0_0(s)}|M^{\mbox {\scriptsize{direct}}}_{\mbox
{\scriptsize{res}}}(s)|$ for
$4m_\pi^2$\,$\leq$\,$s$\,$\leq$\,$4m^2_K$
\cite{AS05,AS07,AS08a,AK06,AK07a}. Taking into account that
Im$\widetilde{I}^{\pi^+}_{\pi^+\pi^-}(s)$\,=\,$\rho_{
\pi^+}(s)M_{00}^{\mbox{\scriptsize{Born}}\,\pi^+}(s)$ one finds
\begin{eqnarray}
&M_{00}(\gamma\gamma\to\pi^+\pi^-;s)=M^{\mbox
{\scriptsize{Born}}\,\pi^+}_{00}(s)+ & \nonumber\\ & +
\widetilde{I}^{\pi^+}_{\pi^+\pi^-}(s)T_{\pi^+\pi^-\to\pi^+\pi^-}(s)+
& \nonumber\\ & +\widetilde{I}^{K^+}_{K^+K^-}(s)T_{K^+K^-
\to\pi^+\pi^-}(s) +M^{\mbox {\scriptsize{direct}}}_{\mbox
{\scriptsize{res}}}(s)= &
\nonumber\\ &=(\mbox{for\ }2m_\pi\leq\sqrt{s}\leq2m_K)= & \nonumber\\
&=\frac{2}{3}e^{i\delta^0_0(s)}A(s)+\frac{1}{3}
e^{i\delta^2_0(s)}B(s),&\label{M00pipi}\end{eqnarray}
\begin{eqnarray} &M_{00}(\gamma\gamma\to\pi^0\pi^0;s)=
\widetilde{I}^{\pi^+}_{\pi^+\pi^-}(s)T_{\pi^+\pi^-\to\pi^0\pi^0}(s)+
& \nonumber\\ &
+\widetilde{I}^{K^+}_{K^+K^-}(s)T_{K^+K^-\to\pi^0\pi^0}(s)+ M^{\mbox
{\scriptsize{direct}}}_{\mbox {\scriptsize{res}}}(s)= &
\nonumber\\ &=(\mbox{for\ }2m_\pi\leq\sqrt{s}\leq2m_K)= & \nonumber\\
&=\frac{2}{3}e^{i\delta^0_0(s)}A(s)-\frac{2}{3}
e^{i\delta^2_0(s)}B(s),&\label{M00pi0pi0}\end{eqnarray}
%----------------------------------------------------------------------------------------------------
where A(s) and B(s) are the real functions.\,\footnote{ $A(s)$\,=\,$
M^{ \mbox{\scriptsize{Born}}\,\pi^+}_{00}(s)\cos\delta^0_0(s)
+(1/\rho_{\pi^+}(s))\mbox{Re}[\widetilde{I}^{\pi^+}_{\pi^+\pi^-}(s)]
\sin\delta^0_0(s)$
$+\frac{3}{2}\widetilde{I}^{K^+}_{K^+K^-}(s)|T_{K^+K^-
\to\pi^+\pi^-}(s)|\pm\frac{3}{2}|M^{\mbox {\scriptsize{direct}}}_{
\mbox{\scriptsize{res}}}(s)|$ and $B(s)$\,= $
M^{\mbox{\scriptsize{Born}}\,\pi^+}_{00}(s)\cos\delta^2_0(s)+
(1/\rho_{\pi^+}(s)) \mbox{Re}[\widetilde{I}^{\pi^+}_{\pi^+\pi^-}(s)
]\sin\delta^2_0(s)$.} Eqs. (\ref{M00pipi}) and (\ref{M00pi0pi0})
show that at one with the Watson theorem the phases of the $S$ wave
amplitudes $\gamma\gamma\to\pi\pi$ with $I$\,=\,0 and 2 coincide
with the phases of the $\pi\pi$ scattering  $\delta^0_0(s)$ and
$\delta^2_0(s)$, respectively, in the elastic region (below the
$K\bar K$ threshold).

We use the following notations and normalizations for the
$\gamma\gamma$\,$\to$\,$\pi\pi$ cross sections:
\begin{eqnarray}&\sigma(\gamma\gamma\to\pi^+\pi^+;|\cos\theta|\leq0.6)
\equiv\sigma=\sigma_0+\sigma_2,& \\ &\sigma(\gamma\gamma
\to\pi^0\pi^0;|\cos\theta|\leq0.8)\equiv\tilde{\sigma}=
\tilde{\sigma}_0+\tilde{\sigma}_2,\label{CS-Notations}\end{eqnarray}
\begin{eqnarray}&\sigma_\lambda=\frac{\rho_{\pi^+}(s)}{64\pi
s}\int^{0.6}_{-0.6}|M_\lambda(\gamma
\gamma\to\pi^+\pi^-;s,\theta)|^2d\cos\theta,\ \ &\\
&\tilde{\sigma}_\lambda=\frac{\rho_{\pi^+}(s)}{128\pi
s}\int^{0.8}_{-0.8}|M_\lambda(\gamma\gamma\to\pi^0\pi^0;s,
\theta)|^2d\cos\theta.\ \ &\label{CS-Normaizations}\end{eqnarray}
Hereinafter the corresponding partial cross sections will be denoted
as $\sigma_{\lambda J}$ and $\tilde{\sigma}_{\lambda J}$.

Before fitting data it is helpful to center on a simplified
(qualitative) scheme of their description.

\begin{figure} % \includegraphics[height=10.2pc]{fig10.eps}
\includegraphics[width=21pc]{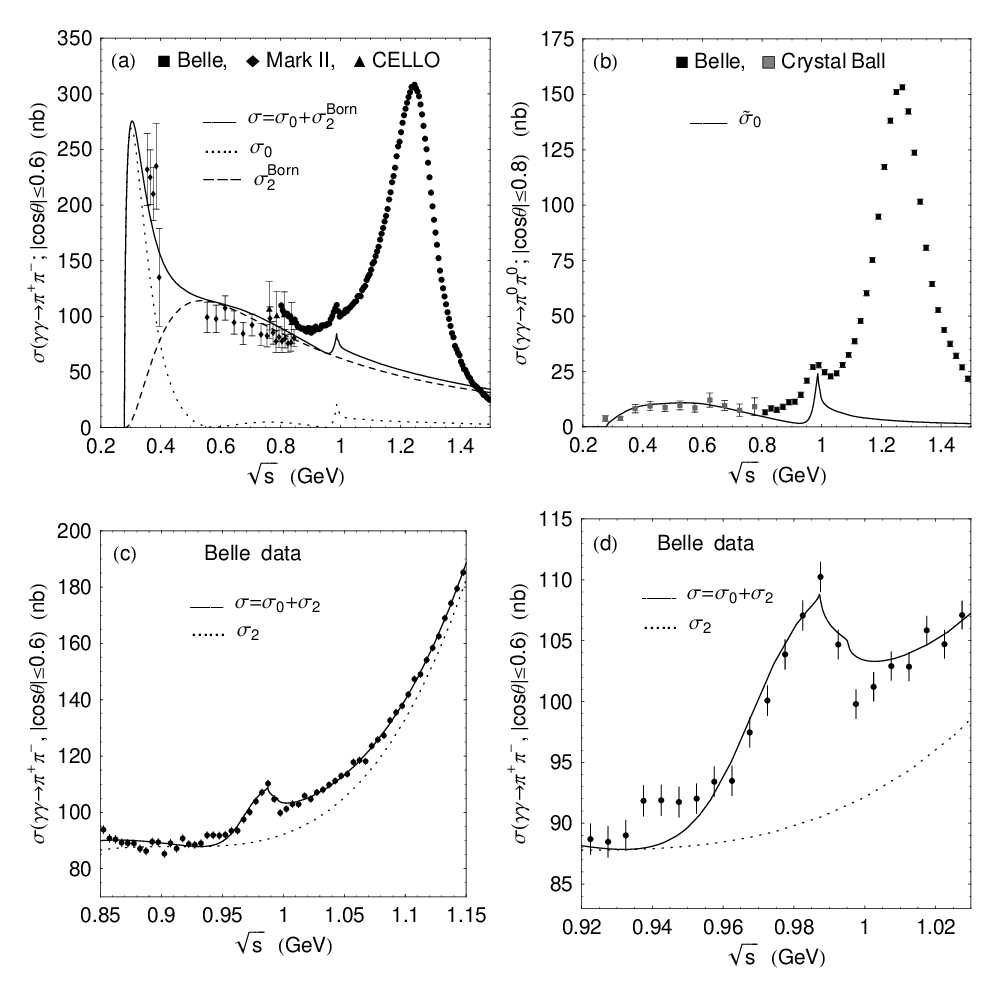}
\caption{{\footnotesize Theoretical curves in plots (a) and (b)
correspond to the simplest model which incorporates only the Born
contributions $\gamma\gamma$\,$\to$\,$\pi^+\pi^-$, from $\pi$
exchange, and $\gamma\gamma$\,$\to$\,$ K^+K^-$, from $K$ exchange,
modified for strong final-state interactions in the $S$ wave. Plot
(c) illustrates the description of the Belle data in the $f_0(980)$
region. (d) The fragment of (c).}}\label{Simple-f0-gg-pipi}
\end{figure}

In Fig. \ref{Simple-f0-gg-pipi}(a) from Ref. \cite{AS08a} are
adduced the theoretical curves for the cross section
$\sigma=\sigma_0+\sigma^{ \mbox{\scriptsize{Born}}}_2$ and its
components $\sigma_0$ and $\sigma^{\mbox{\scriptsize{Born}}}_2$
corresponding to the simplest variant of the above model in which
only the $S$ wave Born amplitudes $\gamma\gamma$\,$\to$\,$\pi^+\pi^-
$ and $\gamma\gamma$\,$\to$\,$ K^+K^-$ are modified by the pion and
kaon strong final-state interactions. As for all higher partial
waves with $\lambda$\,=\,0 and 2, they are taken in the Born
point-like approximation \cite{AS05,AS08a}. This modification
results in appearing the $f_0(980)$ resonance signal in $\sigma_0$,
the value and shape of which agree very well with the Belle data,
see Fig. \ref{Simple-f0-gg-pipi}(a). From comparing  the
corresponding curves in Figs. \ref{Simple-f0-gg-pipi}(a) and
\ref{Belle-gg-pipi-pi0pi0}(a) it follows that the $S$ wave
contribution to $\sigma(\gamma\gamma\to\pi^+\pi^-;|\cos\theta|
\leq0.6)$ is small for $\sqrt{s}>0.5$ GeV. It is clear that the
$f_2(1270)$ resonance contribution is the main element required for
the description of the Belle data on
$\gamma\gamma$\,$\to$\,$\pi^+\pi^-$ in the $\sqrt{s}$ region from
0.8 up to 1.5 GeV. For describing data only near the $f_0(980)$
resonance one can the large non-coherent background under the
resonance, caused by $\sigma_2$, approximate by a polynomial of
$\sqrt{s}$. The result of a such fit is shown in Figs.
\ref{Simple-f0-gg-pipi}(c) and \ref{Simple-f0-gg-pipi}(d)
\cite{AS08a}.

By Fig. \ref{Diagr-gg-pi0pi0} and Eq. (\ref{eq:12}) taking into
account the final-state interactions in the Born
$\gamma\gamma$\,$\to$\,$\pi^+\pi^-$ and $\gamma\gamma$\,$\to$\,$
K^+K^-$ amplitudes leads to the prediction of the $S$ wave amplitude
of the $\gamma\gamma\to\pi^0\pi^0$ reaction \cite{AS05,AS07,AS08a,
AS08b}. In Fig. \ref{Simple-f0-gg-pipi}(b), the $\gamma\gamma$\,$\to
$\,$\pi^0\pi^0$ cross section, evaluated in the outlined above
manner, are compared with the Crystal Ball and Belle
data.\,\footnote{Note that the step of $\sqrt{s}$ for the Crystal
Ball and Belle data, shown in Fig. \ref{Simple-f0-gg-pipi}(b), is 50
MeV and 20 MeV respectively.} In view of the fact that no fitting
parameters are used for the construction of $\tilde{\sigma}_0$, one
should accept that the agreement with the data is rather well at
$\sqrt{s}\leq0.8$ GeV, i.e., in the $\sigma(600)$ resonance region.
It is clear also that at $\sqrt{s}$\,$>$\,0.8 GeV the $f_2(1270)$
resonance responsibility region begins.

So, already at this stage it emerges the following. First, if the
direct coupling constants of $\sigma(600)$ and $f_0(980)$ with
$\gamma\gamma$ are included in fitting their role will be negligible
in agreement with the four-quark model prediction \cite{ADS82a,
ADS82b}. Second, by Eqs. (\ref{eq:10}) and (\ref{eq:12}) the
$\sigma(600) $\,$\to$\,$\gamma\gamma$ and
$f_0(980)$\,$\to$\,$\gamma\gamma$ decays are described by the
triangle loop rescattering diagrams $Resonance $\,$\to$\,$(\pi^+
\pi^-$, $K^+K^-)$\,$\to$\,$\gamma \gamma$ and, consequently, are the
four-quark transitions \cite{AS05,AS07,AS08a,AS08b}.

\begin{figure}
\includegraphics[height=10.2pc]{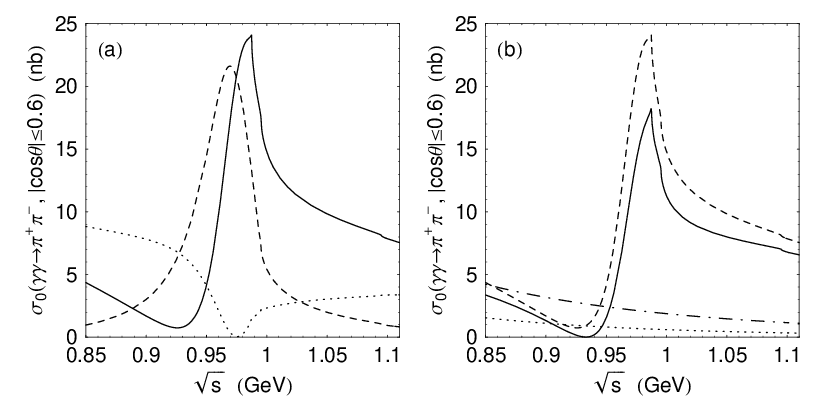}
\caption{{\footnotesize The structure of the $f_0(980)$ signal in
$\sigma_0$. (a) The contributions from the $\gamma\gamma\to
K^+K^-\to\pi^+\pi^-$ (dashed line),
$\gamma\gamma\to\pi^+\pi^-\to\pi^+\pi^-$ (dotted line) rescattering
amplitudes, and  their sum (solid line). (b) The dashed line is
identical to the solid one in (a), the dotted and dot-dashed lines
show the $\sigma^{\mbox{\scriptsize{Born}}}_0$ and
$\sigma^{\mbox{\scriptsize{Born}}}_{00}$ cross sections,
respectively ($\sigma^{\mbox{\scriptsize{Born}}}_0$\,$
<$\,$\sigma^{\mbox{\scriptsize{Born}}}_{00}$ because of the
destructive interference between the $S$ and higher partial waves),
and the solid line corresponds to the resulting $f_0(980)$ signal in
$\sigma_0$.}}\label{Structure-f0}\end{figure}

The interesting and important feature of the $f_0(980)$ signal in
$\gamma\gamma$\,$\to$\,$\pi^+\pi^-$ is its complicated structure
which is shown by Figs. \ref{Structure-f0}(a) and
\ref{Structure-f0}(b). The $\gamma\gamma$\,$\to$\,$K^+K^-$\,$\to
$\,$\pi\pi$ rescattering amplitude plays the determinant role
transferring the $f_0(980)$ peak from the $T_{K^+K^-\to\pi \pi}(s)$
amplitude to the $\gamma\gamma $\,$\to$\,$\pi\pi$ one.\,\footnote{It
provides the natural scale of the $f_0(980)$ production cross
section in $\gamma\gamma$ collisions \cite{AS05}. The maximum of the
cross section $\sigma(\gamma\gamma$\,$ \to$\,$K^+K^-$\,$\to
$\,$f_0(980)$\,$\to$\,$\pi^+\pi^-)$ is controlled by the product of
the ratio of the squares of the coupling constants
$R_{f_0}$\,=\,$g^2_{f_0K^+K^-}/g^2_{f_0 \pi^+\pi^-}$ and the value
$|\widetilde{I}^{K^+}_{K^+K^- }(4m^2_{K^+})|^2$. Its estimate gives
$\sigma(\gamma\gamma$\,$\to$\,$K^+K^-$\,$\to$\,$f_0(980)$\,$
\to$\,$\pi^+\pi^-;|\cos\theta|\leq0.6)\approx0.6\times0.62\alpha^2
R_{f_0}/m^2_{f_0}\approx8$\,nb\,$\times R_{f_0}$, where
$\alpha$\,=\,1/137 and $m_{f_0}$ is the  $f_0(980)$ mass.} The
$\gamma\gamma$\,$\to$\,$\pi^+\pi^-$\,$\to$\,$\pi^+\pi^-$
rescattering in its turn transfers the narrow deep under the $K\bar
K$ threshold from the $T_{\pi\pi\to\pi \pi}(s)$ amplitude, see the
footnote 18, to the $\gamma\gamma $\,$\to$\,$\pi\pi$ one. The
interference of the resonance $\gamma\gamma$\,$\to$\,$K^+K^-
$\,$\to$\,$\pi^+\pi^-$ amplitude with the
$\gamma\gamma$\,$\to$\,$\pi^+ \pi^-$\,$\to$\,$\pi^+\pi^-$
amplitude\,\,\footnote{Note that the relative sign between these
amplitudes is fixed surely \cite{AS05,AS08a}.} exerts the essential
effect on the resulting  shape of the $f_0(980)$ signal, as
indicated by Fig. \ref{Structure-f0}(a). As for the Born
contributions, their influence on the resulting  shape of the
$f_0(980)$ signal is small, see Fig. \ref{Structure-f0}(b).

\begin{figure}
\includegraphics[height=10.2pc]{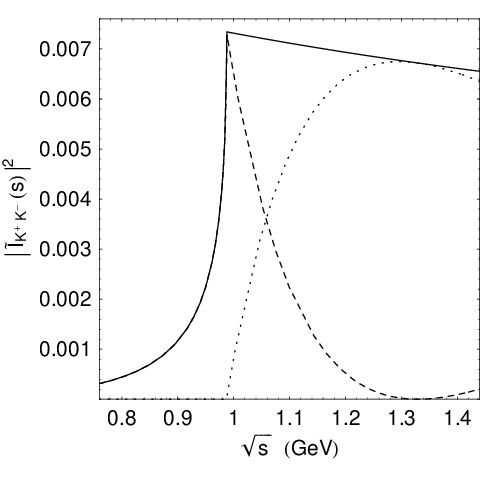}
\caption{{\footnotesize The solid curve shows
$|\widetilde{I}^{K^+}_{K^+K^-}(s)|^2$ as a function of $\sqrt{s}$
(see Appendix 8.3); the dashed and dotted curves above the $K^+K^-$
threshold correspond to the contributions of the real and imaginary
parts of $\widetilde{I}^{K^+}_{K^+K^-}(s)$, respectively.
}}\label{IKKK}\end{figure}

Once more notable fact lies in the drastic change of the $f_0(980)$
production amplitude $\gamma\gamma$\,$\to$\,$K^+K^-$\,$\to$\,$
f_0(980)$ in the $f_0(980)$ peak region \cite{AS05,AS08a} just as
the $\gamma\gamma$\,$\to$\,$K^+K^- $\,$\to$\,$ a_0(980)$ amplitude
in the $a_0(980)$ region \cite{AS88}, see Section 5. In the cross
section its contribution is proportional to
$|\widetilde{I}^{K^+}_{K^+K^-}(s)|^2$, see Eqs. (\ref{eq:10}) and
(\ref{eq:12}). The function $|\widetilde{I}^{K^+}_{K^+K^-}(s)|^2$
decreases drastically immediately under the $K^+K^-$ threshold,
i.e., in the $f_0(980)$ resonance region, see Fig.
\ref{IKKK}.\,\footnote{The function $|\widetilde{I}^{K^+}_{K^+K^-}
(s)|^2$ decreases relatively its maximum at $\sqrt{s}$\,=\,$2m_{K^+}
$\,$\approx$\,0.9873 GeV by 1.66, 2.23, 2.75, 3.27, and 6.33 times
at $\sqrt{s}$\,=\,0.98, 0.97, 0.96, 0.95, and 0.9 GeV,
respectively.} Such a behavior of the $f_0(980)$ two-photon
production amplitude reduces strongly the left slope of the the
$f_0(980)$ peak defined by the resonance amplitude $T_{K^+K^-\to\pi
\pi}(s)$. That is why one cannot approximate the
$f_0(980)$\,$\to$\,$ \gamma\gamma$ decay width by a constant even in
the region $m_{f_0}$\,--\,$\Gamma_{f_0}/2$\,$\leq$\,$\sqrt{s}$\,$
\leq$\,$m_{f_0 }$\,+\,$\Gamma_{f_0}/2$ \cite{AS05}.

So, the above consideration teaches us that all simplest
approximations of the $f_0(980)$ signal shape observed in the
$\gamma\gamma\to\pi^+\pi^-$ and $\gamma\gamma\to\pi^0\pi^0$ cross
sections can give only a rather relative information on the
$f_0(980)$ state two-photon production mechanism and the $f_0(980)$
parameters.

%----------------------------------------------------------------------------------------------------

\begin{figure}
\includegraphics[width=21pc]{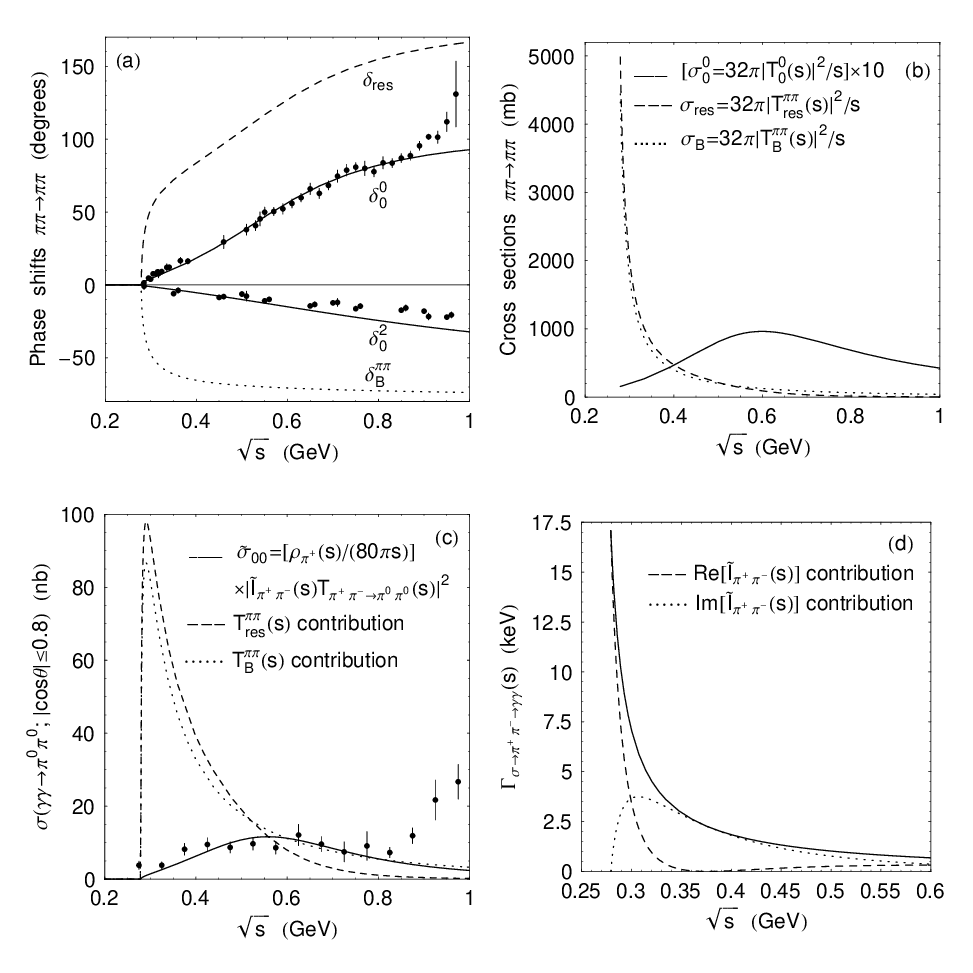}
\caption{{\footnotesize The figure demonstrates the chiral shielding
effect in the reactions $\pi\pi$\,$\to$\,$\pi\pi$ and $\gamma\gamma
$\,$ \to$\,$\pi^0\pi^0$. All the plots have been taken from Ref.
\cite{AS07} dedicated the lightest scalar in the $SU(2)_L\times
SU(2)_R$ linear $\sigma$ model.}} \label{FigPRL}\end{figure}

Fortunately, the already current knowledge of the dynamics of
$T_{\pi^+\pi^-\to\pi\pi}(s)$ [$T^0_0(s)$, $T^2_0(s)$] and
$T_{K^+K^-\to\pi\pi}(s)$ strong interaction amplitudes allows to
advance in understanding the signals about the light scalar mesons
which the data on the $\gamma\gamma\to\pi\pi$ reaction send to us.
In fitting data we use the model for the $T^0_0(s)$ and
$T_{K^+K^-\to\pi\pi}(s)$ amplitudes which was suggested and used for
the joint analysis of the data on the $\pi^0\pi^0$ mass spectrum in
the $\phi$\,$\to$\,$\pi^0\pi^0\gamma$ decay, the $\pi\pi$ scattering
at $2m_\pi<\sqrt{s}<1.6$ GeV, and the $\pi\pi$\,$\to$\,$K\bar K$
reaction \cite{AK06,AK07a}. The $T^0_0(s)$ model takes into account
the contributions of the $\sigma(600)$  and $f_0(980)$ resonances,
their mixing, and the chiral background with the large negative
phase which shields the $\sigma(600)$ resonance (see additionally
\cite{AS94a,AS07,Ac08a,Ac08b}). Eqs. (\ref{eq:10}) and (\ref{eq:12})
transfer the effect of the chiral shielding of the $\sigma(600)$
resonance from the $\pi\pi$ scattering into the
$\gamma\gamma$\,$\to$\,$\pi\pi$ amplitudes. This effect are
demonstrated by Fig. \ref{FigPRL}(a) with the help of the $\pi\pi$
scattering phases $\delta_{\mbox{\scriptsize{res}}}(s)$,
$\delta^{\pi\pi}_B(s)$, and $\delta^0_0(s)$ [see Eqs.
(\ref{T00})--(\ref{Trespipi}) and (\ref{Phase00})], and by Figs.
\ref{FigPRL}(b) and \ref{FigPRL}(c) with the help the corresponding
cross sections of the $\pi\pi$\,$\to$\,$\pi\pi$ and $\gamma\gamma
$\,$\to$\,$\pi^0\pi^0$ reactions. As seen from Fig. \ref{FigPRL}(c),
if it were not for such a shielding the $\gamma\gamma$\,$\to$\,$
\pi^0 \pi^0$ cross section nearby the threshold would be not
$(5-10)$ nb but approximately 100 nb due to the $\pi^+\pi^-$ loop
mechanism of the $\sigma(600)$\,$\to$\,$\gamma\gamma$ decay
\cite{AS07}. The decay width corresponding to this mechanism,
$\Gamma_{\sigma\to \pi^+\pi^-\to\gamma\gamma}(s)$, is shown in Fig.
\ref{FigPRL}(d); see also Eq. (\ref{G-sigma-pipi-gg}) in Appendix
8.1.

According Refs. \cite{AK06,AK07a}, we write
\begin{equation}T^0_0(s)=T^{\pi\pi}_B(s)+e^{2i\delta^{\pi\pi}_{B}(s)}
T^{\pi\pi}_{\mbox{\scriptsize{res}}}(s)\,,\label{T00}\end{equation}
\begin{equation}T^{\pi\pi}_B(s)=\{\exp[2i\delta^{\pi\pi}_{B}(s)]
-1\}/[2i\rho_{\pi^+}(s)]\,,\label{TBpipi}\end{equation}
\begin{equation}T^{\pi\pi}_{\mbox{\scriptsize{res}}}(s)=
\{\eta^0_0(s)\exp[2i\delta_{\mbox{\scriptsize{res}}}(s)]-1\}/
[2i\rho_{\pi^+}(s)]\,,\label{Trespipi}\end{equation}
\begin{equation} T_{K^+K^-\to
\pi^+\pi^-}(s)=e^{i[\delta^{\pi\pi}_{B}(s)+\delta^{K\bar K}_{B}(s)]}
T^{K\bar K\to\pi\pi}_{\mbox{\scriptsize{res}}}(s)\,,
\label{TKKpipi}\end{equation} where $\delta^{\pi\pi}_{B}(s)$ and
$\delta^{K\bar K}_{B}(s)$ are the phase of the elastic $S$ wave
background in the $\pi\pi$ and $K\bar K$ channels with $I$\,=\,0;
the $\pi\pi$ scattering phase
\begin{equation}\delta^0_0(s)=\delta ^{\pi\pi}_B(s)+
\delta_{\mbox{\scriptsize{res}}}(s).\label{Phase00}\end{equation}

The amplitudes of the $\sigma(600)$\,--\,$f_0(980)$ resonance
complex in Eqs. (\ref{eq:10}), (\ref{eq:12}), (\ref{T00}),
(\ref{Trespipi}), and (\ref{TKKpipi}) are \cite{AK06,AK07a}
\begin{equation}T^{\pi\pi}_{\mbox{\scriptsize{res}}}(s)=
3\,\frac{g_{\sigma\pi^+\pi^-}\Delta_{f_0}(s)+
g_{f_0\pi^+\pi^-}\Delta_\sigma(s)} {32\pi[D_\sigma(s)D_{f_0}(s)-
\Pi^2_{f_0\sigma}(s)]}\,,\label{Trespipi-1}\end{equation}
\begin{equation}T^{K\bar K\to\pi\pi}_{\mbox{\scriptsize{res}}}(s)
=\frac{g_{\sigma K^+K^-}\Delta_{f_0}(s)+g_{f_0K^+K^-}
\Delta_\sigma(s)} {16\pi[D_\sigma(s)D_{f_0}(s)-\Pi^2_{f_0\sigma}
(s)]}\,,\label{TresKKpipi}
\end{equation} \begin{equation}
M^{\mbox{\scriptsize{direct}}}_{\mbox{\scriptsize{res}}}(s)=
s\,e^{i\delta^{\pi\pi}_B(s)}\, \frac{g^{(0)}_{\sigma\gamma\gamma}
\Delta_{f_0}(s)+ g^{(0)}_{f_0\gamma\gamma}\Delta_\sigma(s)}
{D_\sigma(s)D_{f_0}(s)-\Pi^2_{f_0\sigma}(s)}\,,\label{MresDir}
\end{equation} where
$\Delta_{f_0}(s)$\,=\,$D_{f_0}(s)g_{\sigma\pi^+\pi^-}$\,+\,$\Pi_{f_0\sigma}
(s)g_{f_0\pi^+\pi^-}$ and $\Delta_\sigma(s)$
=\,$D_\sigma(s)g_{f_0\pi^+\pi^-}$\,+\,$\Pi_{f_0\sigma}
(s)g_{\sigma\pi^+\pi^-}$, $g^{(0)}_{\sigma\gamma \gamma}$ and
$g^{(0)}_{f_0\gamma\gamma}$ are the direct coupling constants of the
$\sigma$ and $f_0$ resonances with the photons. We use the
expressions for the $\delta^{\pi\pi}_B(s)$ and $\delta^{K\bar
K}_B(s)$ phases, the propagators of the $\sigma(600)$ and $f_0(980)$
resonances $1/D_\sigma(s)$ and $1/D_{f_0}(s)$, and the polarization
operator matrix element $\Pi_{f_0\sigma}(s)$ from \cite{AK06} (see
also Appendix). The $m_{f_0}$ was free, the other parameters in the
strong amplitudes ($m_\sigma$, $g_{\sigma\pi^+\pi^-}$,
$g_{f_0K^+K^-}$, etc.) correspond to variant 1 from Table 1 from
this paper.\,\footnote{Removing the misprint in the sign of the
constant $C\equiv C_{f_0\sigma}$ we use $C_{f_0\sigma}=-0.047$ GeV.
Notice that our principal conclusions [the insignificance of the
direct transition $\gamma\gamma\to Light\ Scalar$ and the dominant
role of the four-quark transition $\gamma\gamma\to(\pi^+\pi^+,\,
K^+K^-)\to Light\ Scalar$] are independent on a specific variant
from \cite{AK06,AK07a}.} We also put $\eta^2_0(s)$\,=\,1 for all
$\sqrt{s}$ under consideration and take $\delta^2_0(s)$ from
\cite{AchS03}.

The amplitudes of the $f_2(1270)$ resonance production in Eqs.
(\ref{eq:11}) and (\ref{eq:13}) are
\begin{eqnarray}M_{\gamma\gamma\to
f_2(1270)\to\pi^+\pi^-}(s)=M_{\gamma\gamma\to
f_2(1270)\to\pi^0\pi^0}(s)=\nonumber \\
=\frac{\sqrt{s}\,G_2(s)
\sqrt{(2/3)\Gamma_{f_2\to\pi\pi}(s)/\rho_{\pi^+}(s)}}{m^2_{f_2}-s-i
\sqrt{s}\Gamma^{\mbox{\scriptsize{tot}}}_{f_2}(s)}\,.\mbox{\qquad\ \
} && \label{Mgg-f2-pipi}\end{eqnarray} The main contribution in its
total width $\Gamma^{\mbox{\scriptsize{tot}} }_{f_2}(s)=\Gamma_{
f_2\to\pi\pi}(s)+\Gamma_{f_2\to K\bar K}(s)+\Gamma_{f_2\to4\pi}(s)$
is given by the $\pi\pi$ partial decay width
\begin{eqnarray}\Gamma_{f_2\to\pi\pi}(s)=\Gamma^{\mbox{\scriptsize{tot}}
}_{f_2}(m^2_{f_2})B(f_2\to\pi\pi)\times\nonumber \\
\times\frac{m^2_{f_2}}{s}\frac{q^5_{\pi^+}(s)}{q^5_{\pi^+
}(m^2_{f_2})}\frac{D_2(q_{\pi^+}(m^2_{f_2 })r_{f_2})}{D_2
(q_{\pi^+}(s)r_{f_2})}\,,\label{Gf2-pipi}\end{eqnarray} where
$D_2(x)$\,=\,$9+3x^2+x^4$, $q_{\pi^+}(s)$\,=\,$\sqrt{s}
\rho_{\pi^+}(s)/2$, $r_{f_2}$ is the interaction range and
$B(f_2$\,$\to$\,$\pi\pi)$\,=\,0.848 \cite{PDG08}. The small
contributions of $\Gamma_{f_2\to K\bar K}(s)$ and $\Gamma_{f_2
\to4\pi}(s)$ are the same ones as in \cite{AS08a}. The parameter
$r_{f_2}$ \cite{Ma90,Bo90,Be92,Mo07a,Ue08,AS08a,AS08b} controls the
relative form of the $f_2(1270)$ resonance wings and is very
important especially for fitting data with small errors.

The amplitude $G_2(s)$ in Eq. (\ref{Mgg-f2-pipi}) describes the
coupling of the $f_2(1270)$ resonance with the photons,
\begin{equation}G_2(s)=\sqrt{\Gamma^{(0)}_{f_2\to\gamma\gamma}
(s)}+i\frac{M_{22}^{\mbox{\scriptsize{Born}}\,\pi^+}(s)}{16\pi}
\sqrt{\frac{2}{3}\rho_{\pi^+}(s)\Gamma_{f_2\to\pi\pi}(s)}\,.
\label{G2}\end{equation} The explicit form of the
$M_{22}^{\mbox{\scriptsize{Born}}}(s)$ amplitude is in Appendix 8.1,
Eq. (\ref{M22B}). The $f_2(1270)$\,$\to$\,$\gamma\gamma$ decay width
is \begin{equation}\Gamma_{f_2\to\gamma\gamma} (s)=|G_2(s)|^2
\label{Gf2-gg}\end{equation} and \begin{equation}
\Gamma^{(0)}_{f_2\to\gamma\gamma}(s)=\frac{m_{f_2}}{\sqrt{s}}
\Gamma^{(0)}_{f_2\to\gamma\gamma}(m^2_{f_2})
\frac{s^2}{m^4_{f_2}}\label{G0f2-gg}\end{equation} [here the $s^2$
factor, and also the $s$ factor in Eq. (\ref{MresDir}), is caused by
the gauge invariance requirement]. The second term in  $G_2(s)$
corresponds to the rescattering
$f_2(1270)$\,$\to$\,$\pi^+\pi^-$\,$\to$\,$\gamma\gamma$ with  the
real pions in intermediate state\,\footnote{That is, it corresponds
to the imaginary part of the $f_2(1270)\to\pi^+\pi^-\to\gamma\gamma$
amplitude.} and ensure the fulfillment of the Watson theorem
requirement for the amplitude $\gamma\gamma$\,$\to$\,$\pi\pi$ with
$\lambda$\,=\,$J$\,=\,2 and $I$\,=0 under the first inelastic
threshold. This term gives a small contribution, less then 6\%, in
$\Gamma_{f_2\to\gamma\gamma}( m^2_{f_2})$.\,\footnote{As for a real
part of the $f_2(1270)$\,$\to$\,$\pi^+\pi^-$\,$\to$\,$\gamma\gamma$
amplitude, its modulus is far less than the one of the direct
transition amplitude as different estimations show.}

The simplest approximation (\ref{G0f2-gg}) of the main contribution,
$\Gamma^{(0)}_{f_2\to\gamma\gamma}(s)$, in the
$f_2(1270)$\,$\to$\,$\gamma\gamma$ decay width is completely
adequate to the current state of both theory and experiment. The
parameter $\Gamma^{(0)}_{f_2\to\gamma\gamma}(m^2_{f_2})
$\,=\,$\frac{1}{5}[g^2_{f_2\gamma\gamma}/(16\pi)]m^3_{f_2}$ in Eq.
(\ref{G0f2-gg}) accumulates effectively lack of knowledge of the
values of the amplitudes responsible for the
$f_2(1270)$\,$\to$\,$\gamma\gamma$ decay. By the above reasons, see
Section 3, it is generally agreed that the direct quark-antiquark
transition $q\bar q$\,$\to$\,$\gamma\gamma$ dominates in the
$f_2(1270)$\,$\to$\,$\gamma\gamma$ decay and its amplitude is
characterized by the $g_{f_2\gamma\gamma}$ coupling constant. As
shown in \cite{AS88,AS05,AS07,AS08a,AS08b,AS09,AS10a, AS10b} and as
we state here step by step, the situation is quite different in the
case of the light scalar mesons.

Now everything is ready to come to the discussion of fitting the
Belle data on the $\gamma\gamma$\,$\to$\,$\pi^+\pi^-$ and
$\gamma\gamma$\,$ \to$\,$\pi^0\pi^0$ cross sections which was
carried out in \cite{AS08a,AS08b}.

\begin{figure}
\includegraphics[width=21pc]{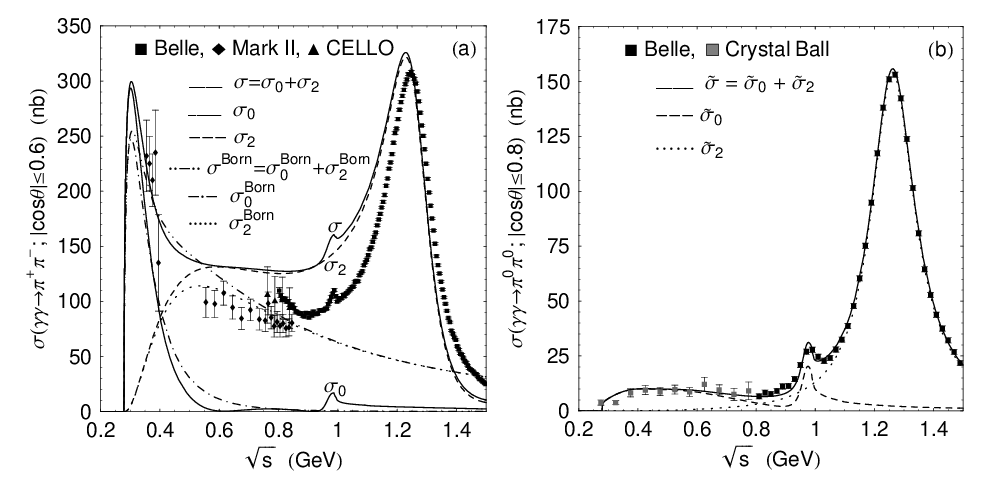}
\caption{{\footnotesize Cross sections for the
$\gamma\gamma\to\pi^+\pi^-$ and $\gamma\gamma\to\pi^0\pi^0$
reactions. Only statistical errors are shown for the Belle data
\cite{Mo07b,Ue08}. The curves in plot (a) are described in the text
and on the figure. The curves in plot (b) are the result of the fit
to the data on the $\gamma\gamma\to\pi^0\pi^0$ reaction.}}
\label{Fit-gg-pi0pi0}
\end{figure}

One considers firstly fitting the
$\gamma\gamma$\,$\to$\,$\pi^0\pi^0$ cross section only, see Fig.
\ref{Fit-gg-pi0pi0}(b), which has the smaller background
contributions under the $f_0(980)$ and $f_2(1270)$ resonances then
the $\gamma\gamma$\,$\to $\,$\pi^+\pi^-$ cross section, compare
Figs. \ref{Fit-gg-pi0pi0}(a) and \ref{Fit-gg-pi0pi0}(b). The solid
curve in Fig. \ref{Fit-gg-pi0pi0}(b), describing these data rather
well, corresponds the following parameters of the model:
$m_{f_2}$\,=\,1.269 GeV,
$\Gamma^{\mbox{\scriptsize{tot}}}_{f_2}(m^2_{f_2})$\,=\,0.182 GeV,
$r_{f_2}$\,=\,8.2 GeV$^{-1}$, $\Gamma_{f_2\to\gamma\gamma
}(m_{f_2})$\,=\,3.62 keV [$\Gamma^{(0)}_{f_2\to\gamma\gamma
}(m_{f_2})$\,=\,3.43 keV], $m_{f_0}$\,=\,0.969 GeV,
$g^{(0)}_{\sigma\gamma\gamma}$\,=\,0.536 GeV$^{-1}$ and
$g^{(0)}_{f_0\gamma\gamma}$\,=\,0.652 GeV$^{-1}$.\,\footnote{The
formally calculated errors in the significant parameters of the
model are negligible due to the high statistical accuracy of the
Belle data. The model dependence of adjustable parameter values is
the main source of their ambiguity.} The fitting indicates smallness
of the direct coupling constants $g^{(0)}_{\sigma\gamma \gamma}$ and
$g^{(0)}_{f_0\gamma \gamma}$:\, $\Gamma^{(0)}_{\sigma \to\gamma
\gamma}(m^2_{\sigma})$\,=\,$|m^2_\sigma
g^{(0)}_{\sigma\gamma\gamma}|^2/(16\pi m_\sigma)$\,=\,0.012 keV and
$\Gamma^{(0)}_{f_0\to\gamma \gamma}(m^2_{f_0})$\,=\,$|m^2_{f_0}
g^{(0)}_{f_0\gamma\gamma}|^2/(16\pi m_{f_0})$\,=\,0.008 keV, in
accordance with the prediction \cite{ADS82a,ADS82b}\,\footnote{One
notes that the small values of these coupling constants are grasped
in fitting due to the interference of the  $M^{\mbox
{\scriptsize{direct}}}_{\mbox {\scriptsize{res}}}(s)$ amplitude, see
Eqs. (\ref{eq:10}), (\ref{eq:12}), and (\ref{MresDir}), with the
contributions of the dominant rescattering mechanisms. In such a
case not the specific above values of $g^{(0)}_{\sigma\gamma
\gamma}$ and $g^{(0)}_{f_0\gamma \gamma}$ are important, but the
fact of their relative smallness, corresponding to
$\Gamma^{(0)}_{\sigma \to\gamma \gamma}(m^2_{\sigma})$ and
$\Gamma^{(0)}_{f_0\to\gamma \gamma}(m^2_{f_0})$ both
$\ll$\,0.1\,keV.} The dominant rescattering mechanisms give the
$\sigma(600)$\,$\to$\,$\pi^+\pi^-$\,$\to$\,$ \gamma \gamma$ width
$\approx$\,(1\,--\,1.75) keV averaged in the region
$0.4<\sqrt{s}<0.5$ GeV \cite{AS07}, see Fig. \ref{FigPRL}(d), and
the $f_0(980)$\,$\to$\,$K^+K^-$\,$\to$\,$\gamma\gamma$ width
$\approx$\,$(0.15-0.2)$ keV averaged over the resonance mass
distribution \cite{AS05}.

But such a fitting of the $\gamma\gamma$\,$\to$\,$\pi^0\pi^0$ cross
section comes into conflict with the data on $\gamma\gamma$\,$\to
$\,$\pi^+\pi^-$, see solid curve for $\sigma$\,=\,$\sigma_0$\,+\,$
\sigma_2$ in Fig. \ref{Fit-gg-pi0pi0}(a). This is connected with the
large Born contribution in $\sigma_2$ and its strong constructive
(destructive) interference with the $f_2(1270)$ resonance
contribution at $\sqrt{s}<m_{f_2}$ ($\sqrt{s}>m_{f_2}$), which are
absent in $\gamma\gamma$\,$\to$\,$\pi^0\pi^0$. We faced this
challenge in \cite{AS08a} and ibidem suggested the following
solution. Matters can be improved by the introduction of the common
cutting form factor $G_{\pi^+}(t,u)$ in the point-like Born
amplitudes $\gamma\gamma$\,$\to$\,$\pi^+ \pi^-$,
$M_\lambda^{\mbox{\scriptsize{Born}}}(s,\theta)\to G_{\pi^+}(t,u)
M_\lambda^{\mbox{\scriptsize{Born}}}(s,\theta)$, where $t$ and $u$
are the Mandelstam variables for the $\gamma\gamma$\,$\to $\,$\pi^+
\pi^-$ reaction.\,\footnote{Such a natural modification of the
point-like Born contribution was discussed in connection with the
data on the $\gamma\gamma$\,$\to$\,$\pi^+\pi^-$ \cite{Po86,Bo90,
Be92,MP87,MP88} and $\gamma\gamma$\,$\to$\,$K^+K^-(K^0\bar K^0)$
reactions \cite{AS92,AS94b}. But only the problem of the consistent
description of the Belle data on $\gamma\gamma$\,$\to$\,$\pi^+\pi^-$
and $\gamma\gamma$\,$\to $\,$\pi^0\pi^0$ indicates the modification
need of the Born sector of the model unambiguously \cite{AS08a,
AS08b}.} To show this we use, as an example, the expression for
$G_{\pi^+}(t,u)$ suggested in Ref. \cite{Po86},
\begin{equation}G_{\pi^+}(t,u)=\frac{1}{s}\left[\frac{m^2_{\pi^+}-t}{1-(u-
m^2_{\pi^+})/x^2_1}+\frac{m^2_{\pi^+}-u}{1-(t-m^2_{\pi^+})/x^2_1}\right],
\label{FF-Poppe}\end{equation} where $x_1$ is a free parameter.
\begin{figure}\begin{tabular}{c}
\includegraphics[height=14pc]{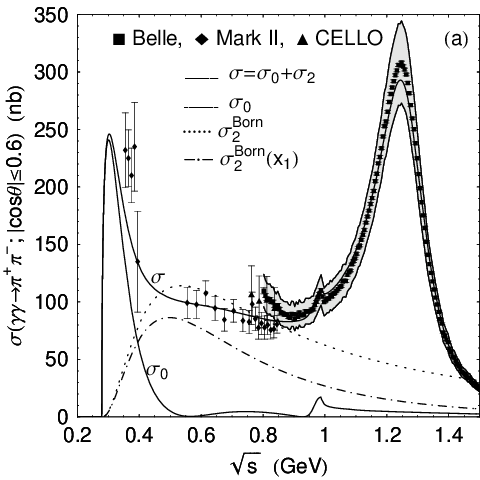}\\
\includegraphics[height=14pc]{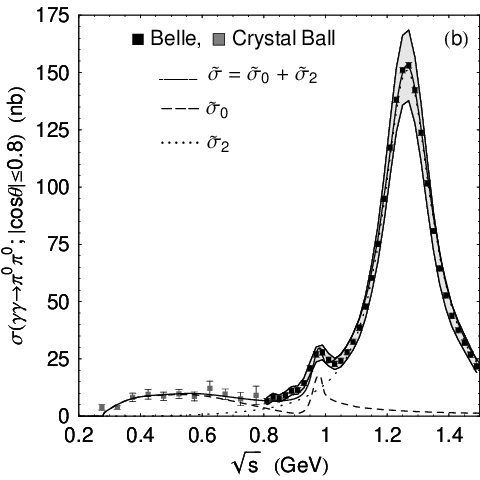}
\end{tabular}\\
\caption{{\footnotesize Joint description of the data on the cross
sections for the reactions $\gamma\gamma\to\pi^+\pi^-$ and
$\gamma\gamma\to\pi^0\pi^0$. The shaded bands correspond to the
Belle data \cite{Mo07b,Ue08} with the statistical and systematic
errors (errors are added quadratically). The curves are described in
the text and on the figures; $\sigma^{\mbox{\scriptsize{
Born}}}_2(x_1)$ in plot (a) is the Born cross section for the
$\gamma\gamma\to\pi^+\pi^-$ reaction with the inclusion of the form
factor.}}\label{Fit-gg-pipi}\end{figure} This ansatz is quite
acceptable in the physical region of the $\gamma\gamma\to\pi^+\pi^-$
reaction. Note that the form factor is introduced by changing the
amplitudes of the elementary one pion exchange
$M_\lambda^{\mbox{\scriptsize{Born}}\,\pi^+}(s,\theta)$ to
$M^{\mbox{\scriptsize{Born}}\,\pi^+}_\lambda(s,\theta;x_1)=G_{\pi^+}(t,u)
M_\lambda^{\mbox{\scriptsize{Born}}\,\pi^+}(s,\theta)$ and does not
break the gauge invariance of the tree approximation \cite{Po86}.
Replacing in (\ref{FF-Poppe}) $m_{\pi^+}$ by $m_{K^+}$ and $x_1$ by
$x_2$ we obtain also the form factor $G_{K^+}(t,u)$ for the Born
amplitudes $\gamma\gamma$\,$\to $\,$K^+K^-$.

The solid curves for $\sigma$\,=\,$\sigma_0$\,+\,$ \sigma_2$ and
$\tilde{\sigma}$\,=\,$ \tilde{\sigma}_0+\tilde{\sigma}_2$ in Figs.
\ref{Fit-gg-pipi}(a) and \ref{Fit-gg-pipi}(b) show the consistent
fitting of the data on the $\gamma\gamma$\,$\to$\,$\pi^+ \pi^-$
cross section in the region $0.85<\sqrt{s}<1.5$\,GeV and on the
$\gamma\gamma$\,$\to$\,$\pi^0\pi^0$ cross section in the region
$2m_\pi<\sqrt{s}<1.5$\,GeV with the form factors modificating the
point-like Born contributions. The obtained description is more than
satisfactory to within the Belle systematic errors which are shown
in Figs. \ref{Fit-gg-pipi}(a) and \ref{Fit-gg-pipi}(b) by means of
the shaded bands. We believe that such a fitting is completely
adequate for the statistic errors of both Belle measurements are so
small that to obtain the formally good enough value of $\chi^2$ in
the combined fitting of the $\pi^+\pi^-$ and $\pi^0\pi^0$ data in
the wide regions of $\sqrt{s}$ without taking the systematic errors
into consideration is practically impossible.\,\footnote{At the same
time we emphasize that the considerable systematic errors, the
sources of which are described in detail in Refs.
\cite{Mo07a,Mo07b,Ue08}, do not depreciate the role of the high
statistics of the data, which allows to resolve the small local
effects connected with the $f_0(980)$ resonance manifestation.}
%-----------------------------------------------------------------------------------------------
The curves in Fig. \ref{Fit-gg-pipi} correspond the following values
of the parameters: $m_{f_2}$\,=\,1.272\,GeV,
$\Gamma^{\mbox{\scriptsize{tot}}}_{f_2}(m^2_{f_2})$\,=\,0.196\,GeV,
$r_{f_2}$\,=\,8.2\,GeV$^{-1}$, $\Gamma_{f_2\to\gamma\gamma
}(m_{f_2})$\,=\,3.83\,keV [$\Gamma^{(0)}_{f_2\to\gamma\gamma
}(m_{f_2})$ =\,3.76\,keV],
%\,\footnote{Our $f_2(1270)$ two-photon
%width (3.83 keV) is more than the PDG fit (2.7 keV). A plausible
%reason of this difference is the fact that we calculate the
%background under the $f_2(1270)$ resonance theoretically. It is
%interesting that our similar treatment \cite{AS88} of the Crystal
%Ball data on the $\gamma\gamma\to\pi^0\eta $ reaction \cite{An86}
%gave for the two-photon width of the $a_2(1320)$ resonance the value
%1.3 keV greater than the PDG average 1 keV keeping the gold $q\bar
%q$ model relation 25:9.}
$m_{f_0}$\,=\,0.969\,GeV, $g^{(0)}_{\sigma\gamma\gamma}$\,=\,$-0.049
$\,GeV$^{-1}$ [$\Gamma^{(0)}_{\sigma\to\gamma \gamma}(m^2_{\sigma})
$\,\,\,miserable], $g^{(0)}_{f_0\gamma\gamma}$\,=\,0.718\,GeV$^{-1}$
[$\Gamma^{(0)}_{f_0\to\gamma\gamma}$ $(m^2_{f_0})$\,$\approx$\,0.01
keV], $x_1$\,=\,0.9 GeV and $x_2$\,=\,1.75 GeV. It is clear from
comparison of Figs. \ref{Fit-gg-pi0pi0}(b) and \ref{Fit-gg-pipi}(b)
that the form factor effect on the $\gamma\gamma\to\pi^0\pi^0$ cross
section is weak in contrast to the  $\gamma\gamma\to\pi^+\pi^-$ one
[compare Figs. \ref{Fit-gg-pi0pi0}(a) and \ref{Fit-gg-pipi}(a)] in
which the $\sigma_2$ contribution is modified mainly. One emphasizes
that all our conclusions about the mechanisms of the two-photon
decays (productions) of the $\sigma(600)$ and $f_0(980)$ resonances
are in force.\,\footnote{Notice that the point-like $\omega$ and
$a_2(1320)$ exchanges in the $\gamma\gamma\to\pi^0\pi^0$ and
$\gamma\gamma\to\pi^+\pi^-$ amplitude, respectively, which give the
contributions (mainly the $S$ wave one) in the cross section,
runaway with increasing the energy and comparable with the
$f_2(1270)$ resonance contribution even in its energy region, are
not observed experimentally. This was puzzled out in our paper
\cite{AS92} with the $\gamma\gamma\to\pi^0\eta$ example (the details
are discussed bellow in Section 5). The proper Reggeization of the
point-like exchanges with the high spins reduces the dangerous
contributions greatly. In addition, the partial cancelations between
the $\omega$ and $h_1(1170)$ exchanges in $\gamma\gamma\to\pi^0
\pi^0$ and the $a_2(1320)$ and $a_1(1260)$ exchanges in
$\gamma\gamma\to\pi^+\pi^-$ take place. As to the $\rho$ exchange in
$\gamma\gamma\to\pi\pi$, its contribution is small for
$g_{\rho\pi\gamma}^2 \approx g_{\omega\pi\gamma}^2/9$ and canceled
additionally by the $b_1(1235)$ one.}

Thus the physics of the  two-photon decays of the light scalar
mesons  acquires the rather clear outline. The mechanism of their
decays into $\gamma \gamma$ does not look like the mechanism of the
classic tensor $q\bar q$ meson decays, which is the direct
annihilation $q\bar q$\,$\to$\,$\gamma\gamma$. The light scalar
meson decays into $\gamma \gamma$ are suppressed in comparison with
the the tensor meson ones. They are caused by the rescattering
mechanisms, i.e., by the four-quark transitions
$\sigma(600)$\,$\to$\,$\pi^+\pi^-$\,$\to$\,$ \gamma\gamma$,
$f_0(980)$\,$\to$\,$K^+K^-$\,$\to$\,$\gamma\gamma$,
$a_0(980)$\,$\to$\,$K^+K^-$\,$\to$\,$\gamma\gamma$, etc. Such a
picture is suggested by experiment and supports the  $q^2\bar q^2$
nature of the light scalars. It is significant that in the scalar
meson case the longing for the exhaustive characteristic of their
coupling with photons via the constant values
$\Gamma_{0^{++}\to\gamma \gamma}(m^2_{0^{++}})$ by analogy with the
tensor mesons cannot be realized for quite a number reasons.

First of all it is clear that when we deal with resonances
accompanied by fundamental background, when two-photon decay widths
change sharply in the resonance region for close inelastic
thresholds, then there is no point in discussing the two-photon
width in the resonance peak.

In this connection it is interesting to consider the cross section
$\gamma\gamma$\,$\to $\,$\pi^+\pi^-$ caused only by the resonance
contributions, i.e.,
\begin{eqnarray}
& \sigma_{\mbox{\scriptsize{res}}}(\gamma\gamma\to\pi^+\pi^-;s)= &
\nonumber \\ & =[\rho_{\pi^+}(s)/(32\pi
s)]\left|\widetilde{I}^{\pi^+}_{\pi^+\pi^-}(s;x_1)
\,e^{2i\delta^{\pi\pi}_{B}(s)} T^{\pi\pi}_{\mbox{\scriptsize{res}
}}(s)\right.& \nonumber\\ & +\left.
\widetilde{I}^{K^+}_{K^+K^-}(s;x_2)\,T_{K^+K^-\to\pi^+\pi^-}(s)+M^{\mbox
{\scriptsize{direct}}}_{\mbox{\scriptsize{res}}}(s)\right|^2 &
\label{CSres-gg-pipi}\end{eqnarray} [see Eqs. (\ref{eq:10}) and
(\ref{Trespipi-1})--(\ref{MresDir})], where the functions
$\widetilde{I}^{\pi^+}_{\pi^+ \pi^-}(s;x_1)$ and
$\widetilde{I}^{K^+}_{K^+K^-}(s;x_2)$ are the analogs of the
$\widetilde{I}^{\pi^+}_{\pi^+\pi^-}(s)$ and
$\widetilde{I}^{K^+}_{K^+K^-}(s)$ functions constructed with taking
the form factors into account, see Appendixes 8.1 and 8.3.
\begin{figure}
\includegraphics[height=11pc]{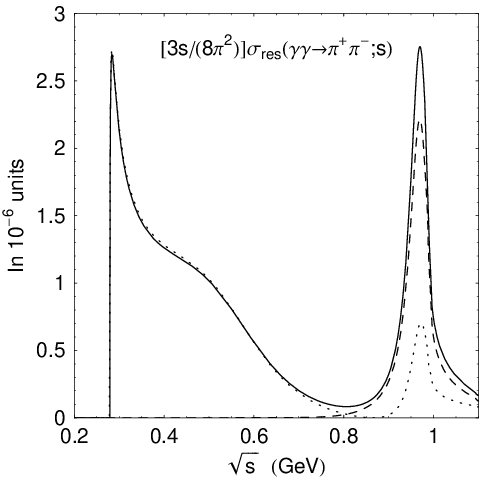}
\caption{{\footnotesize The integrand in Eq. (\ref{Gf0-gg})
corresponding to the joint fit to the $\gamma\gamma\to\pi^+\pi^-$
and $\gamma\gamma\to\pi^0\pi^0$ data (Fig. \ref{Fit-gg-pipi}) is
shown by the solid curve. The dotted and dashed curves show the
contributions from the resonant elastic $\gamma\gamma$\,$\to$\,$
\pi^+\pi^-$\,$\to$\,$\pi^+\pi^-$ and inelastic $\gamma\gamma$\,$
\to$\,$K^+K^-$\,$\to$\,$\pi^+\pi^-$ rescatterings, respectively.}}
\label{Fig-CSres-gg-pipi}\end{figure} Fig. \ref{Fig-CSres-gg-pipi}
shows  the dependence of the $\sigma_{\mbox{\scriptsize{res}}}
(\gamma\gamma$\,$\to$\,$\pi^ +\pi^-;s)$ cross section, multiplied by
the factor $3s/(8\pi^2)$, on the energy. In and around 1 GeV there
is the impressive peak from the $f_0(980)$ resonance due to the
inelastic $\gamma\gamma$\,$\to $\,$K^+K^- $\,$\to$\,$\pi^+\pi^-$
rescattering in the main. Following Refs. \cite{AS88,AS05,AS08b},
one determines the $f_0(980)$\,$\to$\,$\gamma\gamma$ decay width
averaged over the resonance mass distribution in the $\pi\pi$
channel
\begin{equation}\langle\Gamma_{f_0\to\gamma\gamma}\rangle_{\pi\pi}=
\int\limits_{0.8\mbox{\,\scriptsize{GeV}}}^{1.1\mbox{\,\scriptsize{GeV}}}
\frac{3s}{8\pi^2}\sigma_{\mbox{\scriptsize{res}}}
(\gamma\gamma\to\pi^+\pi^-;s)d\sqrt{s}.\label{Gf0-gg}\end{equation}
This value is the adequate functional characteristic of the coupling
of the $f_0(980)$ with $\gamma\gamma$. For the presented combined
fitting, $\langle\Gamma_{f_0\to\gamma\gamma}\rangle_{\pi\pi}
$\,$\approx$ 0.19\,keV \cite{AS08b}. Taking into account that the
wide $\sigma(600)$ resonance dominates in region $2m_\pi<\sqrt{s}
<0.8$\,GeV one obtains by analogy with (\ref{Gf0-gg})
$\langle\Gamma_{\sigma\to\gamma\gamma} \rangle_{\pi\pi}
$\,$\approx$\,0.45\,keV \cite{AS08b}. Note that the cusp near the
$\pi\pi$ threshold in the $[3s/(8\pi^2)]\sigma_{
\mbox{\scriptsize{res}}}(\gamma\gamma $\,$\to$\,$\pi^+\pi^-;s)$
expression, shown in Fig. \ref{Fig-CSres-gg-pipi}, is the
manifestation of the correction for the finite width in the
propagator of the scalar resonance. In Appendix 8.1 there are
adduced the transparent explanation of this phenomenon. In the total
$S$ wave amplitude $\gamma\gamma$\,$\to$\,$\pi\pi$ such a threshold
enhancement is absent due to shielding the resonance contribution in
the amplitude $T^0_0(s)$ by  the chiral background one, see, for
example, Fig. \ref{Fit-gg-pipi}(b).

The above examples, each in its manner give to feel clear the
nontriviality of accessing information about the $\sigma(600)$ and
$f_0(980)$ decays into $\gamma \gamma $. For instance, to determine
$\Gamma_{\sigma\to\gamma \gamma}(m^2_\sigma)$ directly from data is
impossible for the cross section in the $\sigma$ region is formed by
both resonance and compensating background. The specific dynamic
model of the total amplitude needs to their separation. The simple
Breit-Wigner is not enough here.

As for the $f_0(980)$ resonance, experimenters began to take into
consideration two from three important circumstances
\cite{Mo07a,Mo07b,Ue08} (see also \cite{PDG08,PDG10}), for which we
drew  attention in Ref. \cite{AS05}. Firstly, there was taken
account the correction for the finite width due to the coupling of
$f_0(980)$ with $K\bar K$ channel in the $f_0(980)$ resonance
propagator, which effects essentially on the shape of the $f_0(980)$
peak in the $\pi\pi$ channel. Secondly, there was taken into account
the interference of the $f_0(980)$ resonance with the background
though in the simplest form. But no model was constructed for the
$f_0(980)$\,$\to$\,$\gamma\gamma$ decay amplitude which was
approximated simply by a constant \cite{Mo07a,Mo07b,Ue08}. Fitting
data in this way the Belle collaboration extracted the values for
$\Gamma_{f_0\to\gamma \gamma}(m^2_{f_0})$ presented in Table
\ref{TabIV}. But the discussion needs how to understand these
values. First and foremost, one cannot use them for determining a
coupling constant $g_{f_0\gamma\gamma}$ in a effective lagrangian,
i.e., a constant of the direct transition $f_0(980)$\,$\to$\,$
\gamma\gamma$, because such a constant is small and does not
determine the $f_0(980)$\,$\to$\,$ \gamma\gamma$ decay, as shown
above. Until the model of the $f_0(980)$\,$\to$\,$\gamma \gamma$
decay amplitude is not specified, a meaning of the $\Gamma_{f_0\to
\gamma \gamma}(m^2_{f_0})$ values, extracted with the help of the
simplified parametrization, is rather vague.\,\footnote{The above
comments are true also in the $\sigma(600)$ resonance case.} In
principle the $\Gamma_{f_0\to\gamma \gamma}(m^2_{f_0})$ values in
Table \ref{TabIV} can be taken as the preliminary estimations of
$\langle\Gamma_{f_0\to\gamma\gamma}\rangle$, i.e., as the
$f_0(980)$\,$\to$\,$\gamma\gamma$ decay width averaged over the
hadron mass distribution \cite{AS88,AS05,AS08b}.

In the dispersion approach there are introduced usually the pole
two-photon widths $\Gamma_{R\to\gamma\gamma}(pole)$,
$R$\,=\,$\sigma,f_0$ to characterize the coupling $\sigma(600)$ and
$f_0(980)$ resonances with photons (see, for example, \cite{MP88,
MP90,BP99,Pe06,PMUW08}). These widths are determined through the
moduli of the complex pole residues of the $\gamma\gamma$\,$\to
$\,$\pi\pi$ and $\pi\pi$\,$\to$\,$\pi\pi$ partial amplitudes
constructed theoretically. Basing on our investigation \cite{AS07}
we would like to note the following. The residues of the above
amplitudes are essentially complex and cannot be used as any
coupling constants in a hermitian effective lagrangian. These
residues are ``dressed'' by the background for they relate to the
total amplitudes. As our analysis in the $SU(2)_L$\,$\times
$\,$SU(2)_R$ linear $\sigma$-model \cite{AS07} indicated, the
background effects essentially on the values and phases of the
residues. Thus the focus on the values of the $\Gamma_{R\to\gamma
\gamma}(pole)$ type in dispersion approach does not help to reveal
the mechanism of the two-photon decays of the scalar mesons and so
cannot shed light on the nature of the light scalars.

\vspace{0.4cm} \noindent{\large \bf \boldmath 5. Production of the
$a_0(980 )$ resonance in the reaction $\gamma\gamma\to\pi^0\eta$}
\vspace{0.2cm}

\noindent Our conclusions about the important role of the $K^+K^-$
loop mechanism in the two-photon production of the $a_0(980)$
resonance and its possible four-quark nature \cite{AS88,AS91,AS92}
were based on the analysis of the results of the first experiments
Crystal Ball \cite{An86} (see Fig. \ref{CB-gg-pieta}) and JADE
\cite{Oe90} on the $\gamma\gamma$\,$\to $\,$\pi^0\eta$ reaction.
Unfortunately, the large statistical errors in these data and the
rather rough step of the $\pi^0\eta$ invariant mass distribution
(equal 40 MeV in the Crystal Ball experiment and 60 MeV in the JADE
one) left many uncertainties.

%%%%%%%%%%%%%%%%%%%%%%%%%%%%%%%%%%%%%%%%%%%%%%%%%%%%%%%%%%%%%%%%%%%%%%%%%%%%%%%%%%%%%%%%%%
As we have mentioned in Subsection 3.2, recently, the Belle
Collaboration obtained new data on the $\gamma\gamma\to\pi^0\eta$
reaction at the KEKB $e^+e^-$ collider \cite{Ue09}, with statistics
three orders of magnitude higher than those in the preceding Crystal
Ball (336 evens) and JADE (291 events) experiments.

\begin{figure}\includegraphics[height=11pc]{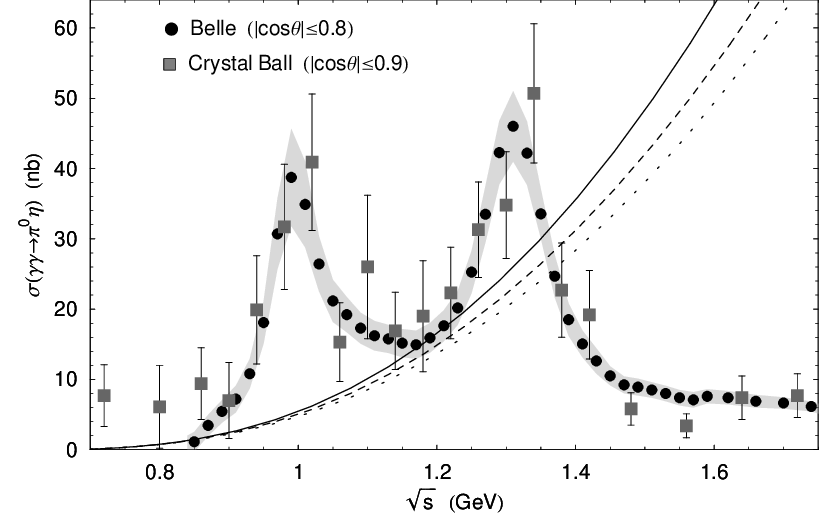}
\caption{{\footnotesize The Belle \cite{Ue09} and Crystal Ball
\cite{An86} data for the $\gamma\gamma\to\pi^0\eta$ cross section.
The average statistical error of the Belle data is approximately
$\pm0.4$ nb, the shaded band shows the size of their systematic
error. The solid, dashed, and dotted lines correspond to the total,
helicity 0, and $S$ wave $\gamma\gamma\to\pi^0\eta$ cross sections
caused by the elementary $\rho$ and $\omega$ exchanges for
$|\cos\theta|\leq0.8$.}}\label{BelleData-pieta}\end{figure}

The experiments revealed a specific feature of the
$\gamma\gamma\to\pi^0\eta$ cross section. It turned out sizable in
the region between the $a_0(980)$ and $a_2 (1320)$ resonances (see
Fig. \ref{BelleData-pieta}),\,\footnote{The JADE data \cite{Oe90} on
$\gamma\gamma\to\pi^0\eta$ are nonnormalized and therefore are not
shown in Fig. \ref{BelleData-pieta}.} which certainly indicates the
presence of additional contributions. These contributions must be
coherent with the resonance ones, because only two lowest $S$ and
$D_2$ partial waves dominate in the $\gamma \gamma\to\pi^0 \eta$
amplitude at the invariant mass of the $\pi^0\eta$ system $\sqrt{s}<
1.4$ GeV \cite{Ue09}. The authors of Ref. \cite{Ue09} performed the
phenomenological fitting of the $\gamma\gamma\to\pi^0\eta$ data
taking into account interference between resonance and background
contributions. It was found that the description of the $S$ wave
requires not only contributions from the $a_0(980)$ resonance and
the possible heavy $a_0(Y)$ resonance, but also a smooth background
whose amplitude is comparable with the amplitude of the $a_0(980)$
resonance at the maximum and has a large imaginary part \cite{Ue09}.
As a result, the background results in almost the quadrupling of the
cross section near the $a_0(980)$ peak and in the filling of the dip
between the $a_0(980)$ and $a_2 (1320)$ resonances. The origin of
such a significant background in the $S$ wave is unknown. Meanwhile,
the imaginary part of the background amplitude is due to the
contributions from real intermediate states, $\pi\eta$, $K\bar K$,
and $\pi\eta'$, and, naturally, requires the distinct dynamical
decoding.

In Refs. \cite{AS10a,AS10b} we shown that the observed experimental
pattern is a result of the interplay of many dynamical factors. To
analyze the data, we significantly developed a model previously
discussed in Refs. \cite{AS88,AS92,AS09}. The basis for this model
is an idea of what the $a_0(980)$ resonance can be as a suitable
candidate in four-quark states. There exists a number of significant
indications in favor of the four-quark nature of the $a_0(980)$;
see, for example, Refs. \cite{ADS80b,ADS82a,ADS82b,ADS84a,AS91,AI89,
Ac98,Ac03a,AK03,Ac08a}.
%---------------------------------------------------------------------------------------------------
The solution obtained by us for the $\gamma\gamma$\,$\to$\,$\pi^0
\eta$ amplitude is in agreement with the expectations of the chiral
theory for the $\pi\eta$ scattering length, with the strong coupling
of the $a_0(980)$ resonance with the $\pi\eta$, $K\bar K$, and
$\pi\eta'$ channels, and with the key role of the $a_0(980)$\,$\to
$\,$(K\bar K+\pi^0\eta+\pi^0\eta')$\,$\to$\,$\gamma \gamma$
rescattering mechanisms in the $a_0(980)$\,$\to$\,$\gamma\gamma$
decay. This picture is much in favor of the $q^2\bar q^2$ nature of
$a_0(980)$ resonance and is consistent with the properties of its
partners, $\sigma_0(600)$ and $f_0(980)$ resonances, in particular,
with those manifested in the $\gamma\gamma$\,$\to$\,$\pi\pi$
reactions. The important role of vector exchanges in the formation
of the non-resonant background in the $\gamma\gamma$\,$\to$\,$\pi^0
\eta$ reaction has been revealed and preliminary information on the
$\pi^0\eta$\,$\to$\,$\pi^0\eta$ reaction has been obtained also in
Refs. \cite{AS10a,AS10b}.

\begin{figure}\includegraphics[width=18pc]
{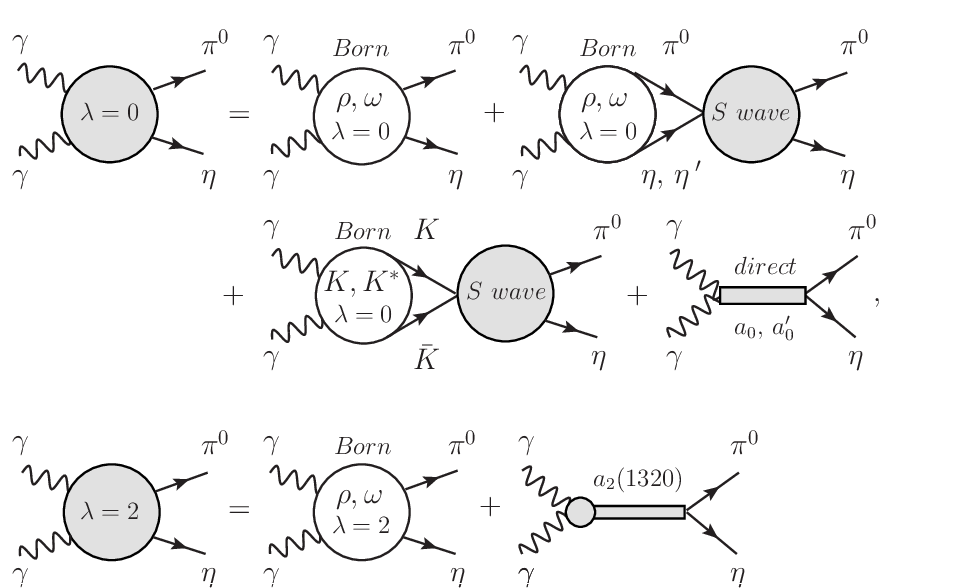} %\vspace{-2mm}
\caption{{\footnotesize The diagrams corresponding to the helicity
amplitudes $\gamma\gamma$\,$\to$\,$\pi^0\eta$, see Eqs.
(\ref{M0pieta}) and (\ref{M2pieta}).}}
\label{Ampl-gg-pieta}\end{figure}

\begin{figure}\begin{tabular}{l}
\includegraphics[width=15pc]{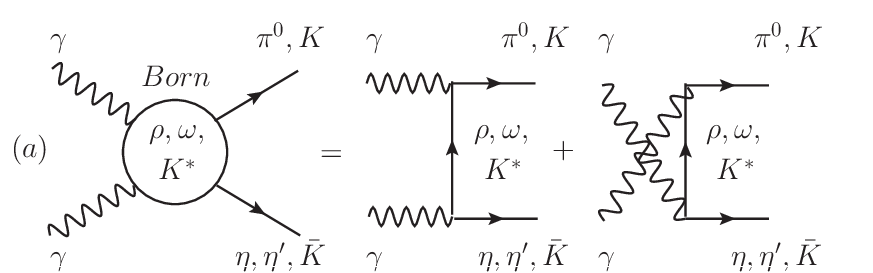}\\
\includegraphics[width=18pc]{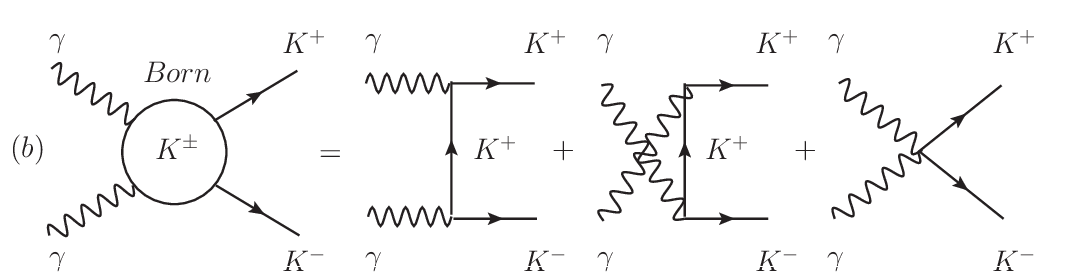}
\end{tabular}\\
\caption{{\footnotesize The Born $\rho$, $\omega$, $K^*$, and $K$
exchange diagrams for $\gamma\gamma$\,$\to$\,$\pi^0\eta$,
$\gamma\gamma$\,$\to$\,$\pi^0\eta'$, and $\gamma\gamma$\,$\to$\,$
K\bar K$.}}\label{BornVK-gg-pieta}\end{figure}

To analyze the Belle data, we constructed the helicity amplitudes
$M_\lambda$ and the corresponding partial amplitudes $M_{\lambda J}$
of the $\gamma\gamma$\,$\to$\,$ \pi^0\eta$ reaction, where the
electromagnetic Born contributions from $\rho$, $\omega$, $K^*$, and
$K$ exchanges modified by the form factors and strong elastic and
inelastic final-state interactions in the $\pi^0\eta$, $\pi^0\eta'$,
$K^+K^-$, and $K^0\bar K^0$ channels, and the contributions from the
direct interaction of the resonances with photons are taken into
account:
\begin{eqnarray}
& M_0(\gamma\gamma\to\pi^0\eta;s,\theta)=M^{\mbox{\scriptsize{Born}}
\,V}_0(\gamma\gamma\to\pi^0\eta;s,\theta)+ & \nonumber\\
& +\widetilde{I}^V_ {\pi^0\eta}(s)\,T_{\pi^0\eta\to\pi^0\eta}(s)+
\widetilde{I}^V_ {\pi^0\eta'}(s)\,T_{\pi^0\eta'\to\pi^0\eta}(s)+ &
\nonumber\\ & +\left(\widetilde{I}^{K^{*+}}_{K^+K^-}(s)-
\widetilde{I}^{K^{*0}}_{K^0\bar K^0}(s)+\widetilde{I}^{K^+}_{K^+K^-}
(s;x_2)\right)\times & \nonumber\\
& \times\,T_{K^+K^-\to \pi^0\eta}(s)+\widetilde{M}^{\mbox
{\scriptsize{direct}}}_{\mbox{\scriptsize{res}}}(s), &
\label{M0pieta} \\[0.14cm]
& M_2(\gamma\gamma\to\pi^0\eta;s,\theta)=M^{\mbox{
\scriptsize{Born}}\,V}_2(\gamma\gamma\to\pi^0\eta;s,\theta)+&
\nonumber\\ & +80\pi d^2_{20}(\theta)M_{\gamma\gamma\to
a_2(1320)\to\pi^0\eta}(s), & \label{M2pieta}
\end{eqnarray}
where $\theta$ is the polar angle of the produced $\pi^0$ (or
$\eta$) meson in the $\gamma\gamma$ center-of-mass system. Figs.
\ref{Ampl-gg-pieta} and \ref{BornVK-gg-pieta} show the diagrams
corresponding to these amplitudes.

The first terms in the right-hand parts of Eqs. (\ref{M0pieta}) and
(\ref{M2pieta}) represent the real Born helicity amplitudes, which
are the sums of the $\rho$ and $\omega$ exchange contributions equal
in magnitude and are written in the form \cite{AS88,AS92}
\begin{eqnarray}\label{MBornV0} && \mbox{\qquad\ \ \ }
M^{\mbox{\scriptsize{Born}}
\,V}_0(\gamma\gamma\to\pi^0\eta;s,\theta)=\nonumber \\
&& =2g_{\omega\pi\gamma}g_{\omega\eta\gamma}\frac{s}{4}\left[
\frac{tG_\omega(s,t)}{t-m^2_\omega}+\frac{uG_\omega(s,
u)}{u-m^2_\omega}\right],\mbox{\ } \\[0.14cm]
\label{MBornV2} && \mbox{\qquad\ \ \ }M^{\mbox{\scriptsize{Born}}
\,V}_2(\gamma\gamma\to\pi^0\eta;s,\theta)=\nonumber \\
&& =2g_{\omega\pi\gamma}g_{\omega\eta\gamma}\frac{m^2_\pi
m^2_\eta-tu}{4}\left[
\frac{G_\omega(s,t)}{t-m^2_\omega}+\frac{G_\omega(s,
u)}{u-m^2_\omega}\right],\mbox{\ \ \ \ } \end{eqnarray} where
$g_{\omega\eta\gamma}$\,=\,$\frac{1}{3}g_{\omega\pi
\gamma}\sin(\theta_i-\theta_P)$, $g^2_{\omega\pi\gamma}$\,=\,$12\pi
\Gamma_{\omega\to\pi\gamma} [(m^2_\omega-m^2_\pi)/(2m_\omega)
]^{-3}\approx0.519$ GeV$^{-2}$ \cite{PDG08,PDG10}, the ``ideal''
mixing angle $\theta_i$\,=\,35.3$ ^\circ$, $\theta_P$ is the mixing
angle in the pseudoscalar nonet, which is a free parameter; $t$ and
$u$ are the Mandelstam variables for the reaction $\gamma\gamma
\to\pi^0\eta$, $G_\omega(s,t)$ and $G_\omega(s, u)$ are the $t$ and
$u$ channel form factors [for the elementary $\rho$ and $\omega$
exchanges $G_\omega(s,t)$\,=\,$G_\omega(s, u)$\,=\,1]. In the
corresponding Born amplitudes for $\gamma\gamma\to\pi^0\eta'$,
$g_{\omega\eta'\gamma} $\,=\,$\frac{1}{3}g_{\omega\pi\gamma} \cos
(\theta_i-\theta_P)$, and for $\gamma\gamma\to K\bar K$ with the
$K^*$ exchange, $g^2_{K^{*+}K^+\gamma}\approx0.064$ GeV$^{-2}$ and
$g^2_{K^{*0} K^0\gamma}\approx0.151$ GeV$^{-2}$ \cite{PDG08,PDG10}.

Note that information on the bare Born sources of the $\gamma\gamma
\to\pi^0\eta$ reaction corresponding to the exchanges with the
quantum numbers of the $\rho$ and $\omega$ mesons (as well as the
b1(1235) and h1(1170) mesons) is very scarce in the nonasymptotic
energy range of interest. It is known certainly only that the
elementary $\rho$ and $\omega$ exchanges, whose contributions to the
$\gamma\gamma\to\pi^0\eta$ cross section (primarily to the S wave)
increase very rapidly with the energy, are not observed
experimentally (see Fig. \ref{BelleData-pieta}).\,\footnote{These
contributions are weakly sensible to the $\theta_P$ values under
discussion \cite{PDG10}. The curves in Fig. \ref{BelleData-pieta}
correspond to $\theta_P$\,=\,$-22^\circ$.} This fact was explained
in Ref. \cite{AS92} by the Reggeization of the elementary exchanges,
which suppresses dangerous contributions even in the range of 1-1.5
GeV. For this reason, we use the Regge type form factors
$G_\omega(s,t)$\,=\,$\exp[(t-m^2_\omega)b_\omega (s)]$, $G_\omega(s,
u)$\,=\,$\exp[(u-m^2_\omega)b_\omega(s)]$, where
$b_\omega(s)$\,=\,$b^0_\omega+(\alpha'_\omega/4) \ln[1+(s/s_0)^4]$,
$b^0_\omega$\,=\,0, $\alpha'_\omega$\,=\,0.8 GeV$^{-2}$ and
$s_0$\,=\,1 GeV$^2$ (and similar for the $K^*$ exchange).

As for the $b_1(1235)$ and $h_1(1170)$ exchanges, their amplitudes
have the form similar to Eqs (\ref{MBornV0}) and (\ref{MBornV2})
except for the common sign in the amplitude with helicity 0. The
estimates show that the axial-vector exchange amplitudes are at
least five times smaller than the corresponding vector exchange
amplitudes and we neglect their contributions.\,\footnote{The
exchanges with high spins in the $\gamma\gamma$\,$\to$\,$\pi^0\eta$
reaction are the correction against the background of the $K$
exchange contribution. This correction is required to describe the
data for $\gamma\gamma$\,$\to$\,$\pi^0\eta$, as shown bellow. As to
the $\gamma\gamma$\,$\to$\,$\pi\pi$ reactions, the corrections from
the high spin exchanges prove to be less significant against the
background of the summary contribution of the $\pi$ and $K$
exchanges, and we are unable to catch them at this stage.}
                                                                %%%% <----------\footnote on the high spin exchanges

The terms in Eq. (\ref{M0pieta}) proportional to the $S$ wave hadron
amplitudes $T_{\pi^0\eta\to\pi^0\eta}(s)$, $T_{\pi^0\eta'\to\pi^0
\eta}(s)$, and $T_{K^+K^-\to\pi^0\eta}(s)$ are attributed to the
rescattering mechanisms. In these amplitudes, we take into account
the contribution from the mixed $a_0(980)$ and heavy $a_0(Y)$
resonances (bellow, for brevity, they are denoted as $a_0$ and
$a'_0$, respectively) and the background contributions:
\begin{eqnarray} T_{\pi^0\eta\to\pi^0\eta}(s)=T_0^1 (s)=
\frac{\eta^1_0(s)e^{2i \delta^1_0(s)}-1}{2i\rho_{\pi\eta}(s)}=\ \
&& \nonumber \\ =T_{\pi\eta}^{bg}(s)+e^{2i\delta_{\pi\eta}^{bg}
(s)}T^{res}_{\pi^0\eta\to\pi^0\eta}(s)\,,\ \ \mbox{\qquad}&&
\label{Tpietapieta}\\
T_{\pi^0\eta'\to\pi^0\eta}(s)=T^{res}_{\pi^0\eta'
\to\pi^0\eta}(s)\,e^{i[\delta_{\pi\eta'}^{bg}(s)+\delta_{\pi\eta}^{bg}
(s)]},\ \ \ &\label{Tpieta1pieta} \\  T_{K^+K^-\to\pi^0\eta}(s)=
T^{res}_{K^+K^-\to\pi^0 \eta}(s)\,e^{i[\delta_{K\bar
K}^{bg}(s)+\delta_{\pi\eta}^{bg}(s)]},&\label{TKKpieta}
\end{eqnarray}
where $T_{\pi\eta}^{bg}(s)=(e^{2i\delta_{\pi\eta}^{bg}(s)}-1)/(2i
\rho_{\pi\eta}(s))$, $T^{res}_{\pi^0\eta\to\pi^0\eta}(s)
=(\eta^1_0(s)e^{2i\delta_{\pi\eta}^{res}(s)}-1)/(2i\rho_{
\pi\eta}(s))$, $\delta^1_0(s)=\delta_{\pi\eta}^{bg}(s)+
\delta_{\pi\eta}^{res}(s)$,
$\rho_{ab}(s)$\,=\,$\sqrt{s-m_{ab}^{(+)\,2}}\sqrt{
s-m_{ab}^{(-)\,2}}\Big/s$, $m_{ab}^{(\pm)}$\,=\,$m_b\pm m_a$,
$ab$\,=\,$\pi\eta$, $K^+K^-$, $K^0\bar K^0 $, $\pi\eta'$;
$\delta_{\pi\eta}^{bg}(s)$, $\delta_{\pi\eta'}^{bg}(s)$ and
$\delta_{K\bar K}^{bg}(s)$ are the phase shifts of the elastic
background contributions in the channels $\pi\eta$, $\pi\eta'$, and
$K\bar K$ with isospin $I=1$, respectively (see Appendix 8.2).

The amplitudes of the $a_0$\,--\,$a'_0$ resonance complex in Eqs.
(\ref{Tpietapieta})--(\ref{TKKpieta}) have the form analogous to
Eqs. (\ref{Trespipi-1}), (\ref{TresKKpipi}) \cite{AS10a,AS10b,AK06,
AS98}
\begin{equation}\label{Tres}
T^{res}_{ab\to\pi^0\eta}(s)=\frac{g_{a_0ab}\Delta_{a'_0}(s)+g_{a'_0ab
}\Delta_{a_0}(s)}{16\pi[D_{a_0}(s)D_{a'_0}(s)-\Pi^2_{a_0a'_0}(s)]}\,,
\end{equation} where
$\Delta_{a'_0}(s)$\,=\,$D_{a'_0}(s)g_{a_0\pi^0\eta}+\Pi_{a_0a'_0}
(s)g_{a'_0\pi^0\eta}$ and $\Delta_{a_0}(s)$
=\,$D_{a_0}(s)g_{a'_0\pi^0\eta}+\Pi_{a_0a'_0}(s)g_{a_0\pi^0\eta}$;
$g_{a_0ab}$ and $g_{a'_0ab}$ are the coupling constants; and
$1/D_{a_0}(s)=1/(m^2_{a_0} -s+\sum_{ab}[\mbox{Re}\Pi^{ab}_{a_0}(
m^2_{a_0})-\Pi^{ab}_{a_0}(s) ])$ is the propagator for the $a_0$
resonance (and similar for the $a'_0$ resonance), where
$\mbox{Re}\Pi^{ab}_{a_0}(s)$ is determined by a singly subtracted
dispersion integral of $\mbox{Im}\Pi^{ab}_{a_0}
(s)=\sqrt{s}\Gamma_{a_0\to ab}(s)=g^2_{a_0ab}\rho_{ab}(s)/(16\pi)$,
$\Pi_{a_0a'_0}(s)$\,=\,$C_{a_0a'_0}+\sum_{ab}(g_{a'_0ab}/g_{a_0ab})
\Pi^{ab}_{a_0}(s)$, and $C_{a_0a'_0}$ is the resonance mixing
parameter; the explicit form of the polarization operators
$\Pi^{ab}_{a_0}(s)$ \cite{ADS80a,AK03,AK04YF,AS09} see in Appendix
8.2. The amplitude
\begin{equation}
\label{Mdirpieta} \widetilde{M}^{\mbox{\scriptsize{direct}}}_
{\mbox{\scriptsize{res}}}(s)= s\frac{g^{(0)}_{a_0\gamma\gamma}\Delta
_{a'_0}(s)+ g^{(0)}_{a'_0\gamma\gamma}\Delta_{a_0}(s)} {D_{a_0}(s)
D_{a'_0}(s)-\Pi^2_{a_0a'_0}(s)}e^{i\delta_{\pi\eta}^{bg}(s)}
\end{equation} in Eq. (\ref{M0pieta})
describes the $\gamma\gamma\to\pi^0\eta$ transition caused by the
direct coupling constants $g^{(0)}_{a_0\gamma\gamma}$ and
$g^{(0)}_{a'_0\gamma\gamma }$ of the $a_0$ and $a'_0$ resonances
with the photons; the factor $s$ appears due to the gauge
invariance.

Equation (\ref{M0pieta}) implies that the amplitudes
$T_{ab\to\pi^0\eta}(s)$ in the $\gamma\gamma\to ab\to\pi^0\eta$
rescattering loops (see Fig. \ref{Ampl-gg-pieta}) are on the mass
shell. In so doing, the functions $\widetilde{I}^V_{\pi^0\eta}(s)$,
$\widetilde{I}^V_{\pi^0\eta'}(s)$, $\widetilde{I}^{K^*}_{K\bar
K}(s)$, and the above mentioned function $\widetilde{I}^{K^+}_{K^+
K^-}(s;x_2)$, are the amplitudes of the triangle loop diagrams
describing the transitions $\gamma\gamma$\,$\to$\,$ab$\,$\to$\,({\it
scalar state with a mass}\,=\,$\sqrt{s}$), where the meson pairs
$\pi^0\eta$, $\pi^0\eta'$, and $K\bar K$ are created by
electromagnetic Born sources (see Fig. \ref{BornVK-gg-pieta});
corresponding formulae see in Appendixes 8.2 and 8.3. The
constructed amplitude $M_0(\gamma\gamma\to\pi^0\eta;s,\theta)$
satisfies the Watson theorem in the elastic region.

For the $a_2(1320)$ production amplitude in (\ref{M2pieta}), we use
the parametrization similar to (\ref{Mgg-f2-pipi}) and
(\ref{Gf2-pipi}):
\begin{eqnarray}\label{Ma2}
M_{\gamma\gamma\to a_2(1320)\to\pi^0\eta}(s)=\mbox{\qquad\qquad}
\nonumber \\ =\frac{\sqrt{s\Gamma_{a_2\to\gamma\gamma}(s)
\Gamma^{\mbox{\scriptsize{tot}}}_{a_2}(s)B(a_2\to\pi\eta)
/\rho_{\pi\eta}(s)}}{m^2_{a_2}-s-i
\sqrt{s}\Gamma^{\mbox{\scriptsize{tot}}}_{a_2}(s)}\,, &&
\end{eqnarray} where
\begin{equation}
\label{Ga2tot}\Gamma^{\mbox{\scriptsize{tot}}}_{a_2}(s)
=\Gamma^{\mbox{\scriptsize{tot}}}_{a_2}\frac{m^2_{a_2}}{s}
\frac{q^5_{\pi\eta}(s)} {q^5_{\pi\eta}(m^2_{a_2})}
\frac{D_2(q_{\pi\eta}(m^2_{a_2 })r_{a_2})}{D_2(q_{\pi
\eta}(s)r_{a_2})}\,,\end{equation}
$q_{\pi\eta}(s)$\,=\,$\sqrt{s}\rho_{\pi\eta}(s)/2$, $D_2(x)
$\,=\,$9+3x^2+x^4$, $r_{a_2}$ is the interaction radius, and
$\Gamma_{a_2\to\gamma\gamma}(s)=(\frac{\sqrt{s}}{m_{a_2}})^3
\Gamma_{a_2\to\gamma\gamma}$. Recall that the $f_2(1270)$\,$\to
$\,$\gamma \gamma$ and $a_2(1320)$\,$\to$\,$\gamma\gamma$ decays
widths rather well satisfy the relation $\Gamma_{f_2\to\gamma\gamma
}/\Gamma_{a_2\to \gamma\gamma }$\,=\,$25/9$ \cite{PDG08,PDG10,
MPW94}, which is valid in the naive $q\bar q$ model for the direct
transitions $q\bar q$\,$\to$\,$\gamma\gamma$.

\begin{figure}\includegraphics[height=14pc]{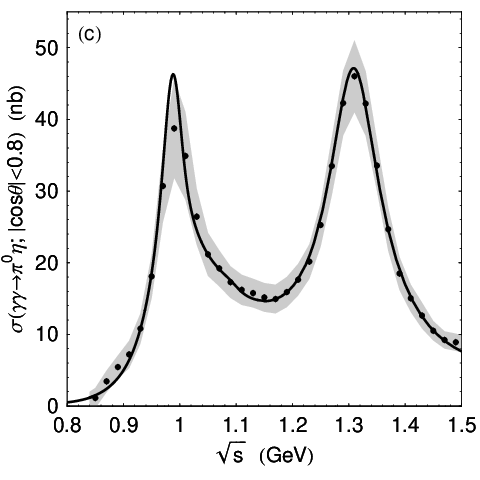}
\caption{{\footnotesize The fit to the Belle data on the
$\gamma\gamma$\,$\to$\,$\pi^0\eta$ reaction cross section. The
resulting solid line corresponds to the solid line 1 in Fig.
\ref{FigAB-gg-pieta}(a) (or in Fig \ref{FigAB-gg-pieta}(b)), folded
with a Gaussian with $\sigma$\,=\,10\,MeV mass resolution; the
shaded band shows the size of the systematic error of the
data.}}\label{FigC-gg-pieta}
\end{figure}

\begin{figure}\begin{tabular}{c}
\includegraphics[height=14pc]{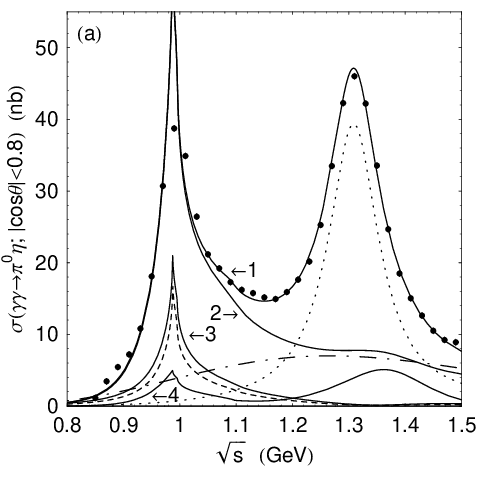}\\
\includegraphics[height=14pc]{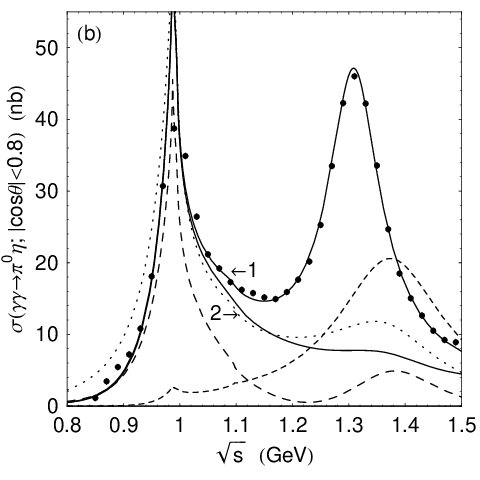}\\
\end{tabular}\\
\caption{{\footnotesize The fit to the Belle data. (a) Solid line 1
is the total $\gamma\gamma$\,$\to$\,$\pi^0\eta$ cross section, solid
line 2 and the dotted line are the helicity 0 and 2 components of
the cross section, solid line 3 is the contribution from the
$\gamma\gamma$\,$\to$\,$K^+K^-$\,$\to$\,$\pi^0\eta$ rescattering
with the intermediate $K+K-$ pair created due to the Born $K$
exchange, the dashed line is the contribution from the $\gamma\gamma
$\,$\to$\,$K\bar K$\,$\to$\,$\pi^0\eta$ with the intermediate $K\bar
K$ pairs created due to the Born $K$ and $K^*$ exchanges, the
dash-dotted line is the contribution from the Born $\rho$ and
$\omega$ exchanges with $\lambda$\,=0, and solid line 4 is the joint
contribution from these exchanges and the S wave rescattering
$\gamma\gamma$\,$\to$\,$(\pi^0\eta+\pi^0\eta')$\,$\to$\,$\pi^0\eta$.
(b) Solid lines 1 and 2 are the same as in panel (a), the
short-dashed line corresponds to the contribution of the amplitude
$\widetilde{M}^{\mbox {\scriptsize{direct}}}_{
\mbox{\scriptsize{res}}}(s)$ caused by the direct decays of the
$a_0$ and $a'_0$ resonances into photons, the dotted line is the
total contribution from the $a_0-a'_0$ resonance complex, and the
long-dashed line is the helicity 0 cross section without the
contribution of the direct transition amplitude $\widetilde{M}^{
\mbox{\scriptsize{direct}}}_{\mbox{\scriptsize{res}}}(s)$.}}
\label{FigAB-gg-pieta}\end{figure}

The results of our fit to the Belle data on the $\gamma
\gamma\to\pi^0\eta$ reaction cross section are shown in Figs.
\ref{FigC-gg-pieta} and \ref{FigAB-gg-pieta}. The corresponding
values of the model parameters are quoted in Appendix 8.2. The good
agreement with the experimental data, see Fig. \ref{FigC-gg-pieta},
allows for definite conclusions on the main dynamical constituents
of the $\gamma\gamma$\,$\to$\,$\pi^0\eta$ reaction mechanism whose
contributions are shown in detail in Figs. \ref{FigAB-gg-pieta}(a)
and \ref{FigAB-gg-pieta}(b).

%-----------------------------------------------------------------------------------------------

Let us begin with the contribution from the inelastic rescattering
$\gamma\gamma$\,$\to$\,$K^+K^-$\,$\to$\,$\pi^0\eta$, where the
intermediate $K^+K^-$ pair is create due to the charge one kaon
exchange (see Fig. \ref{BornVK-gg-pieta}(b)). This mechanism, as in
the case of the $f_0(980)$ production in the $\gamma\gamma$\,$\to
$\,$\pi\pi$ reactions \cite{AS08a,AS08b}, specifies the natural
scale for the $a_0(980)$ resonance production cross section in
$\gamma\gamma$\,$\to$\,$\pi^0\eta$, and leads also to the narrowing
$a_0(980)$ peak in this channel \cite{AS88,AS09}. The maximum of the
cross section $\sigma(\gamma\gamma$\,$\to$\,$K^+K^-$\,$\to$\,$a_0
(980)$\,$\to $\,$\pi^0\eta)$ is controlled by the product of the
ratio of the squares of the coupling constants $R_{a_0}$\,=\,$
g^2_{a_0K^+K^-}/g^2_{a_0 \pi\eta}$ and the value
$|\widetilde{I}^{K^+}_{K^+K^- }(4m^2_{K^+};x_2)|^2$. Its estimate
gives $\sigma(\gamma\gamma$\,$\to$\,$ K^+K^-$\,$\to$\,$ a_0(980)$\,$
\to$\,$\pi^0\eta;|\cos\theta|\leq0.8) \approx0.8\times1.4\alpha^2
R_{a_0}/m^2_{a_0}\approx24$\,nb\,$\times R_{a_0}$ (here we neglect
the heavy $a'_0 $ resonance contribution). Bellow the $K^+K^-$
threshold, the function $|\widetilde{I}^{K^+}_{K^+K^-}(s;x_2)|^2$
decreases sharply, resulting in the narrowing of the $a_0(980)$ peak
in the $\gamma\gamma $\,$\to$\,$K^+K^-$\,$\to$\,$a_0(980)$\,$\to
$\,$\pi^0\eta$ cross section \cite{AS88,AS09}. The $\gamma\gamma
$\,$\to$\,$K^+K^-$\,$\to$\,$\pi^0\eta$ rescattering contribution to
the $\gamma\gamma\to\pi^0\eta$ cross section is shown by solid line
3 in Fig. \ref{FigAB-gg-pieta}(a). The $K^*$ exchange also slightly
narrows the $a_0(980)$ peak (see the dashed line under solid line 3
in Fig. \ref{FigAB-gg-pieta}(a)).

One $\gamma\gamma$\,$\to$\,$K\bar K$\,$\to$\,$\pi^0\eta$
rescattering mechanism is evidently insufficient to describe the
data in the region of the $a_0(980)$ resonance. The addition of the
Born contribution from the $\rho$ and $\omega$ exchanges, which is
modified by the $S$ wave $\gamma\gamma$\,$\to$\,$(\pi^0\eta+\pi^0
\eta')$\,$\to$\,$\pi^0\eta$ rescattering, and the amplitude
$\widetilde{M}^{\mbox{\scriptsize {direct}}}_{
\mbox{\scriptsize{res}}}(s)$, which is due to the direct transitions
of the $a_0$ and $a'_0$ resonances into photons, makes it possible
to obtain the observed cross section magnitude. The contributions of
these two mechanisms themselves are small in the region of the
$a_0(980)$ resonance (see solid line 4 in Fig.
\ref{FigAB-gg-pieta}(a) for the first of them and the short-dashed
line in Fig. \ref{FigAB-gg-pieta}(b) for the second), but their
coherent sum with the contribution from the $\gamma\gamma$\,$\to$\,$
K\bar K$\,$\to$\,$\pi^0\eta$ inelastic rescattering (see the
diagrams for the amplitude with $\lambda$\,=\,0 in Fig.
\ref{Ampl-gg-pieta}) results in the considerable enhancement of the
$a_0(980)$ resonance (see solid line 2 in Fig.
\ref{FigAB-gg-pieta}(a)). Recall that all the $S$ wave contributions
to the $\gamma\gamma $\,$\to$\,$\pi^0\eta$ amplitude below the
$K^+K^-$ threshold have the same phase according to the Watson
theorem.

%-------------------------------------------------------------------------------------

\begin{figure}
\includegraphics[width=21pc]{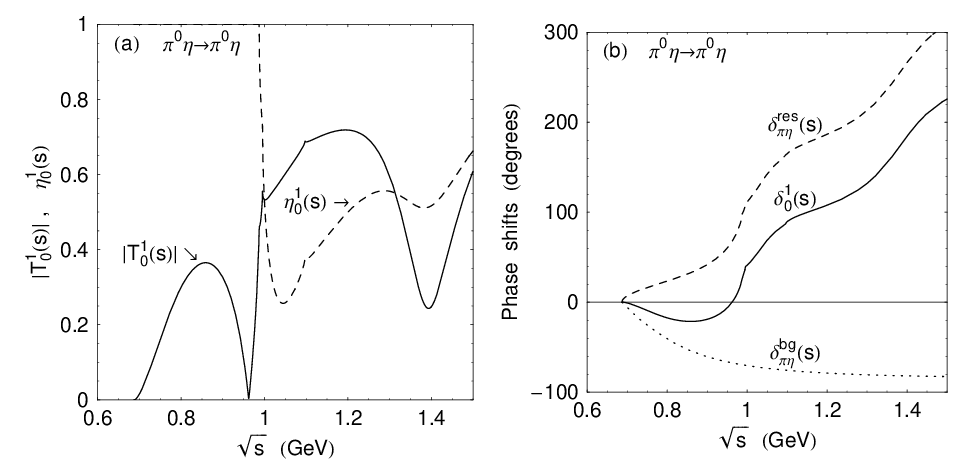}\\
\caption{{\footnotesize The $S$ $\pi^0\eta\to\pi^0\eta$ amplitude.
(a) $|T^1_0(s)|$ and inelasticity $\eta^1_0(s)$; (b) phase shifts
($a^1_0$\,=\,0.0098).}} \label{Fig-pieta-pieta}\end{figure}

Note that, as a by-product, we extracted from the fitting of the
$\gamma\gamma$\,$\to$\,$\pi^0\eta$ data the preliminary information
on the $S$ wave amplitude of the $\pi^0\eta$\,$\to$\,$\pi^0\eta$
reaction, which is important for the low-energy physics of
pseudoscalar mesons. The characteristics for the $S$ wave amplitude
$\pi^0\eta$\,$\to$\,$\pi^0\eta$ are represented in Fig.
\ref{Fig-pieta-pieta}. Here worth noting is the important role of
the background $\pi^0\eta$ elastic amplitude $T^{bg}_{\pi\eta}(s)$,
see Eq. (\ref{Tpietapieta}). First, the choice of the negative
background phase $\delta^{bg}_{\pi\eta}(s)$ (see Fig.
\ref{Fig-pieta-pieta}(b)) in $T^{bg}_{\pi\eta}(s)$ makes it possible
to fit the $\pi\eta$ scattering length in the model under
consideration to the estimates based on the current algebra
\cite{Os70,Pe71} and chiral perturbation theory \cite{BKM91,BFS00},
according to which $a^1_0$\,(in units of
$m^{-1}_\pi$)\,$\approx$\,$0.005-0.01$. The resonance contribution
($\approx$\,0.3) to $a^1_0$ is compensated by the background
contribution. Second, the significant negative value of
$\delta^{bg}_{\pi\eta}(s)$ near 1 GeV ensures the resonance-like
behavior of the cross section shown by solid line 4 in Fig.
\ref{FigAB-gg-pieta}(a).

We now turn to Fig. \ref{FigAB-gg-pieta}(b) and discuss the
contribution from the possibly existing heavy $a'_0$ resonance
\cite{PDG10}) with the mass $m_{a'_0}\approx$\,1.4 GeV. The cross
section corresponding to the amplitude $\widetilde{M}^{
\mbox{\scriptsize{direct}}}_ {\mbox{\scriptsize{res}}}(s)$ (see the
short-dashed line) exhibits a pronounced enhancement near 1.4 GeV.
In the cross section corresponding to the total contribution from
the resonances (see the dotted line), i.e., from the amplitude
$\widetilde{M}^{\mbox{\scriptsize{direct }}}_
{\mbox{\scriptsize{res}}}(s)$ and rescattering amplitudes
proportional to the amplitudes of the resonance transitions
$ab\to\pi^0\eta$ ($ab$\,=\,$\pi\eta$, $K^+K^-$, $K^0\bar K^0 $,
$\pi\eta'$), this enhancement is transferred to a shoulder. Finally,
in the total cross section $\sigma_0$ (see solid line 2)
additionally including the Born $\gamma\gamma$\,$\to$\,$\pi^0\eta$
contribution and the $\gamma\gamma$\,$\to$\,$\pi^0\eta$\,$\to$\,$
\pi^0\eta$ rescattering caused by the background $\pi^0\eta$\,$\to
$\,$ \pi^0\eta$ elastic amplitude, any resonance attributes near 1.4
GeV are absent. Thus, a strong destructive interference exists
between different contributions and masks the $a'_0$ resonance in
the $\gamma\gamma$\,$\to$\,$\pi^0\eta$ cross section. Nevertheless,
in many respects owing to the $a'_0$, we succeed in modeling a
significant smooth background under the $a_2(1320)$ and between
$a_0(980)$ and $a_2(1320)$ resonances, which is required by the
Belle data \cite{Ue09}. Note that due to the resulting
compensations, the wide interval of (1.28--1.42)\,GeV is allowed for
the mass of the $a'_0$ resonance (see Ref. \cite{AS10a} for
details).\,\footnote{Recall, that in the previous Section it was not
required to introduce any heavy scalar isoscalar resonance for the
theoretical description of the $\gamma\gamma\to \pi^+\pi^-$ and
$\gamma\gamma\to\pi^0\pi^0$ processes, as well as in Refs.
\cite{Mo07b,Ue08} for the phenomenological treatment of the
experimental data. In principle, it could be the $f_0(1370)$
resonance \cite{PDG10}. As a matter of fact, situation with the
heavy scalar resonances with the masses $^>_\sim$\,1.3 GeV has been
strongly tangled for a long time. For example, the authors of the
review \cite{KZ07} seriously doubt in the existence of such a state
as the $f_0(1370)$ (in this connection see also Refs. \cite{AS96,
Oc10}). It is possible that the wish to see the scalar resonances
with the masses of (1.3--1.4)\,GeV as the partners of the well
established $b_1(1235)$, $h_1(1170)$, $a_1(1260)$, $f_1(1285)$,
$a_2(1320)$, and $f_2(1270)$ states, belonging to the lower $P$ wave
$q\bar q$ multiplet, is not realized in the naive way. In any case,
this question remains open and requires further experimental and
theoretical investigations.}

Let us consider now the $\gamma\gamma$\,$\to$\,$\pi^0\eta$ cross
section due only to the resonance contributions and, by analogy with
Eq. (\ref{Gf0-gg}), determine  the width of the $a_0(980)$\,$\to
$\,$\gamma\gamma$ decay averaged over the resonance mass
distribution in the $\pi\eta$ channel \cite{AS88,AS09}:
\begin{equation}\langle\Gamma_{a_0\to\gamma\gamma}\rangle_{\pi\eta}=
\int\limits_{0.9\mbox{\,\scriptsize{GeV}}}^{1.1\mbox{\,\scriptsize{GeV}}}
\frac{s}{4\pi^2}\sigma_{\mbox{\scriptsize{res}}}(\gamma\gamma\to
\pi^0\eta;s)d\sqrt{s}\label{Ga0-gg}\end{equation} (the integral is
calculated over the region of the $a_0(980)$ resonance). Taking into
account the contributions from all of rescattering processes and
direct decays into $\gamma\gamma$ to the $\sigma_{
\mbox{\scriptsize{res}}}$ cross section, we obtain
$\langle\Gamma_{a_0\to(K\bar K+\pi\eta+\pi\eta'+\mbox
{\scriptsize{direct}})\to\gamma\gamma} \rangle_{\pi\eta}$\,$
\approx$\,0.4 keV. Taking into account the contributions from only
the rescattering processes, $\langle\Gamma_{a_0\to(K\bar K+\pi\eta
+\pi\eta')\to\gamma\gamma}\rangle_{\pi\eta}$\,$\approx $\,0.23 keV,
and, taking into account the contributions from only the direct
decays, $\langle\Gamma^{ \mbox{\scriptsize{direct}}}_{a_0\to
\gamma\gamma} \rangle_{ \pi\eta}$\,$\approx$\,0.028 keV.

The performed analysis indicates that the $a_0(980)$\,$\to$\,$(K\bar
K+\pi^0\eta+\pi^0\eta')$\,$\to$\,$\gamma \gamma$ rescattering
mechanisms, i.e., the four-quark transitions, dominate in the
$a_0(980)$\,$\to $\,$\gamma\gamma$ decay. This picture is evidence
of the $q^2\bar q^2$ nature of the $a_0(980)$ resonance and is in
agreement with the properties of the $\sigma_0(600)$ and $f_0(980)$
resonances, which are its partners. As to the ideal $q\bar q$ model
prediction for the two-photon decay widths of the $f_0(980)$ and
$a_0(980)$ mesons, $\Gamma_{f_0\to\gamma\gamma }/\Gamma_{a_0\to
\gamma\gamma}=25/9$, it is excluded by experiment. \,\footnote{As
already mentioned in Ref. \cite{AS09}, the model of nonrelativistic
$K\bar K$ molecules is unjustified, because the momenta in the kaon
loops describing the $\phi\to K^+K^-\to \gamma(f_0/a_0)$ and
$f_0/a_0\to K^+K^-\to\gamma\gamma$ decays are high
\cite{Ac08a,AK07b,AK08}. Our analysis gives an additional reason
against the molecular model. The point is that the $a_0(980)$
resonance is strongly coupled with the $K\bar K$ and $\pi\eta$
channels, which are equivalent in the $q^2\bar q^2$ model. A weakly
bound $K\bar K+\pi\eta$ molecule seems to be impossible. Moreover,
the widths of the two-photon decays of the scalar resonances in the
molecular model are calculated at the resonance point
\cite{Ka06,Ha07}, but this is insufficient for describing the
$\gamma\gamma\to\pi^+\pi^-$, $\gamma\gamma\to\pi^0\pi^0$, and
$\gamma\gamma\to\pi^0\eta$ reactions. Attempts of the description of
the data on these processes in the framework of the molecular model
are absent and, therefore, the results obtained in this model have
the academic character.}

\vspace{0.4cm} \noindent{\large \bf 6. Preliminary summary}
\vspace{0.2cm}

\noindent Results of the theoretical analysis of the experimental
achievements in the low energy region, up to 1 GeV, can be
formulated in the following way.

\begin{enumerate}
\item{} Naive consideration of the mass spectrum of the light scalar
mesons, $\sigma(600)$, $\kappa(800)$, $f_0(980)$, and $a_0(980)$,
gives an idea of their $q^2\bar q^2$ structure.

\item{} Both intensity and mechanism of the $a_0(980)/f_0(980)$
production in the $\phi(1020)$ meson radiative decays, the
four-quark transitions $\phi(1020)\to K^+K^-\to\gamma [a_0(980)/f_0
(980)]$, indicate the $q^2\bar q^2$ nature of the $a_0(980)$ and
$f_0(980)$ states.

\item{} Intensities and mechanisms of the two-photon production of the
light scalars, the four-quark transitions $\gamma\gamma\to\pi^+\pi^-
\to\sigma(600)$, $\gamma\gamma\to\pi^0\eta\to a_0(980)$, and $\gamma
\gamma\to K^+K^-\to f_0(980)/a_0(980)$, also indicate their $q^2\bar
q^2$ nature.

\item{} In addition, the absence of the
$J/\psi$\,$\to$\,$\gamma f_0(980)$, $\rho a_0(980)$, $\omega
f_0(980)$ decays in contrast to the intensive $J/\psi$\,$\to$
$\gamma f_2(1270)$, $\,\gamma f'_2(1525)$, $\rho a_2(1320)$, $\omega
f_2(1270)$ decays intrigues against the $P$ wave $q\bar q$ structure
of the $a_0(980)$ and $f_0(980)$ resonances.

\item{} It seems also undisputed that in all respects the $a_0(980)$ and
$f_0(980)$ mesons are strangers in the company of the well
established $b_1(1235)$, $h_1(1170)$, $a_1(1260)$, $f_1(1285)$,
$a_2(1320)$, and $f_2(1270)$ mesons, which are the members of the
lower $P$ wave $q\bar q$ multiplet.
\end{enumerate}

\vspace{0.4cm} \noindent{\large \bf 7. Future Trends} \vspace{0.2cm}

\noindent{\bf \boldmath 7.1. The $f_0(980)$ and $a_0(980)$
resonances near $\gamma\gamma\to K^+K^-$ and $\gamma\gamma\to
K^0\bar K^0$ reaction thresholds}

\noindent The Belle Collaboration investigated the $\gamma\gamma$\,$
\to$\,$\pi^+\pi^-$, $\gamma \gamma$\,$\to$\,$\pi^0\pi^0$, and
$\gamma\gamma$\,$\to$\,$\pi^0\eta$ reactions with the highest
statistics.\,\footnote{Note that high precision measurements of the
$\gamma\gamma$\,$\to$\,$\pi^+\pi^-$ cross section for
0.28\,GeV\,$<$\,$\sqrt{s}$\,$<$\,0.45\,GeV are planned for the
KLOE-2 detector at upgraded DA$\Phi$NE $\phi$ factory in Frascati
\cite{A-C10,Cz10}; the existing MARK II data \cite{Bo90} have in
this region very large error-bars, see Fig. \ref{MII-CB-gg-pipi}(a).
Measurements of the integral and differential cross sections for
$\gamma\gamma$\,$\to$\,$\pi^+\pi^-$ and
$\gamma\gamma$\,$\to$\,$\pi^0 \pi^0$ in the $\sqrt{s}$ region from
0.45 GeV to 1.1 GeV \cite{A-C10,Cz10}, which will complete the
information from previous experiments on the $\sigma(600)$ and
$f_0(980)$ resonance production, are also planned. In particular,
the statistical uncertainty in the $\gamma\gamma$\,$\to$\,$\pi^0
\pi^0$ cross section in the $\sigma(600)$ meson region (see Fig.
\ref{MII-CB-gg-pipi}(b)) will be reduced to 2\%.} In July 2010, the
Belle Collaboration reported also the first data on the $\gamma
\gamma$\,$\to$\,$\eta\eta$ reaction \cite{Ueh10}. The $\gamma
\gamma$\,$\to$\,$\eta\eta$ cross section for $\sqrt{s}$\,$>$\,1.2
GeV is dominated by the contributions from the tensor resonances
$f_2(1270)$, $a_2(1320)$, and $f'_2(1525)$. But near the threshold,
$2m_\eta$\,=\,1.0957 GeV\,$<$\,$\sqrt{s}$\,$<$\,1.2 GeV, there is a
noticeable $S$ wave contribution, $\approx(1.5\pm0.15\pm0.7)$ nb,
which indicates the presence of some subthreshold resonance strongly
coupled with the $\eta\eta$ channel. Such a resonance in the
$q^2\bar q^2$ model is the $f_0(980)$. Unfortunately, the
$\gamma\gamma$\,$\to$\,$\eta\eta$ reaction is not so good for its
investigation, because here only the end of the tail of this
resonance can be seen.

High statistics information is still lacking for the reactions
$\gamma\gamma$\,$\to$\,$K^+K^-$ and $\gamma\gamma$\,$\to $\,$K^0\bar
K^0$ in the 1 GeV region. It is expected that the four-quark nature
of the $a_0(980)$ and $f_0(980)$ resonances shows itself in these
channels very originally \cite{AS92,AS94b}.

\begin{figure}
\includegraphics[width=21pc]{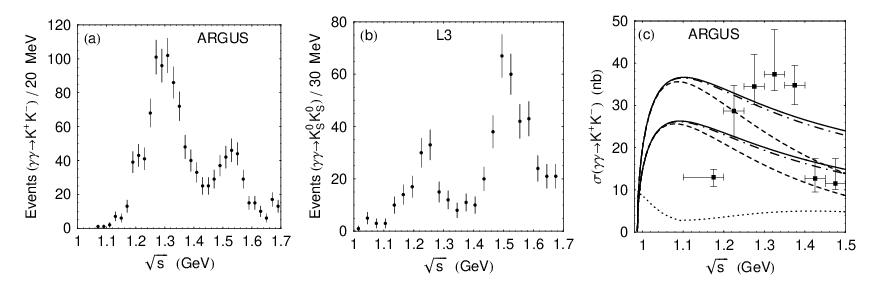}
\caption{{\footnotesize The $K^+K^-$ (a) and $K^0_SK^0_S$ (b) mass
spectra measured by ARGUS \cite{Alb90} and L3 \cite{Acc01},
respectively. (c) This plot illustrates the scale of the $K\bar K$
production cross section in $\gamma\gamma$ collisions. The
experimental points show the cross section for $\gamma\gamma\to
K^+K^-$ with allowed contributions from $\lambda J$\,=\,[22, 02, 00]
\cite{Alb90}. The upper dashed, dot-dashed, and solid curves
correspond to the $\gamma\gamma\to K^+K^-$ Born cross section with
$\lambda J$\,=00, $\lambda J$\,=\,[00, 22], and the total one,
respectively (the $\lambda J$\,=02 contribution is negligible). The
lower dashed, dot-dashed, and solid curves correspond to the same
cross sections but modified by the form factor (see Sec. 4 and
Appendix 8.3). The dotted curve shows the estimate of the the $S$
wave $\gamma\gamma\to K^+K^-$ cross section in our model.}}
\label{gg-KK}\end{figure}

As the experiments show \cite{Al83,Alt85,Jo86,Ber88,Beh89,FH91,
Alb90,Bra00,Acc01,Ab04}, the $\gamma\gamma $\,$\to$\,$K^+K^-$ and
$\gamma\gamma $\,$\to$\,$K^0_SK^0_S$ cross sections in the region
1\,$<$\,$\sqrt{s}$\,$<$\,1.7 GeV are saturated in fact with the
contributions of the classical tensor $f_2(1270)$, $a_2(1320)$, and
$f'_2(1525)$ resonances, creating in the helicity 2 states, see Fig.
\ref{gg-KK}. The constructive and destructive interference between
the $f_2(1270)$ and $a_2(1320)$ resonance contributions is observed
in $\gamma\gamma$\,$\to$\,$K^+K^-$ and $\gamma\gamma$\,$\to$\,$
K^0\bar K^0$, respectively, in agreement with the $q\bar q$ model
\cite{FLR75}. Notice that the region of the $K\bar K$ thresholds,
$2m_K<\sqrt{s}<1.1$ GeV, sensitive to the $S$ wave contributions is
not investigated in fact. The sensitivity of the ARGUS experiment to
the $K^+K^-$ events for $2m_{K^+}<\sqrt{s}<1.1$ GeV was negligible
\cite{Alb90}, see Fig. \ref{gg-KK}(a), and the total statistics of
the L3 \cite{Acc01}, see Fig. \ref{gg-KK}(b), and CLEO \cite{Bra00}
experiments on $\gamma\gamma $\,$\to$\,$K^0_SK^0_S$ for $2m_{K^0}<
\sqrt{s}<1.1$ GeV are within 60 events.

%--------------------------------------------------------------------------------------------------

The absence of the considerable non-resonance background in the
$\gamma\gamma$\,$\to$\,$K^+K^-$ cross section seems at first sight
rather surprising for the one kaon exchange Born contribution
comparable with the tensor resonance contributions, see Fig.
\ref{gg-KK}(c). As seen from this figure, the $S$ wave contribution
dominates in the Born cross section at $\sqrt{s}<1.5$ GeV. One would
think that the large non-coherent background should be under the
tensor meson peaks in the $K^+K^-$ channel. But taking into account
of the resonance interaction between the $K^+$ and $K^-$ mesons in
the final state results in the cancelation of the considerable part
of this background \cite{AS92,AS94b}. The principal point is that
the $S$ wave Born $\gamma\gamma$\,$\to$\,$K^+K^-$ amplitude acquires
the $\xi(s)$\,=\,$[1$\,+\,$i\rho_{K^+}(s)T_{K^+K^-\to K^+K^-}(s)]$
factor due to the $\gamma\gamma$\,$\to$\,$K^+K^-$\,$ \to$\,$K^+K^-$
rescattering amplitude with the real kaons in the intermediate
state. The $a_0(980)$  and $f_0(980)$ resonance contributions
dominate in the $T_{K^+K^-\to K^+K^-}(s)$ amplitude near the
$K^+K^-$ threshold and provide it with the considerable imaginary
part for the strong coupling with the $K\bar K$ channels in the
four-quark scheme. As a result the $|\xi(s)|^2$ factor is
considerably less than 1 just above the $K^+K^-$ threshold and the
seed $S$ wave Born contribution is considerably reduced in the wide
region of $\sqrt{s}$. The dotted curve in Fig. \ref{gg-KK}(c)
represents the estimation of the $S$ wave $\gamma\gamma\to K^+K^-$
cross section obtained in the model under consideration (see details
in Appendix 8.3). This estimation agrees with those obtained earlier
\cite{AS92,AS94b}.

So one can hope to detect in the partial wave analysis of the
$\gamma\gamma$\,$\to$\,$K^+K^-$ reaction at $2m_{K^+}< \sqrt{s}<1.1$
GeV the scalar contributions at the rate of about 5--10 nb.  As for
the $\gamma\gamma$\,$\to$\,$K^0\bar K^0$ reaction, its amplitude has
not the Born contribution and the $a_0(980)$ resonance contribution
has the opposite sign in comparison with the $\gamma\gamma$\,$\to
$\,$K^+K^-$ channel. As a result the contributions of the $S$ wave
$\gamma\gamma\to K^+K^-\to K^0\bar K^0$ rescattering amplitudes with
the isotopic spin $I$\,=\,0 and 1 practically cancel each other and
the the corresponding cross section should be at the rate of about
$\lesssim$\,1 nb.

\vspace{0.2cm} \noindent{\bf \boldmath 7.2. The $\sigma(600)$,
$f_0(980) $, and $a_0(980)$ resonances in $\gamma\gamma^*$
collisions}

\noindent The investigations of light scalar mesons in the $\gamma
\gamma^*(Q^2)$ collisions are promising. If $\sigma(600)$,
$f_0(980)$, and $a_0(980)$ resonances are four-quark states, their
contributions to the $\gamma\gamma^*(Q^2)$\,$\to$\,$\pi^0\pi^0$ and
$\gamma\gamma^*(Q^2)$\,$\to$\,$\pi^0\eta$ cross sections should
decrease with increasing $Q^2$ more rapidly than the contributions
from the classical tensor mesons $f_2(1270)$ and $a_2(1320)$. A
similar behavior of the contribution from the exotic $q^2\bar q^2$
resonance state with $I^G(J^{PC}) $\,=\,$2^+(2^{++})$
\cite{ADS82a,ADS82b,AS91} to the $\gamma\gamma^*$\,$\to
$\,$\rho^0\rho^0$ and $\gamma\gamma^*$\,$\to$\,$\rho^+\rho^-$ cross
sections was recently observed by the L3 Collaboration
\cite{L31,L32,L33,L34}.

\vspace{0.2cm} \noindent{\bf \boldmath 7.3. Searches for the
$J/\psi\to\omega f_0(980)$ and $J/\psi\to\rho a_0(980)$ decays}

\noindent These decays are important to elucidate the nature of the
$f_0(980)$ and $a_0(980)$ resonances \cite{Ac98,Ac02,Ac03c}. The
$J/\psi\to\rho a_0(980)$ decay has not been discovered as yet,
$B(J/\psi\to\rho a_0(980))<4.4\times10^{-4}$ \cite{Ac98}. As for the
information on $B(J/\psi\to\omega f_0(980))=(1.4\pm0.5)\times 10^{-4
}$ \cite{PDG10}, it would be more correctly replaced by a suitable
upper limit \cite{Ac02,Ac03c}.

\vspace{0.2cm} \noindent{\bf \boldmath 7.4. Inelasticity of $\pi\pi$
scattering and $f_0(980)-a_0(980)$ mixing}

\noindent By now considerable progress has been made in the
experimental investigations of the $f_0(980)$ and $a_0(980)$ mesons
in various reactions. Nevertheless, it turns out that equally good
descriptions of the available data can be obtained for appreciably
different sets of the coupling constants $g_{f_0K^+K^- }$,
$g_{f_0\pi^+\pi^-}$, etc. (see, for example, Refs. \cite{ADS84a,
ADS80a,ADS84b,AK06,AK07a}). Certainly, it would be highly desirable
to fix their values. In respect of the coupling constants
$g_{f_0K^+K^-}$ and $g_{ f_0\pi^+\pi^-}$, this question could be
elucidated by precise data on the inelasticity of $\pi\pi$
scattering near the $K\bar K$ threshold, that have not been updated
since 1975 \cite{Pr73,Hy73,Gr74,Hy75}. It is very likely that such
data in the raw form are in hand of the VES Collaboration, which was
performing measurements of the $\pi^-p\to \pi^+\pi^-n$ reaction at
IHEP (Protvino). Moreover, the product of the coupling constants
$g_{a_0K^+K^-}g_{f_0K^+K^-}$ may be fixed from data on the
$f_0(980)-a_0(980)$ mixing, that are expected from the BESIII
detector \cite{Har10}.

Exclusive information on $g_{a_0K^+K^-}g_{f_0K^+K^-}$ can result
from investigations of the spin asymmetry jump, due to the $f_0
(980)-a_0(980)$ mixing, in the $\pi^-p\to f_0(980)n\to a_0(980)
n\to\pi^0\eta n$ reaction \cite{AS04a}.

\vspace{0.2cm}

This work was supported in part by the RFFI Grant No. 10-02-00016
from the Russian Foundation for Basic Research.

\vspace{0.4cm} \noindent{\large \bf 8. Appendix}

\vspace{0.2cm}\noindent{\bf \boldmath 8.1.  $\gamma\gamma\to\pi\pi$}

\noindent Below there is the list of the expressions for the Born
helicity amplitudes corresponding to the charged one pion exchange
mechanism and for the triangle loop integrals $\widetilde{I}^{\pi^+
}_{\pi^+\pi^-}(s)$ and $\widetilde{I}^{\pi^+}_{\pi^+\pi^-}(s;x_1)$
used in Section 4. In addition, a few useful auxiliary formulae for
the solitary scalar resonance are adduced.

The Born helicity amplitudes for the elementary one pion exchange in
the $\gamma\gamma$\,$\to$\,$\pi^+\pi^-$ reaction have the form
\begin{equation}
M_0^{\mbox{\scriptsize{Born}}\,\pi^+}(s,\theta)=\frac{4m^2_{\pi^+}}
{s}\frac{ 8\pi\alpha}{1-\rho^2_{\pi^+}(s)\cos^2\theta}\,,
\label{M0born}\end{equation}
%-----------------------------------------------------------------------------------------------
\begin{equation} M_2^{
\mbox{\scriptsize{Born}}\,\pi^+}(s,\theta)=\frac{8\pi\alpha \rho^2_
{\pi^+}(s)\sin^2\theta}{1-\rho^2_{\pi^+}(s)\cos^2 \theta}\,,
\label{M2born}\end{equation}
($\rho_{\pi^+}(s)=\sqrt{1-4m^2_{\pi^+}/s}$). Their partial wave
expansions are
\begin{equation}
M_\lambda^{\mbox{\scriptsize{Born}}\,\pi^+}(s,\theta)=\sum_{J\geq
\lambda}(2J+1)M_{\lambda J}^{\mbox{\scriptsize{Born}}\,\pi^+}(s)
d^J_{\lambda0}(\theta)\,,\label{PWA}\end{equation} where
$d^J_{\lambda0}(\theta)$ are usual $d$-functions (see, for example,
\cite{PDG08,PDG10}).  Three lower partial waves have the form
\begin{equation}
M^{\mbox{\scriptsize{Born}}\,\pi^+}_{00}(s)=4\pi\alpha\frac{1-
\rho^2_{\pi^+}(s)}{\rho_{\pi^+}(s)}\,\ln\frac{1+\rho_{\pi^+}(s)}{1-
\rho_{\pi^+}(s)}\,,\label{M00B}\end{equation}
\begin{eqnarray}
M_{02}^{\mbox{\scriptsize{Born}}\,\pi^+}(s)=4\pi\alpha\frac{1-\rho^2_{
\pi^+}(s)}{\rho^2_{\pi^+}(s)}\left[\frac{3-\rho^2_{\pi^+}(s)}{2\rho_{
\pi^+}(s)}\times\right. && \nonumber \\
\times\left.\ln\frac{1+\rho_{\pi^+}(s)}{1-\rho_{\pi^+}(s)}
-3\right]\,, \qquad\qquad && \label{M02B}\end{eqnarray}
\begin{eqnarray}
M_{22}^{\mbox{\scriptsize{Born}}\,\pi^+}(s)=4\pi\alpha\sqrt{\frac{3}{2}}
\left[\frac{(1-\rho^2_{\pi^+}(s))^2}{2\rho^3_{\pi^+}(s)}\ln\frac{1+\rho_{
\pi^+}(s)}{1-\rho_{\pi^+}(s)}-\right. && \nonumber\\
\left.-\frac{1}{\rho^2_{\pi^+}(s)}+ \frac{5}{3}\right]\,.
\qquad\qquad\qquad && \label{M22B}\end{eqnarray}

The amplitude of the triangle loop diagram, describing the
transition $\gamma\gamma$\,$\to$\,$\pi^+\pi^-$\,$\to$\,({\it scalar
state with a mass}\,=\,$\sqrt{s}$), is defined by
\begin{equation}\widetilde{I}^{\pi^+}_{\pi^+\pi^-}(s)=\frac{s}{\pi}
\int\limits^\infty_{4m^2_{\pi^+}}\frac{\rho_{\pi^+}(s')
M^{\mbox{\scriptsize{Born}}\,\pi^+}_{00}(s')}{s'(s'-s-i\varepsilon)}ds'
\,.\label{Ipipi}\end{equation} The behavior
$\widetilde{I}^{\pi^+}_{\pi^+\pi^-}(s)$\,$\propto$\,$s$, when
$s$\,$\to$\,0, is the gauge invariance consequence. For
$s$\,$\geq$\,$4m^2_{\pi^+}$
\begin{equation}\widetilde{I}^{\pi^+}_{\pi^+\pi^-}(s)=8\alpha(
\frac{m^2_{\pi^+}}{s}[\pi-2\arctan|\rho_{\pi^+}(s)|]^2-1)\,,
\label{Ipipi1}\end{equation} for $s$\,$\geq$\,$4m^2_{\pi^+}$
\begin{equation}\widetilde{I}^{\pi^+}_{\pi^+\pi^-}(s)=8\alpha \left\{
\frac{m^2_{\pi^+}}{s}\left[\pi+i\ln\frac{1+\rho_{\pi^+}(s)}
{1-\rho_{\pi^+}(s)}\right]^2-1\right\}\,.\label{Ipipi2}\end{equation}

The form factor, see Eq. ((\ref{FF-Poppe}),
$$G_{\pi^+}(t,u)=\frac{1}{s}\left[\frac{m^2_{\pi^+}-t}{1-(u-m^2_{\pi^+}
)/x^2_1}+\frac{m^2_{\pi^+}-u}{1-(t-m^2_{\pi^+})/x^2_1}\right]$$
(here $t$\,=\,$m^2_{\pi^+}$\,$-$\,$s[1$\,$-$\,$\rho_{\pi^+}(s)
\cos\theta]/2$ and $u$\,=\,$m^2_{\pi^+}$\,$-$\,$s[1$ +\,$\rho_{
\pi^+}(s)\cos\theta]/2$) modifies the Born partial wave amplitudes.
Let us introduce the notations:
\begin{equation}M^{\mbox{\scriptsize{Born}}\,\pi^+}_{0J}(s)=
\frac{1-\rho^2_{\pi^+}(s)}{\rho_{\pi^+}(s)}F^{
\mbox{\scriptsize{Born}}\,\pi^+}_{0J}(\rho_{\pi^+}(s))\,,\label{MB0J}
\end{equation}
\begin{equation}M^{\mbox{\scriptsize{Born}}\,\pi^+}_{2J}(s)
=\rho_{\pi^+}(s)F^{\mbox{\scriptsize{Born}}\,\pi^+}_{2J}(\rho_{\pi^+}
(s))\,.\label{MB2J}\end{equation} Then the modified amplitudes can
be represented in the form
\begin{eqnarray} M^{\mbox{\scriptsize{Born}}\,\pi^+}_{0J}(s;x_1)=
\frac{1-\rho^2_{\pi^+}(s)}{\rho_{\pi^+}(s)}\times\qquad\ && \nonumber \\
\times\left[F^{\mbox{\scriptsize{Born}}\,\pi^+}_{0J}
(\rho_{\pi^+}(s))-F^{\mbox{\scriptsize{Born}}\,\pi^+}_{0J}(\rho_{\pi^+}
(s;x_1))\right], &&\label{MB0Jff}\end{eqnarray}
\begin{eqnarray}M^{\mbox{\scriptsize{Born}}\,\pi^+}_{2J}(s;x_1)
=\rho_{\pi^+}(s)\times\qquad\qquad && \nonumber \\
\times\left[F^{\mbox{\scriptsize{Born}}\,\pi^+}
_{2J}(\rho_{\pi^+}(s))-F^{\mbox{\scriptsize{Born}}\,\pi^+}_{2J}
(\rho_{\pi^+}(s;x_1))\right], & \label{MB2Jff}\end{eqnarray} where
\begin{equation}\rho_{\pi^+}(s;x_1)=\rho_{\pi^+}(s)/(1+
2x^2_1/s)\,.\label{rhopiff}\end{equation} The function
$\widetilde{I}^{\pi^+}_{\pi^+\pi^-}(s)$, see Eqs. (\ref{Ipipi})--
(\ref{Ipipi2}), is replaced, with taking into account the form
factor, by
\begin{equation}\widetilde{I}^{\pi^+}_{\pi^+\pi^-}(s;x_1)=\frac{s}{\pi}
\int\limits^\infty_{4m^2_{\pi^+}}\frac{\rho_{\pi^+}(s')
M^{\mbox{\scriptsize{Born}}\,\pi^+}_{00}(s';x_1)}{s'(s'-s-i\varepsilon)}ds'\,.
\label{Ipipiff}\end{equation} In this case the numerical integration
needs certainly.

To make easy understanding the structure and normalization of the
sufficiently complicated expressions used in fitting data, one
adduces the formulae for the production cross section of the
$\sigma$ resonance and for its two-photon decay width due to the
rescattering mechanism, $\gamma\gamma$\,$\to$\,$\pi^+\pi^-$\,$
\to$\,$\sigma $\,$\to$\,$\pi^+\pi^-$, in the imaginary case of the
solitary scalar $\sigma$ resonance coupled only to the $\pi\pi$
channel.

The corresponding resonance cross section has the familiar form
\begin{eqnarray} \sigma_{\mbox{\scriptsize{res}}}(\gamma\gamma\to
\pi^+\pi^-;s)= \qquad\qquad & \nonumber \\
=\frac{8\pi}{s}\,\frac{\sqrt{s}\Gamma_{\sigma\to\pi^+\pi^-\to
\gamma\gamma}(s)\,\sqrt{s}\Gamma_{\sigma\to
\pi^+\pi^-}(s)}{|D_{\sigma}(s)|^2}\,, &&
\label{gg-sigma-pipi}\end{eqnarray} where \begin{eqnarray} \Gamma_{
\sigma\to\pi^+\pi^-\to\gamma\gamma}(s)=\frac{1}{16\pi\sqrt{s}
}|M_{\sigma\to\pi^+\pi^-\to\gamma\gamma}(s)|^2= && \nonumber \\
=\left|\frac{1}{16\pi}\,\widetilde{I}^{\pi^+}_{\pi^+\pi^-}(s)\right|^2
\,\frac{g^2_{\sigma\pi^+\pi^-}}{16\pi\sqrt{s}}\,.\qquad\qquad &&
\label{G-sigma-pipi-gg}\end{eqnarray} If the $\sigma$ can else
directly transit into $\gamma\gamma$ with the amplitude
$sg^{(0)}_{\sigma\gamma\gamma}$ then the width
$\Gamma_{\sigma\to\pi^+\pi^-\to\gamma\gamma}(s)$ in Eq.
(\ref{gg-sigma-pipi}) should be replaced by
\begin{equation}\Gamma_{\sigma\to\gamma\gamma}(s)=\frac{1}{16\pi
\sqrt{s}}|M_{\sigma\to\gamma\gamma}(s)|^2\,,\label{G-sigma-gg}
\end{equation} where
\begin{equation}M_{\sigma\to\gamma\gamma}(s)=
M_{\sigma\to\pi^+\pi^-\to\gamma\gamma}(s)+sg^{(0)}_{\sigma\gamma\gamma}
\,.\label{M-sigma-gg}\end{equation} The propagator of the $\sigma$
resonance with the $m_\sigma$ Breit-Wigner mass in Eq.
(\ref{gg-sigma-pipi}) has the form
\begin{equation}\frac{1}{D_{\sigma}(s)}=\frac{1}{m^2_{\sigma}-s+
\mbox{Re}\Pi^{\pi\pi}_{\sigma}(m^2_{\sigma})-\Pi^{\pi\pi}_{\sigma}(s)}
\label{Prop-sigma}\,,\end{equation} where $\Pi^{\pi\pi}_{\sigma}(s)$
is the polarization operator of the $\sigma$ resonance for the
contribution of the $\pi^+\pi^-$ and $\pi^0\pi^0$ intermediate
states. For $s\geq4m^2_{\pi^+}$\,(=\,$4m^2_{\pi^0}$)
\begin{equation} \Pi^{\pi\pi}_{\sigma}(s)=\frac{3}{2}\frac{g^2_{\sigma
\pi^+\pi^-}}{16\pi}\rho_{\pi^+}(s)\left[i-\frac{1}{\pi}\ln\frac{1+
\rho_{\pi^+}(s)}{1-\rho_{\pi^+}(s)}\right]\,.\label{Pi-sigma1}\end{equation}
If 0\,$<$\,$s$\,$<$\,$4m^2_{\pi^+}$ then $\rho_{\pi^+}(s)$\,$\to
$\,$i|\rho_{\pi^+}(s)|$ and
\begin{equation} \Pi^{\pi\pi}_{\sigma}(s)=-\frac{3}{2}\frac{g^2_{\sigma
\pi^+\pi^-}}{16\pi}|\rho_{\pi^+}(s)|\left[1-\frac{2}{\pi}\arctan
|\rho_{\pi^+}(s)|\right].\label{Pi-sigma2}\end{equation}  The
$\sigma\to\pi\pi$ decay width is
\begin{equation}\Gamma_{\sigma\to\pi\pi}(s)=\frac{1}{\sqrt{s}}
\mbox{Im}\Pi^{\pi\pi}_{\sigma}(s)=\frac{3}{2}\frac{g^2_{\sigma
\pi^+\pi^-}}{16\pi}\frac{\rho_{\pi^+}(s)}{\sqrt{s}}\,.
\label{G-sigma-pipi}\end{equation}

\begin{figure}
\includegraphics[height=12.5pc]{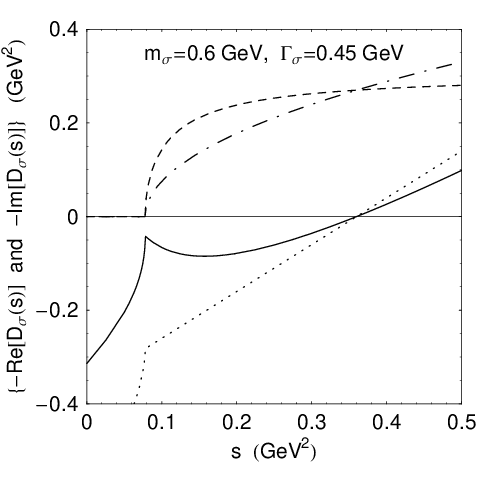}  % 10.2pc
\caption{{\footnotesize Demonstration of the finite width correction
with an example of the single $\sigma$ resonance. The curves are
described in the text.}}\label{Casp}\end{figure}

The function $\mbox{Re}[\Pi^{\pi\pi}_{\sigma}(m^2_{\sigma})$\,$-$\,$
\Pi^{\pi\pi}_{\sigma}(s)]$  in the denominator of Eq.
(\ref{Prop-sigma}) is the correction for the finite width of the
resonance. In Fig. \ref{Casp} the real and imaginary parts of the
inverse propagator $D_\sigma(s)$ (taken with the sign minus) are
shown by the solid and dashed curves in the case of the resonance
with the mass $m_\sigma$\,=\,0.6 GeV and the width
$\Gamma_\sigma$\,=\,$\Gamma_{\sigma\to\pi\pi}(m^2_\sigma) $\,=\,0.45
GeV. As may be inferred from this figure, $\mbox{Re}[D_\sigma(s)]$
can be close to 0 at $s$\,=\,$4m^2_{\pi^+}$ due to the correction
for the finite width in the case of the large one. Then this results
in the threshold cusp in the amplitudes proportional to
$|1/D_\sigma(s)|$.\,\footnote{The references to papers, in which the
the finite width corrections and the analytic properties of the
propagators of the realistic $f_0(980)$, $a_0(980)$, and
$\sigma(600)$ resonances have been investigated, are pointed out in
Section 2. In connection with the $\gamma\gamma$\,$\to$\,$\pi^0\eta$
and $\gamma\gamma$\,$\to$\,$\pi\pi$ reactions these corrections are
discussed also in the papers \cite{AS88,AS05}.} For reference, in
Fig. \ref{Casp} the  real and imaginary parts of the inverse
propagator $D_\sigma(s)$\,=\,$m^2_{\sigma}$\,$-$\,$s$\,$-$\,$
im_\sigma\Gamma_\sigma \sqrt{(s-4m^2_{\pi^+})/(m^2_{\sigma}
-4m^2_{\pi^+})}$ without the correction for the finite width
\cite{Fl76} (also taken with the sign minus) are shown by the dotted
and dot-dashed curves, respectively, at the same values of
$m_\sigma$ and $\Gamma_\sigma$.

\vspace{0.2cm}\noindent{\bf \boldmath 8.2. $\gamma\gamma\to\pi^0
\eta$}

\noindent The polarization operators of the $a_0$ resonance
$\Pi^{ab}_{a_0}(s)$ ($ab$\,=\,$\pi\eta$, $K^+K^-$, $K^0\bar K^0$,
$\pi\eta'$), introduced in Section 4 (see the paragraph with Eqs.
(\ref{Tres}) and (\ref{Mdirpieta})), have the following form: for
$s$\,$\geq$\,$m_{ab}^{(+)\,2}$ ($m_{ab}^{(\pm)}$\,=\,$m_b\pm m_a$,
$m_b\geq m_a$)
\begin{eqnarray}\label{Pi-ab-a01}\Pi^{ab}_{a_0}(s)=\frac{g^2_{a_0\to
ab}} {16\pi}\left[\frac{m_{ab}^{(+)}m_{ab}^{(-)}}{\pi
s}\ln\frac{m_a}{m_b}+\rho_{ab}(s)\times\right.\ \nonumber\\
\left.\times\left(i-\frac{1}{\pi}\,\ln\frac{
\sqrt{s-m_{ab}^{(-)\,2}}+\sqrt{s-m_{ab}^{(+)\,2}}}
{\sqrt{s-m_{ab}^{(-)\,2}}-\sqrt{s-m_{ab}^{(+)\,2}}} \right)\right],
\end{eqnarray} where $\rho_{ab}(s)$\,=\,$\sqrt{s-m_{ab}^{(+)\,2}}\sqrt{
s-m_{ab}^{(-)\,2}}\Big/s$, for
$m_{ab}^{(-)\,2}$\,$<$\,$s$\,$<$\,$m_{ab}^{(+)\,2}$
\begin{eqnarray}\label{Pi-ab-a02}\Pi^{ab}_{a_0}(s)=\frac{g^2_{a_0\to ab}}
{16\pi}\left[\frac{m_{ab}^{(+)}m_{ab}^{(-)}}{\pi
s}\ln\frac{m_a}{m_b}-\right.\nonumber\\
\left.-\rho_{ab}(s)\left(1-\frac{2}{\pi}\arctan\frac{\sqrt{
m_{ab}^{(+)\,2}-s}}{\sqrt{s-m_{ab}^{(-)\,2}}}\right)\right],\end{eqnarray}
where $\rho_{ab}(s)$\,=\,$\sqrt{m_{ab}^{(+)\,2}-s}
\sqrt{s-m_{ab}^{(-)\,2}}\Big/s$, and for $s\leq m_{ab}^{(-)\,2}$
\begin{eqnarray}\label{Pi-ab-a03}\Pi^{ab}_{a_0}(s)=\frac{g^2_{a_0\to
ab}} {16\pi}\left[\frac{m_{ab}^{(+)}m_{ab}^{(-)}}{\pi
s}\ln\frac{m_a}{m_b}-\right.\ \ \nonumber\\
\left.-\rho_{ab}(s)\frac{1}{\pi}\,\ln\frac{
\sqrt{m_{ab}^{(+)\,2}-s}+\sqrt{m_{ab}^{(-)\,2}-s}}
{\sqrt{m_{ab}^{(+)\,2}-s}-\sqrt{m_{ab}^{(-)\,2}-s}}\right],\end{eqnarray}
where $\rho_{ab}(s)$\,=\,$\sqrt{m_{ab}^{(+)\,2}-s}\sqrt{
m_{ab}^{(-)\,2}-s}\Big/s$.

The triangle loop integral in Eq. (\ref{M0pieta}) is
\begin{equation}\widetilde{I}^{V}_{\pi\eta}(s)=\frac{s}{\pi}
\int\limits^\infty_{(m_\pi+m_\eta)^2}\frac{\rho_{\pi\eta}(s')
M^{\mbox{\scriptsize{Born}}\,V}_{00}(\gamma\gamma\to\pi^0\eta;s')}{s'
(s'-s-i\varepsilon)}ds'\,,\label{IVpieta}\end{equation} where
\begin{eqnarray} M^{\mbox{\scriptsize{Born}}\,V}_{00}(\gamma\gamma\to\pi^0
\eta;s)=\qquad & \\ \nonumber
=\frac{1}{2}\int\limits^1_{-1}M^{\mbox{\scriptsize{Born}}\,V}_0(\gamma
\gamma\to\pi^0\eta;s,\theta)d\cos\theta\,, && \label{MB00pieta}
\end{eqnarray} is the $S$ wave Born amplitude, and the amplitude
$M^{\mbox{\scriptsize{Born}}\,V}_0(\gamma \gamma\to\pi^0\eta;s,
\theta)$ is defined by Eq. (\ref{MBornV0}). The functions
$\widetilde{I}^V_{\pi^0\eta'}(s)$ and $\widetilde{I}^{K^*}_{K\bar
K}(s)$ in Eq. (\ref{M0pieta}) are calculated similarly and the
function $\widetilde{I}^{K^+}_{K^+K^-}(s;x_2)$ is calculated with
Eq. (\ref{IKKKff}) in Appendix 8.3.

For the background phase shifts we use the simplest
parametrizations, which are suitable in the physical region of the
$\gamma\gamma $\,$\to$\,$\pi^0\eta$ reaction:
\begin{eqnarray}
e^{i\delta^{bg}_{ab}(s)}=[(1+iF_{ab}(s))/(1-iF_{ab}(s))]^{1/2}, && \\
F_{\pi\eta}(s)=\frac{\sqrt{1-m^{(+)\,2}_{\pi\eta}/s}\left(c_0+c_1
\left(s-m^{(+)\,2}_{\pi\eta}\right)\right)}
{1+c_2\left(s-m^{(+)\,2}_{\pi\eta}\right)^2}, && \\
F_{K\bar K}(s)=f_{K\bar
K}\sqrt{s}\left(\rho_{K^+K^-}(s)+\rho_{K^0\bar K^0}(s)\right)/2,
&& \\
F_{\pi\eta'}(s)=f_{\pi\eta'}\sqrt{s-m^{(+)\,2}_{\pi\eta'}}.\qquad\qquad
&& \end{eqnarray}

The curves in Figs. \ref{FigC-gg-pieta}, \ref{FigAB-gg-pieta}, and
\ref{Fig-pieta-pieta} correspond to the following model parameters:
($m_{a_0},\ g_{a_0\pi\eta},\ g_{a_0K^+K^-},\
g_{a_0\pi\eta'}$)\,=\,(0.9845, 4.23, 3.79, $-2.13$) GeV;
($m_{a'_0},\ g_{a'_0\pi\eta},\ g_{a'_0K^+K^-},\
g_{a'_0\pi\eta'}$)\,=\,(1.4, 3.3, 0.28, 2.91) GeV;
($g_{a_0\gamma\gamma},\,g_{a_0\gamma\gamma}$)\,=\,(1.77,
$-11.5$)$\times10^{-3}$ GeV$^{-1}$; $C_{a_0a'_0}$\,=\,0.06 GeV$^2$,
$c_0$\,=\,$-0603$, $c_1$\,=\,$-6.48$ GeV$^{-2}$, $c_2$\,=\,0.121
GeV$^{-4}$; ($f_{K\bar K},\ f_{\pi\eta'}$)\,=\,($-0.37$, 0.28)
GeV$^{-1}$; ($m_{a_2},\ \Gamma^{\mbox{\scriptsize{tot}}}_{a_2}
$)\,=\,(1.322, 0.116) GeV; $\Gamma^{(0)}_{a_2\to\gamma\gamma}
$\,=\,1.053 keV, $r_{a_2}$\,=\,1.9 GeV$^{-1}$, $\theta_{P}
$\,=\,$-24^\circ$ (see Ref. \cite{AS10a} for details).

\vspace{0.2cm}\noindent{\bf \boldmath 8.3.  $\gamma\gamma\to K\bar
K$}

\noindent The Born amplitude of the reaction $\gamma\gamma$\,$\to
$\,$K^+K^-$, caused by the elementary one kaon exchange,
$M_\lambda^{\mbox{\scriptsize{Born}}\,K^+}(s,\theta)$ and
$M_{\lambda J}^{\mbox{\scriptsize{Born}}\,K^+}(s)$ result from the
corresponding $\gamma\gamma$\,$\to$\,$\pi^+\pi^-$ Born amplitudes
$M_\lambda^{\mbox{\scriptsize{Born}}\,\pi^+}(s,\theta)$ and
$M_{\lambda J}^{\mbox{\scriptsize{Born}}\,\pi^+}(s)$ by the
substitution of $m_{K^+}$ for $m_{\pi^+}$ and of $\rho_{K^+}(s)=
\sqrt{1-4m^2_{K^+}/s}$ for $\rho_{\pi^+}(s)$ in Eqs. (\ref{M0born}),
(\ref{M2born}), and (\ref{M00B})--(\ref{M22B}):
\begin{equation}
M_0^{\mbox{\scriptsize{Born}}\,K^+}(s,\theta)=\frac{4m^2_{K^+}}{s}\frac{
8\pi\alpha}{1-\rho^2_{K^+}(s)\cos^2\theta}\,,\label{M0bornK}\end{equation}
\begin{equation}
M_2^{\mbox{\scriptsize{Born}}\,K^+}(s,\theta)=\frac{8\pi\alpha\rho^2_{K^+}
(s)\sin^2\theta}{1-\rho^2_{K^+}(s)\cos^2\theta}\,,\label{M2bornK}
\end{equation} \begin{equation}
M^{\mbox{\scriptsize{Born}}\,K^+}_{00}(s)=4\pi\alpha\frac{1-\rho^2_{K^+}
(s)}{\rho_{K^+}(s)}\,\ln\frac{1+\rho_{K^+}(s)}{1-\rho_{K^+}(s)}\,,
\label{M00BK}\end{equation}
\begin{eqnarray}M_{02}^{\mbox{\scriptsize{Born}}\,K^+}(s)=4\pi\alpha\frac{1-
\rho^2_{K^+}(s)}{\rho^2_{K^+}(s)}\left[\frac{3-\rho^2_{K^+}(s)}{2\rho_{K^+}
(s)}\times\right. && \nonumber \\
\times\left.\ln\frac{1+\rho_{K^+}(s)}{1-\rho_{K^+}(s)} -3\right]\,,
\qquad\qquad && \label{M02BK}\end{eqnarray}
\begin{eqnarray}M_{22}^{\mbox{\scriptsize{Born}}\,K^+}(s)=4\pi\alpha\sqrt{
\frac{3}{2}}\left[\frac{(1-\rho^2_{K^+}(s))^2}{2\rho^3_{K^+}(s)}\ln\frac{1
+\rho_{K^+}(s)}{1-\rho_{K^+}(s)}-\right. && \nonumber \\
-\left.\frac{1}{\rho^2_{K^+}(s)}+ \frac{5}{3}\right]\,.
\qquad\qquad\qquad && \label{M22BK}\end{eqnarray}

The function $\widetilde{I}^{K^+}_{K^+K^-}(s)$ results from
$\widetilde{I}^{\pi^+}_{\pi^+\pi^-}(s)$ by the substitution in Eqs.
(\ref{Ipipi1}) and (\ref{Ipipi2}) of $m_{K^+}$ for $m_{\pi^+}$ and
of $\rho_{K^+}(s)$ for $\rho_{\pi^+}(s)$, and thus for
0\,$<$\,$s$\,$<$\,$4m^2_{K^+}$
\begin{equation}\widetilde{I}^{K^+}_{K^+K^-}(s)=8\alpha\left\{
\frac{m^2_{K^+}}{s}\left[\pi-2\arctan|\rho_{K^+}(s)|\right]^2-1
\right\}\label{IKKK1}\end{equation} and for $s\geq4m^2_{K^+}$
\begin{equation}\widetilde{I}^{K^+}_{K^+K^-}(s)=8\alpha \left\{
\frac{m^2_{K^+}}{s}\left[\pi+i\ln\frac{1+\rho_{K^+}(s)}
{1-\rho_{K^+}(s)}\right]^2-1\right\}.\label{IKKK2}\end{equation}

Taking account of the form factor
\begin{equation}
G_{K^+}(t,u)=\frac{1}{s}\left[\frac{m^2_{K^+}-t}{1-
(u-m^2_{K^+})/x^2_2}+\frac{m^2_{K^+}-u}{1-(t-m^2_{K^+})/x^2_2}\right]
\label{FF2-Poppe}\end{equation} (here
$t$\,=\,$m^2_{K^+}$\,$-$\,$s[1$\,$-$\,$\rho_{K^+}(s)\cos\theta]/2$
and $u$\,=\,$m^2_{K^+}$\,$-$\,$s[1$ $+$\,$\rho_{
K^+}(s)\cos\theta]/2$), the partial amplitudes $M_{\lambda
J}^{\mbox{\scriptsize{Born}}\,K^+}(s)$ are replaced by $M_{\lambda
J}^{\mbox{\scriptsize{Born}}\,K^+}(s;x_2)$. Substituting
$\rho_{K^+}(s)$ instead $\rho_{\pi^+}(s)$ and
$\rho_{K^+}(s;x_2)=\rho_{K^+}(s)/(1+ 2x^2_2/s)$ instead
$\rho_{\pi^+}(s;x_1)$ in Eqs. (\ref{MB0J})--(\ref{MB2Jff}), one gets
\begin{equation}M^{\mbox{\scriptsize{Born}}\,K^+}_{0J}(s)=
\frac{1-\rho^2_{K^+}(s)}{\rho_{K^+}(s)}F^{\mbox{\scriptsize{Born}}
\,K^+}_{0J}(\rho_{K^+}(s)),\label{MB0J-K0}\end{equation}
\begin{equation}M^{\mbox{\scriptsize{Born}}\,K^+}_{2J}(s)
=\rho_{K^+}(s)F^{\mbox{\scriptsize{Born}}\,K^+
}_{2J}(\rho_{K^+}(s)),\label{MB2J-K2}\end{equation}
\begin{eqnarray}M^{\mbox{\scriptsize{Born}}\,K^+}_{0J}(s;x_2)=\frac{
1-\rho^2_{K^+}(s)}{\rho_{K^+}(s)}\left[F^{\mbox{\scriptsize{Born}}\,K^+
}_{0J}(\rho_{K^+}(s))-\right. && \nonumber
\\ \left.-F^{\mbox{\scriptsize{Born}}\,K^+
}_{0J}(\rho_{K^+}(s;x_2))\right],\qquad\qquad
&&\label{MB0Jff-K0J}\end{eqnarray}
\begin{eqnarray}M^{\mbox{\scriptsize{Born}}\,K^+}_{2J}(s;x_2)=
\rho_{K^+}(s)\left[F^{\mbox{\scriptsize{Born}}\,K^+
}_{2J}(\rho_{K^+}(s))-\right. & \nonumber
\\ \left.-F^{\mbox{\scriptsize{Born}}\,K^+
}_{2J}(\rho_{K^+}(s;x_2))\right]. \qquad\qquad
&\label{MB2Jff-K2J}\end{eqnarray} Correspondingly, with taking into
account the form factor, the function
$\widetilde{I}^{K^+}_{K^+K^-}(s)$ is replaced by
\begin{equation}\widetilde{I}^{K^+}_{K^+K^-}(s;x_2)=\frac{s}{\pi}
\int\limits^\infty_{4m^2_{K^+}}\frac{\rho_{K^+}(s')M^{
\mbox{\scriptsize{Born}}\,K^+}_{00}(s';x_2)}{s'(s'-s-i\varepsilon)}ds'
.\label{IKKKff}\end{equation} Note that $0.68\times|
\widetilde{I}^{K^+}_{K^+K^-}(s)|^2$ coincides with
$|\widetilde{I}^{K^+}_{K^+K^-}(s;x_2)|^2$ within an accuracy better
than 3\% in the range 0.8 GeV\,$<$\,$ \sqrt{s}$\,$<$\,1.2 GeV at
$x_2$\,=\,1.75 GeV.

The $S$ wave amplitudes of the reactions $\gamma\gamma$\,$\to$\,$
K^+K^-$ and $\gamma\gamma$\,$\to$\,$K^0\bar K^0$, which we used for
estimates in the region of the $K\bar K$ thresholds, have the form
\begin{eqnarray}
& M_{00}(\gamma\gamma\to
K^+K^-;s)=M^{\mbox{\scriptsize{Born}}\,K^+}_{00}(s;x_2)+ & \nonumber\\
& +\widetilde{I}^{\pi+}_{\pi^+\pi^-}(s;x_1)T_{\pi^+\pi^-\to
K^+K^-}(s)+\widetilde{I}^{K^+}_{K^+K^-}(s;x_2)\times & \nonumber\\
& \times\,T_{K^+K^-\to K^+K^-}(s)+ M^{\mbox
{\scriptsize{direct}}}_{\mbox{\scriptsize{res}};+}(s), &
\label{MKPKM00}
\end{eqnarray}
\begin{eqnarray}
& M_{00}(\gamma\gamma\to K^0\bar K^0;s)= & \nonumber\\
& =\widetilde{I}^{\pi+}_{\pi^+\pi^-}(s;x_1)\,T_{\pi^+\pi^-\to K^0
\bar K^0}(s)+\widetilde{I}^{K^+}_{K^+K^-}(s;x_2)\times & \nonumber\\
& \times\,T_{K^+K^-\to K^0\bar K^0}(s)+ M^{\mbox
{\scriptsize{direct}}}_{\mbox{\scriptsize{res}};-}(s). &
\label{MK0K000}\end{eqnarray} The corresponding cross section are
\begin{equation}
\sigma_{00}(\gamma\gamma\to K^+K^-)=\frac{\rho_{K^+}(s)}{32\pi
s}|M_{00}(\gamma\gamma \to K^+K^-;s)|^2,\label{CSKPKM}\end{equation}
\begin{equation} \sigma_{00}(\gamma\gamma\to
K^0_SK^0_S)=\frac{\rho_{K^0}(s)}{64\pi s}|M_{00}(\gamma\gamma\to
K^0\bar K^0;s)|^2.\label{CSK0K0} \end{equation} The amplitudes of
the $\pi\pi$\,$\to$\,$K\bar K$ reactions, $T_{\pi^+\pi^-\to
K^+K^-}(s)$ =\,$ T_{\pi^+\pi^-\to K^0 \bar
K^0}(s)$\,=\,$T_{K^+K^-\to\pi^+\pi^-}(s)$, are defined by Eqs.
(\ref{TKKpipi}) and (\ref{TresKKpipi}). The $K^+K^-$\,$\to$\,$K^+
K^-$ and $K^+K^-$\,$\to$\,$K^0\bar K^0$ reaction amplitudes are
given by
\begin{equation}T_{K^+K^-\to K^+K^-}(s)=[t^0_0(s)+t^1_0(s)]/2,
\label{TKPKM}\end{equation} \begin{equation}T_{K^+K^-\to K^0\bar
K^0}(s)=[t^0_0(s)-t^1_0(s)]/2, \label{TK0K0}\end{equation} where
$t^I_0(s)$ are the $S$ wave $K\bar K$\,$\to$\,$K\bar K$ reaction
amplitudes with isospin $I$\,=\,0 and 1;
\begin{equation}t^0_0(s)=\frac{e^{2i\delta^{K\bar K}_{B}(s)}
-1}{2i\rho_{K^+}(s)}+e^{2i\delta^{K\bar K}_{B}(s)}T^{K\bar
K}_{\mbox{\scriptsize{res}};0}(s)\,,\label{tK00}\end{equation}

\begin{equation}t^1_0(s)=\frac{e^{2i\delta_{K\bar
K}^{bg}(s)}-1}{2i\rho_{K^+}(s)}+e^{2i\delta_{K\bar
K}^{bg}(s)}T^{K\bar K}_{\mbox{\scriptsize{res}};1} (s)\,,
\label{tK01}\end{equation} where $\delta^{K\bar K}_B(s)$ and
$\delta_{K\bar K}^{bg}(s)$ are the phases in the channels with
$I$\,=\,0 and 1, respectively.

\begin{equation} T^{K\bar
K}_{\mbox{\scriptsize{res}};0}(s)=\frac{g_{\sigma
K^+K^-}\overline{\Delta}^0_{f_0}(s)+g_{f_0K^+K^-}\overline{\Delta}^0_\sigma(s)}
{8\pi[D_\sigma(s)D_{f_0}(s)-\Pi^2_{f_0\sigma}(s)]}\,,\label{TresKKKK0}
\end{equation}

\begin{equation} T^{K\bar K}_{\mbox{\scriptsize{res}};1}(s)=\frac{g_{a_0
K^+K^-}\overline{\Delta}^1_{a'_0}(s)+g_{a'_0K^+K^-}\overline{\Delta}^1_{a_0}(s)}
{8\pi[D_{a_0}(s)D_{a'_0}(s)-\Pi^2_{a_0a'_0}(s)]}\,,\label{TresKKKK1}
\end{equation}
where\vspace{0.15cm}

$\overline{\Delta}^0_{f_0}(s)=D_{f_0}(s) g_{\sigma
K^+K^-}+\Pi_{f_0\sigma} (s)g_{f_0K^+K^-},$

$\overline{\Delta}^0_\sigma(s)=D_\sigma(s)g_{f_0K^+K^-}+
\Pi_{f_0\sigma}(s)g_{\sigma K^+K^-},$

$\overline{\Delta}^1_{a'_0}(s)=D_{a'_0}(s)g_{a_0
K^+K^-}+\Pi_{a_0a'_0} (s)g_{a'_0K^+K^-},$

$\overline{\Delta}^1_{a_0}(s)=D_{a_0}(s)g_{a'_0
K^+K^-}+\Pi_{a_0a'_0} (s)g_{a_0K^+K^-}.$\\[0.2cm]
The amplitudes of the direct resonance transitions into photons are
given by
\begin{eqnarray} M^{\mbox{\scriptsize{direct}}}_{
\mbox{\scriptsize{res}};\pm}(s)=s\,e^{i\delta^{K\bar
K}_B(s)}\,\frac{g^{(0)}_{\sigma\gamma\gamma}
\overline{\Delta}^0_{f_0}(s)+g^{(0)}_{f_0\gamma\gamma}
\overline{\Delta}^0_\sigma(s)}{D_\sigma(s)D_{f_0}(s)-\Pi^2_{f_0\sigma}(s)}
\,&& \nonumber
\\ \pm\,s\,e^{i\delta_{K\bar
K}^{bg}(s)}\,\frac{g^{(0)}_{a_0\gamma\gamma}
\overline{\Delta}^1_{a'_0}(s)+g^{(0)}_{a'_0\gamma\gamma}
\overline{\Delta}^1_{a_0}(s)}{D_{a_0}(s)D_{a'_0}(s)-\Pi^2_{a_0a'_0}(s)}\,.\
\ && \end{eqnarray} \vspace{0.2cm}

%-----------------------------------------------------------------------------------------
%\newpage

\end{document}